\documentclass[12pt]{JHEP3}
\usepackage{graphicx}
\usepackage{amsmath,epsfig}
\usepackage{amssymb,amsfonts}
\usepackage{latexsym}
\usepackage{slashed}
\usepackage{epsfig}
\usepackage{xcolor}
\usepackage{subfigure}
\usepackage{subfig}
\usepackage{cancel}

\usepackage[vcentermath,enableskew]{youngtab}
\def\hri#1#2{\href{http://arxiv.org/abs/#1}{[ArXiv:#1]#2}}
\def\hre#1#2{\href{http://arxiv.org/abs/#1/#2}{[ArXiv:#1/#2]}}

\def\hrj#1#2{\href{www.doi.org/#1}{#2}}
\relax
\renewcommand{\ap}{{\alpha^\prime}}
\renewcommand{\theequation}{\arabic{section}.\arabic{equation}}

\def\be{\begin{equation}}
\def\ee{\end{equation}}
\def\qslash{{q\hspace{-5.3pt}/}}
\def\pslash{{p\hspace{-5.3pt}/}}

\def\lslash{{\ell\hspace{-5.3pt}/}}

\newcommand{\bear}{\begin{eqnarray}}
\newcommand{\bea}{\begin{eqnarray}}
\newcommand{\eear}{\end{eqnarray}}
\newcommand{\eea}{\end{eqnarray}}
\def\hre#1#2{\href{http://arxiv.org/abs/#1/#2}{[ArXiv:#1/#2]}}

\newbox\pippobox

\def\a{\alpha}		\def\b{\beta}		\def\g{\gamma}		
\def\e{\varepsilon}			\def\h{\eta}			\def\q{\theta}
				\def\l{\lambda}		\def\m{\mu}
\def\n{\nu}									\def\r{\rho}	
\def\s{\sigma}

		\def\D{\Delta}				
						\def\S{\Sigma}		
			\def\Y{\Psi}

\renewcommand{\b}[1]{\textbf{#1}}

\def\II{\relax{\rm I\kern-.18em I}}

\def\e{\epsilon}
\def\m{\mu}
\def\n{\nu}
\def\r{\rho}
\def\s{\sigma}
\def\pa{\partial}

\def\sp{\;\;\;,\;\;\;}

\def\a{\alpha}
\def\b{\beta}

\def\tr{\ensuremath{\mathrm{Tr}}}

\def\l{\lambda}
\def\g{\gamma}

\def\ie{{\it i.e.}}
\def\eg{{\it e.g.}}
\def\<{\big\langle}
\def\>{\big\rangle}

\def\calA{{\cal A}}

\def\calF{{\cal F}}

\def\II{{\cal I}}
\def\JJ{{\cal J}}

\def\WW{{\cal W}}

\def\nn{\nonumber}

\newcommand{\der}{{\partial}}

\renewcommand{\h}[1]{{\widehat{\rm #1}}}
\newcommand{\vis}[1]{{\rm{#1}}}
\def\aa{{\widehat{A}}}

\def\eqp{{\;\;.}}

\title{String (gravi)photons, ``dark brane photons'', holography and the hypercharge portal}

\author{\large P.~Anastasopoulos$^{1}$\footnote{pascal.anastasopoulos@univie.ac.at }, M.~Bianchi$^{2}$\footnote{massimo.bianchi@roma2.infn.it}, D.~Consoli$^{1}$\footnote{dario.consoli@univie.ac.at}, E.~Kiritsis$^{3}$\\
 ~\\
{$^1$ \href{https://mathphys.univie.ac.at}{Mathematical Physics Group, University of Vienna}, \\
Boltzmanngasse 5, 1090 Vienna, Austria}\\
 ~\\
{$^2$ \href{
https://www-en.fisica.uniroma2.it/sezioni/research/research-areas/theoretical-physics-of-fundamental-interactions/}{Dipartimento  di  Fisica,  Universit\`a  di  Roma  ``Tor  Vergata'' \\  \& \  I.N.F.N.  Sezione  di  Roma  ``Tor  Vergata''}, \\
Via della Ricerca Scientifica, 00133 Roma, Italy}\\
 ~\\
{$^3$ \href{http://hep.physics.uoc.gr}{Crete Center for Theoretical Physics}, Institute for Theoretical and Computational Physics,
Department of Physics,
University of Crete, 70013, Heraklion, Greece}\\
  \centerline{and}\\
\href{http://www.apc.univ-paris7.fr}{Universite de Paris, CNRS, Astroparticule et Cosmologie,  F-75006 Paris, France}\\
}

\preprint{UWThPh 2020-15\\ CCTP-2020-7\\ITCP-IPP-2020/7}

\abstract{The mixing of graviphotons and dark brane photons to the Standard Model hypercharge is analyzed in full generality, in weakly-coupled string theory. Both the direct mixing as well as effective terms that provide mixing after inclusion of SM corrections are estimated to lowest order. The results are compared with Effective Field Theory (EFT) couplings, originating in a hidden large-N theory coupled to the SM where the dark photons are composite. The string theory mixing terms are typically subleading compared with the generic EFT couplings. The case where the hidden theory is a holographic theory is also analyzed, providing also suppressed mixing terms to the SM hypercharge.}

\begin{document}

\section{Introduction}

Vector particles lying outside the Standard Model (SM) of particle interactions have been entertained since the origin of the field, for various reasons. The motivations evolved over the years, following the deeper and deeper understanding of particle interactions inside the SM.
In particle physics, such particles come under the name of $Z'$ and appear in many (unified or not) extensions of the SM \cite{Leike}-\cite{Lang}.
They also appear in many effective actions descending from string theory, especially the heterotic string. The reasons are as for their QFT counterparts: they are low-energy remnants of a larger gauge group at high energy, being it unified or not.

There is however, a second type of such U(1) vectors appearing in string theory vacua with open and closed unoriented strings, \cite{Bianchi:1990yu}-\cite{Angelantonj:1996uy}, which are known as type II `orientifolds'. Such string theories allow local brane constructions that the SM can be embedded in, \cite{AKT,Ib}, and have been analyzed extensively.
The new type of U(1)'s appearing, are known as anomalous U(1)'s and have very different properties from other low-energy U(1)'s that appear in string theory, \cite{rev}. In particular, they are always massive\footnote{They become massless only in the trivial limit $g_s=0$.}, but their masses are typically much lower than the ones of other stringy states as they are (parametrically) a one-loop effect due to anomalies \cite{AKR}. Such U(1) vector bosons have been advocated to be interesting low-energy extensions of the SM in orientifold constructions, \cite{AKT,muon}, and have been advocated to be part of dark matter, \cite{AL}.

It was shown that at least two anomalous U(1)'s and typically three, are part of the SM brane stack in its orientifold realizations, \cite{ADKS,Bachas}, and such symmetries are typically  global symmetries of the SM that are gauged\footnote{Even U(1) symmetries that have zero charge traces, $Tr[Q]=0$ can have massive gauge bosons, as the mass is also affected by higher-dimensional anomalies, \cite{AKR}. For example, if in higher-dimensional decompactification limits, the theory is anomalous in the higher dimensions, it will have a non-zero mass in four-dimensions, \cite{anasta}.}, \cite{IQ}.
The low-energy extensions of the SM in this case are richer than typical Z' extensions \cite{irges}-\cite{AbelGoodsell}, due to the non-trivial anomaly structure of such symmetries and the appearance of generalized Chern-Simons terms in the effective field theory (EFT) \cite{bianchi}.

In all the above cases, the U(1)'s involved had weak but not ultra-weak coupling, and masses that may be light, but not too light compared to SM masses.
Cosmological considerations however motivated the study of extra U(1) vectors that may have very weak couplings, down to gravitational strengths, as well as light masses, compared with SM mass scales, \cite{Fayet}-\cite{Fabbri}.
These may play the role of dark matter (rare) or more often mediate dark matter interactions or interactions between the dark matter and the SM.
They may also be around in a theory and then they may also create phenomenological problems. It is customary to call such vectors, ``dark photons", and it this name we shall also use in this paper to mean a U(1) outside the SM .
Note, that as we shall explain later on, we could have the SM containing a number of anomalous U(1)s, with couplings that are not tiny, but these will be considered as parts of the SM and will not be call ``dark photons". In orientifold realizations of the SM, the distinction can be made sharp: U(1)'s associated to the SM are those localized on the branes of the SM stack.
Eventually, we shall distinguish the ``dark" U(1)s into two groups in string theory. Those that arise in the closed string sector, we shall call them collectively {\em ``graviphotons"}. Those arising from branes, other than the brane stack we shall call ``{\em dark brane photons}".

Part of the definition of dark photons is that in the beginning (i.e. before taking into account possible mixings and other couplings), SM particles are not charged under such dark U(1)'s. Therefore, the dominant coupling of such (dark) U(1) vectors to the SM is via the so-called hypercharge portal, i.e. a dimension four coupling mixing the
field strength of the dark photon to the hypercharge field strength, $F_{Y}^{\mu\nu}\hat F_{\mu\nu}$.
Such a coupling is gauge invariant and therefore allowed by symmetries.

As was shown early on, $\cite{h}$, the presence of such a coupling, even with a tiny coefficient, is modifying the EFT significantly and introduces important changes in electromagnetism that are strongly constrained by experiment \cite{ship}.
Moreover, such a coupling is dimensionless, and it can therefore receive quantum contributions from heavy states that are enhanced logarithmically.
Therefore, such a coupling is a very sensitive SM portal, that can easily destroy the viability of a SM extension.

Each massive state, charged under both hypercharge and the dark photon, contributes to the mixing, when both couplings are weak, as
\be
L_\textup{mix}=-{Q Y\over 24\pi^2}\log{M^2\over \Lambda^2}~~~
F_Y^{\mu\nu}\hat F_{\mu\nu}
\label{1_1} \ee
where $Y$ is the hypercharge of the massive particle, $Q$ is its charge under the dark photon, $M$ is the mass of the particle and $\Lambda$ the cutoff of the theory.
However, if the dark photon theory is strongly coupled, this calculation is complicated by the fact that the coupling is strong, and that magnetic monopole states also contribute to the mixing. In such cases, the calculation has only been done in QFTs with approximate N=2 supersymmetry \cite{Heck}.

In a renormalizable QFT, the mixing of dark photons to the hypercharge, as well as their masses can be theoretically adjusted more or less at will, as in most cases, the photons are treated as fundamental. This is what the majority of dark photon models, i.e. renormalizable QFTs with spontaneously broken dark U(1), do.
The experimental constraints are important, but allow many orders of magnitude for the relevant parameters, (see for example \cite{ship}, figure 2.6, on page 26).
One may bring-in  swampland constraints,  and a set is discussed in \cite{Reece}. There are many loopholes for  these arguments, that are already mentioned in this paper. However, even taking into account such constraints:

a) There is no known swampland constraint on the mixing coupling other than the destabilization criterion, which is weak as a constraint.

b) The other swampland constraints, \cite{Reece} allow many orders of magnitude of the parameter space that is not constrained by current data, \cite{ship}.

As mentioned,  in most of the literature, dark photons are treated as fundamental and therefore their parameters are theoretically unconstrained. Composite vectors have been discussed in the literature, and an extensive list of references appears in the introduction of \cite{u1} starting with the work of Bj\"orken in the 60's. Most of this past work is theoretical and based of variants of the Nambu-Jona Lasinio four-Fermi interaction.
They have been extensively discussed also in the context of technicolor models and the composite Higgs context, \cite{Contino}.
Although, composite dark matter models are abundant today, there is only a single attempt to use a composite U(1) from the dark sector that will be mixing with the hypercharge in \cite{Barbieri}. Moreover, the mixing term is assumed rather than calculated or estimated from the underlying theory.
There are also two works, in the Randall-Sundrum context that can be related to composite model via a (crude) holographic correspondence, \cite{Gh,Morriss}.

In string theory, however, the freedom of choosing almost at will couplings and mixings seems limited, as among other things, there are typically infinite towers of states contributing to the mixing. In the heterotic string, such mixing has been analyzed qualitatively in \cite{Dienes} and calculated for simple orbifolds in \cite{Goodsell}.

In the other model building arena, orientifolds, and for U(1)'s arising on branes, (``dark brane U(1)'s" in our terminology), the bifundamentals contributing to the mixing are the states associated with strings that stretch between the respective D-branes. The leading one-loop mixing contributions come from the annulus diagrams of such strings and their orientation reversal cousins, \cite{Abel,Wit}. In particular, the kinetic mixing of abelian gauge bosons in toroidal orbifolds with D3-branes at orbifold singularities, was studied in \cite{Wit}, and shown to depend only on the closed-string complex structure moduli rather than on the K\"ahler moduli.

In \cite{Camara}, different mechanisms for RR graviphotons mixing with D-brane vector bosons were studied with a focus on Type IIA unoriented models on CY manifolds and their M-theory lift. Geometric criteria were derived for the mixing to occur either via standard kinetic mixing or via mass terms induced by St\"uckelberg couplings.
In \cite{Marchesano}, the kinetic mixing between brane U(1)'s was considered and studied. This type of mixing arises in part as a byproduct of RR (closed string) U(1) mixing with D-brane U(1)'s.

\subsection{The field theory setup\label{out1}}

In almost all descriptions of dark photons\footnote{The only exception to our knowledge is \cite{Barbieri} and indirectly, \cite{Gh,Morriss}.}, elementary, weakly-coupled vectors are used. There are, however, other contexts in which dark photons may be composite vectors, where the fundamental underlying particles are strongly coupled. This is a generic avatar of the framework  that provides emergent gravity coupled to the SM \cite{1,grav}.

The relevant framework that we shall therefore assume in this paper is
\be
{\rm (hidden~~ large-N~~ gauge~~ theory)} \times {\rm (messengers)}\times {\rm \widetilde {SM}}
\label{i2}\ee
where the  hidden large-$N$ gauge theory is assumed to be at strong coupling.
The messengers are bi-fundamentals under the hidden gauge theory gauge group and the SM gauge group. They are massive with mass $M$ that is assumed to be much larger than all the scales of the SM.
Finally ${\rm \tilde {SM}}$  is a UV extension of the SM.

The complete theory is a standard four-dimensional quantum field theory (QFT) defined on a flat Minkowski background metric and it is assumed to be UV-complete\footnote{This means that the UV limit of that theory is a strongly coupled CFT.}.
We shall not fix the details of the hidden theory, neither that of the messengers, nor the particular UV Extension of the SM, as our goal will be to study general properties of this framework.

There is another picture of the same theory that is valid well below the messenger mass M. In this regime, we can integrate out the messengers and obtain an effective theory with the structure

\be
{\rm (hidden~~ large-N~~ gauge~~ theory)} \times {\rm  \widetilde {SM}}
\label{i1}
\ee
where now the coupling between the hidden theory and the SM is given by products of gauge invariant operators of the two theories. Almost all such couplings are ``irrelevant" in the IR, but are crucial in communicating the low-energy physics of the hidden theory and coupling it to the low-energy SM operators.

Various aspects of this framework and its IR avatars, where studied in \cite{1,grav,axion,u1}.
In such a theory, an emergent ``graviton" appears as a composite of the hidden theory energy-momentum tensor and couples to the SM \cite{1}.
In the generic case, this composite ``graviton" looks nowhere near what we usually call a graviton. However, if the hidden theory is holographic, then this graviton is a massless graviton in a higher dimension, and in a non-trivial background as we learn from the AdS/CFT correspondence.
The emerging gravitational interaction in this case has the standard features of gravity (it is weakly coupled, has diffeomorphism invariance and is accompanied by a small number of additional fields, the analogue of graviphotons and light/massless scalars).

When the gravitational interaction is reduced back to four-dimensions, the graviton acquires massive features due to the non-trivial gravitational background in the higher dimensions.

A holographic dual picture of the setup in (\ref{i1}) is given by a bulk gravitational theory living in more than four dimensions\footnote{The number of extra dimensions as understood in holography is related to the adjoint degrees of freedom in the hidden holographic theory.}, dual to the holographic hidden QFT and coupled to a four-dimensional ``SM" brane, embedded in the bulk geometry\footnote{The bulk geometry is non-trivial and asymptotically AdS.} and coupled to it.
It resembles the Randall-Sundrum picture, \cite{RS} but with important differences.  The main difference from RS realizations of this idea is that there is no UV cutoff for the bulk description of the hidden holographic QFT and no RS $Z_2$ boundary conditions for the brane.
There is however a  UV (radial) cutoff on the bulk description at the position of the messenger mass scale $M$, \cite{grav}.

In this gravitational description of the combined theory, one can accommodate the self-tuning of the SM (brane) cosmological constant, \cite{self}, and calculate the effective four-dimensional graviton mass as a function of the theory parameters. The presence of large $N$ in the holographic theory is important, as it implies weak effective coupling between the composites, in particular the graviton. It is also responsible for a suppression of the graviton effective four-dimensional mass \cite{self}.

This framework produces two other classes of light emergent (composite) particles. The first is a universal axion, emerging from the instanton density of the hidden QFT \cite{axion}. It couples to the SM instanton densities and has mass contributions both from the hidden theory and the SM, unlike fundamental axion fields. It also has a compositeness scale above which its interactions are non-local but controllable. Such a scenario can also provide a new portal to dark matter, as argued in  \cite{Anastasopoulos:2020gbu}.

Another avatar of this framework  are emergent (composite)  vectors, originating in the hidden theory, that eventually couple to the SM, \cite{u1}.
Such vectors are associated with exact global symmetries of the hidden theory, and the composite vector is generated from the hidden conserved U(1) current.
Generically, such a composite vector is weakly-coupled and massive but as in the case of gravitons and axions, the mass can vary substantially.
Like the axion case, in certain energy regimes, the compositeness will become visible and the interactions become non-local.

\subsection{Results and outlook\label{out}}
 In this paper, we focus on the physics of emergent (composite) vectors arising in the framework described in (\ref{i2}) and (\ref{i1}).
A general study of emergent vectors in \cite{u1} has dealt with theoretical issues.
In this paper, we study the most important coupling of emergent dark photons to the SM\footnote{The associated couplings of the graviton in the same context were studied in \cite{grav} and those of emergent axions in \cite{axion}.} in the absence of minimal couplings  with SM fields. This is known as the hypercharge mixing portal\footnote{We therefore assume that the SM fields, in the standard frame (before mixing) are not charged under the emergent U(1).}.
We also remind the reader that the four-dimensional QFTs in (\ref{i2}) and (\ref{i1}) are non-gravitational and defined on a flat space-time.

\begin{enumerate}

\item We treat the hidden  theory in (\ref{i1}) first as a large-N theory (with strong coupling) but not necessarily holographic.
We use effective field theory techniques to classify its exact global symmetries and then determine the possible couplings of an emergent vector originating in the hidden theory to SM operators. Such couplings are chosen so that  they contribute, once SM quantum corrections are included, to the hypercharge mixing with the emergent vector. To our knowledge, this analysis of the various sources of mixing has not been done before, as almost invariably the dark sector is considered at weak coupling and the dark photon elementary\footnote{The only exception being \cite{Barbieri}.}.

What we are interested in, in this context, is the large-N dependence of the induced mixing, and this is determined from generic EFT arguments.

The classification of the global symmetries of the hidden theory and the estimates of the couplings can be found  in subsections \ref{s21}-\ref{s25}.

\item In section \ref{s26}, we then treat the hidden theory as a holographic theory, by using the gravitational description of a (generic) holographic bulk that describes the hidden theory, coupled to a brane that describes the SM.
    The bulk is always higher-dimensional, and we consider the concrete example of a five-dimensional bulk. The bulk geometry is now non-trivial and asymptotically AdS. This is dual to the flat space QFT in (\ref{i1})

    The reason this analysis is done separately, is because in the holographic case, there are two important differences with the generic large N case. First, the vector is higher dimensional, and second, SM quantum corrections generate a DGP-like effect due to a localized kinetic term generated on the brane.

To our knowledge, such a setup was never before analyzed in relation to dark vectors (and probably vectors in general).

\item
The third framework,  analyzed in section \ref{stringsetup}, is that of standard perturbative string theory around four-dimensional flat space times some internal compactification manifold. We analyze the couplings of graviphotons and other brane photons to the hypercharge on a putative SM-stack of D-branes.
This is a priori independent from the previous two items 1 and 2 above.

In this context, we shall study various sources of dark U(1)s and their mixing with the hypercharge of the SM-stack.
There are several ways that our analysis goes beyond the state of the art on this topic. In string theory, we consider the couplings of RR U(1)'s and their mixing to branes U(1)'s also in the presence of RR fluxes, going therefore beyond the analysis of \cite{Camara},\cite{Marchesano}.
We also consider possible NSNS U(1)s that may arise in non-CY compactifications. On the other hand, our analysis is perturbative in the string coupling, as we would like eventually to  compare via the map to holographic theories.

The second reason that we do this analysis, is to qualitatively compare the dark-photon/hypercharge mixing in the  string theory framework, with the field theory framework described in items 1 and 2 above.
The rational for doing this is the string-theory/gauge theory correspondence.

The reason that such a comparison is interesting, is because the AdS/CFT correspondence indicates that weakly-coupled string theories are dual to strongly coupled, large N (and holographic) QFTs.
Therefore, we expect to learn about strongly coupled QFTs from weakly coupled string theory and about strongly coupled string theory from weakly coupled QFTs.

The vectors emerging from global symmetries of the hidden QFT, are similar to graviphotons and other bulk U(1)'s like the RR U(1)'s of closed string theory.
Vectors emerging from flavour sectors, like the messenger sector, are qualitatively similar to  U(1)'s appearing on D-branes distinct from the SM D-brane stack in orientifold constructions.
Therefore in string theory  there will be two types of ``dark" U(1)s. The first type involves U(1)s originating in the closed string sector, that we shall call {\em graviphotons}. The second involves U(1)'s that emerge on other brane stacks, that have weak couplings because they wrap large manifolds. These are the {\em dark brane photons} in our nomenclature.

This holographic map, allows a correspondence between calculations in a strongly coupled large $N$-theory and related calculations in string theory.

Of course, the string theory calculations are done in string theory around flat space, while the string theory dual of the field theory framework in (\ref{i1}) is expected to be in a non-trivial asymptotically AdS gravitational background. In particular, we consider the effect of bulk fluxes in the RR and NSNS sector at linear order. This represents a first step towards a full fledged computation in flux vacua that is beyond the scope of the present investigation, due to the lack of viable world-sheet formulations.

\end{enumerate}

On the field theory side, we analyze not only direct kinetic mixing of a dark photon, but other emergent couplings to the SM fields, which can generate kinetic mixing once SM quantum corrections are included. The most important such couplings were classified and studied.
This was done when the hidden theory is a generic large $N$ theory, using Effective Field Theory (EFT).
A similar analysis is also done in the particular case where the hidden theory is a holographic QFT, by using the holographic dual picture, where the SM is represented as a brane inserted in the holographic bulk.

Our analysis will not be confined to a concrete model or set of models. It aspires to give general, broad-brush estimates that are generically valid.

We shall consistently use the following terms for U(1)'s in the rest of the paper.
\begin{itemize}

\item {\em Dark photon}: any U(1) beyond the SM model and not belonging to the SM stack. A detailed definition of the SM stack involves the minimal set of gauge groups, so that all SM fields can be written as bifundamentals. This is explicitly realized in SM implementations in the context of orientifolds, \cite{ADKS}. It is also necessary in coupling the SM to a hidden holographic theory, \cite{1}. It can be shown that at least two extra anomalous U(1)'s are necessary for this realization, \cite{ADKS}.

\item {\em Anomalous U(1)}: any U(1) beyond those of the SM model, but belonging to the SM stack. Such U(1)s are broken and massive as usual in string theory.

\item {\em Graviphoton}: In string theory, any U(1) arising in the closed string sector of type II orientifolds. In a holographic (hidden) theory, it is a global U(1) arising in the adjoint sector.

\item {\em Dark brane photon}: In string theory, any U(1) arising from D-branes beyond the SM stack. In a holographic (hidden) theory, a global U(1) arising in the fundamental (messenger) sector.

\end{itemize}

The holographic and EFT analysis is performed in section \ref{holosetup}, and is complemented by appendices \ref{mixing}, \ref{ggc} and \ref{gau12}.
What we find is as follows.

\begin{itemize}

\item Based on general EFT principles and the large $N$ expansion, we establish that the leading mixing term of a dark photon to the hypercharge, if non-zero, is generically suppressed by a single power of $1/N$.
  This estimate is non-perturbative and is valid in the absence of bi-charged particles under both U(1)'s in question. It also includes
the quantum corrections of the SM.

\item If the hidden theory is also holographic, then the mixing intertwines non-trivially with the localized kinetic term and the mixing term of the bulk graviphoton.
  The interaction among  hypercharges is modified. The interaction between bulk charges and hypercharges is also modified.
Moreover, in this case, there is an extra suppression of the effects of mixing, as both the strength of the dark photon interaction as well as its mixing are of order $1/N$.

In \cite{u1}, it was shown that the interaction between bulk charges mediated by the emergent vector, is five-dimensional at intermediate distances but becomes four-dimensional at short and long distances. Moreover, at long distances it is massive.
Here, because of the mixing, the same applies to the interaction between bulk charges and hypercharges, as well as among hypercharges alone.

\end{itemize}

We then study similar effects in weakly-coupled type II string theory around flat space, involving the coupling of graviphotons or dark brane photons to the hypercharge and other SM operators on the SM stack of branes, in section \ref{stringsetup} and appendix \ref{Mixing_string}. We finally compare the string theory and QFT results in section \ref{comparison}. The summary of this comparison is presented in table \ref{tin} below.
What we find is as follows.

\begin{enumerate}

\begin{table}[h!]
\begin{tabular}{ccccc}
{\bf EFT}& {\bf EFT } & {\bf graviphoton} ~~& {\bf graviphoton}
& {\bf dark photon} \\
{\bf coupling}&{\bf estimate} & {\bf \cancel{bulk fluxes}} & {\bf +bulk fluxes} &~~
 \\
~~~&~ & ~~& &~~
 \\
$F\hat F$& ${\cal O}\left({1\over N}\right)$& ${\cal O}\left(g_s^2\right)$
&${\cal O}(g_s^{3/2})$ &${\cal O}\left(g_s\right)$   \\
$\phi~F\hat F$& ${\cal O}\left({1\over N}\right)$& ${\cal O}\left(g_s\right)$
& &   \\
$DHDH^{\dagger}\hat F$& ${\cal O}\left({1\over N}\right)$& ${\cal O}\left(g_s^2\right)$
&${\cal O}\left(g_s^{2}\right)$ &${\cal O}(g_s^{3/2})$   \\
$ H H^{\dagger}F\hat F$& ${\cal O}\left({1\over N^{3/2}}\right)$& ${\cal O}(g_s^{5/2})$
&${\cal O}(g_s^{5/2})$ &${\cal O}\left(g_s^{ 2}\right)$   \\
$\bar \psi H\gamma^{\mu\nu}\psi\hat F_{\mu\nu}$& ${\cal O}\left({1\over N^{3/2}}\right)$& ${\cal O}(g_s^{3/2})$
&${\cal O}(g_s^{5/2})$ &${\cal O}\left(g_s^{ 2}\right)$   \\
$\bar \psi H\psi F\hat F_{\mu\nu}$& ${\cal O}\left({1\over N^{ 2}}\right)$& ${\cal O}\left(g_s^{2}\right)$
&${\cal O}\left(g_s^{3}\right)$ &${\cal O}(g_s^{5/2})$   \\
\end{tabular}
\caption{Comparison of the kinetic mixing contributions and other generating terms in large-$N$ EFT and weakly coupled string theory.
$F$ denotes the hypercharge field strength, $\hat F$ the dark photon field strength, $H$ is the Higgs doublet, $\phi$ is an adjoint scalar field on the SM stack and $\psi$ collectively denotes SM fermions. }
\label{tin}
\end{table}

\item In the large-$N$, EFT case, the direct mixing of the dark photon to the hypercharge is of order ${\cal O}(N^{-1})$.

  In string theory, when the dark photon is a graviphoton, such a mixing is further suppressed as ${\cal O}(g_s^2)$ in the absence of RR fluxes, and as ${\cal O}(g_s^{3/2})$ in the presence of bulk fluxes.
Only in a particular case, when the mixing term is multiplied by an adjoint scalar in the SM stack, is the EFT estimate matching the string theory estimate, ${\cal O}(g_s)$. However, as analyzed in section \ref{css}, the local phenomenology of the SM makes such a coupling impossible, as
such a scalar is beyond the strict spectrum of the SM. It may however exist in mild extensions of the SM model. Its presence is certainly connected to the presence of extra anomalous U(1)'s in the SM stack, \cite{ADKS,irges}.
 The phenomenological implications in such a case remain to be investigated.

In all other cases, such a mixing is at best ${\cal O}(g^{3/2}_s)\sim {\cal O}(N^{-{3/2}})$.

 The direct mixing term of a dark brane photon to the hypercharge is of order ${\cal O}\left({N^{-1}}\right)\sim {\cal O}(g_s)$ and appears at one loop mediated by bi-charged massive particles. This has been extensively studied by many authors, as mentioned earlier.

\item There are indirect couplings involving the dark photon that upon SM quantum corrections can generate a mixing term with the hypercharge.
  The leading such type of coupling is a dipole interaction of the Higgs with the dark photon. Such a term in EFT is generically of order ${\cal O}\left({N^{-1}}\right)$ and upon SM quantum corrections gives an order
${\cal O}\left({ N^{-1}}\right)$ mixing term between the extra photon and the hypercharge.

For graviphotons, this same coupling is subleading both in the absence and presence of bulk fluxes, scaling at best as ${\cal O}\left({ N^{-2}}\right)\sim {\cal O}(g^2_s)$

For dark brane photons, the same coupling can appear, but it is also subleading in $N$, $\sim {\cal O}(N^{-{3/2}})$ and is also proportional to two scalar vevs.

Therefore in all cases, the string theory estimates are subleading both to the EFT, and the leading ${\cal O}\left({N^{-1}}\right)$ estimate of the direct mixing case.

\item A coupling of the mixing term $F\hat F$ to the Higgs vev squared and its higher derivative avatars, where $\hat F$ is the field strength of a dark photon, is of order $\sim {\cal O}(N^{-{3/2}})$ in EFT. Via the SM quantum corrections, it generates a mixing term $F\hat F$ with the hypercharge that is of order $\sim {\cal O}(N^{-{3/2}})$.

For graviphotons, both without and with bulk fluxes, it is of order ${\cal O}(g^{5/2}_s)\sim {\cal O}(N^{-{5/2}})$ and therefore subleading compared to the EFT estimate.

If $\hat F$ is a dark brane photon, a similar coupling is also ${\cal O}(g_s^2)\sim {\cal O}\left({1/N^2}\right)$, and therefore again subleading compared to EFT.

\item

The dipole coupling of the dark photon to the fermions of the SM, $\hat F_{\mu\nu}\bar \psi~H\gamma^{\mu\nu}\psi$, (combined
with the Higgs for gauge invariance) is of order ${\cal O}\left(N^{-{3/2}}\right)$ in EFT. Upon SM quantum corrections, it generates a mixing term $F\hat F$ that is of a similar order, ${\cal O}\left(N^{-{3/2}}\right)$.

 For string theory graviphotons, the same coupling is of the same order, $\sim {\cal O}(g_{s}^{3/2})$ in the absence of bulk fluxes and subleading, $\sim {\cal O}(g_{s}^{5/2})$ in the presence of bulk fluxes.

For dark brane photons, the same coupling can appear, but it is also subleading in $N$, $\sim {\cal O}(N^{-{2}})$.

 \item A coupling of the mixing term $F\hat F$ to the fermion mass term $\bar\psi H \psi$, is $\sim {\cal O}(N^{-{ 2}})$ in EFT.
   Upon SM quantum corrections, it generates a mixing term, $F\hat F$ to the hypercharge that is of order $\sim {\cal O}(N^{-{ 2}})$.

   When $\hat F$ is the field strength of a graviphoton, the same coupling appears at the same order $\sim {\cal O}(g_s^2)$ in the absence of bulk RR fluxes. When there are bulk RR fluxes, the same coupling appears at order $\sim {\cal O}(g_s^3)$.

If $\hat F$ is a dark brane photon, a similar coupling appears at order $\sim {\cal O}(N^{-{5/2}})$.

\item If instead of the hypercharge, we consider couplings to the anomalous U(1)'s accompanying the SM, the same estimates as above apply for the mixing to a dark photon.

  \end{enumerate}

The weakly-coupled string theory results can be translated via the AdS/CFT correspondence to estimates for the coupling of the hypercharge of the SM to a dark photon arising on a holographic theory, along the lines described in \cite{1,u1} with $g_s\sim N^{-1}$.

What we have learned is that the mixing terms in such a case are generically subleading to those predicted by standard EFT. More precisely, the mixing term of a graviphoton to the hypercharge is of order ${\cal O}\left({1/N}\right)$ only in the presence of an appropriate adjoint scalar with a vev that could appear in some extensions of the SM. Such extensions are interesting to investigate in view of the enhancement of the hypercharge portal they provide.

In all other cases, such a mixing is at least of order ${\cal O}({ N^{-{3/2}}})$ and by appropriate tuning, it can be made of order ${\cal O}\left({N^{-2}}\right)$.

The fact that quantum corrections originating in holographic theories, are not generic, is not new. In \cite{Luk}, quantum corrections to the cosmological constant have been investigated in holographic theories, among other things.
Although for generic theories, such corrections are known to be both positive and negative, for holographic theories, they can only be {\em negative}.

The case where the hidden theory is a holographic theory, holds another particularity. The dark photon comes together with a tower of KK states that can be either discrete or continuous, depending on the nature of the hidden theory. Moreover, it typically has an extra contribution to its kinetic term as well as a mixing term due to SM corrections. This provides a more exotic structure to its phenomenology, that has not been analyzed in detail, to our knowledge.

We conclude that emergent (composite) dark photons arising from large-N strongly-coupled or holographic hidden theories provide new physics associated to the hypercharge portal of the SM.
We have estimated that the strength of the  hypercharge mixing is naturally small, and can be of order ${\cal O}(N^{-3/2})$ and in tuned cases ${\cal O}(N^{-2})$. This implies that one can evade today's experimental constraints with relative ease. Moreover, such physics may have important repercussions in the dark matter arena.
There is also the exceptional case of an extension of the SM by an adjoint scalar, whose vev can enhance the mixing to ${\cal O}(N^{-1})$. An analysis of this case, may also be interesting.

\section{Holography and emergent (dark) photons}
\label{holosetup}

The starting point of our analysis is described in \cite{1}, namely two QFTs interacting with each other, defining a UV-complete QFT. We define as the ``visible" QFT the theory whose dynamics and correlation functions we shall be interested in. This stands for the Standard Model (SM) and its variants. On the contrary, we call ``hidden" QFT, and we denote it as $\h{QFT}$, the other theory. Both theories are defined on a flat and not dynamical space-time background $g_{\m\n}\equiv \eta_{\m\n}$ and they interact through a set of massive bifundamental messenger fields of mass $M$. $M$ is assumed to be much larger compared to any other characteristic scale of the two QFTs.

At energy scales much smaller than the messenger mass, $M$, the messengers can be integrated-out, leaving the hidden $\h{QFT}$ interacting with the visible one, via a series of non-renormalizable interactions. Of interest in this paper are the effective induced interactions that are related to global $U(1)$ symmetries. The study of other types of symmetries is undertaken in \cite{grav,axion} and results in effective theories of emergent gravity and axions. We shall now give a few more details and review the precise setup originally described in~\cite{1,grav}.

Our starting point is a local relativistic quantum field theory as in (\ref{i2}) on a fixed space-time background, which we take to be four-dimensional and its metric $\eta_{\mu\nu}$ to be the flat Minkowski metric. We assume that this quantum field theory has the features:
\begin{itemize}

\item[$(a)$] It possesses a large scale $M$, the messenger mass, and all the other characteristic mass scales, $m_i$, are such that $m_i \ll M$.

\item[$(b)$] At energies $E\gg M$ the dynamics is described by a well-defined ultraviolet theory. For example, this could be a UV fixed point described by a four-dimensional conformal field theory.\footnote{We could also envisage a more exotic UV behaviour involving higher dimensional QFTs or some form of string theory.}
  The scale M, is defining the mass scale of bifundamentals that couple (in a UV-complete way) the hidden theory to a UV extension of the  SM. It is plausible that $M$ could be due to a vev of the combined theory.

\item[$(c)$] At energies $E\ll M$, there is an effective description of the low-energy dynamics in terms of two separate sets of distinct quantum field theories communicating to each other via irrelevant interactions\footnote{If the hidden theory has a scalar operator of dimension $\leq 2$, an interaction of relevant operators may appear in (\ref{2_2}).}  as in (\ref{i1}). We shall call the first quantum field theory the {\it visible} QFT (essentially the SM) and shall denote all quantities associated with that theory with normal font notation. We shall call the second quantum field theory the {\it hidden} $\h{QFT}$ and shall denote all its quantities with a hat notation. Schematically, we have the following low-energy description in terms of an effective action
\be
\label{2_1}
S_{IR} = S_{visible}(\Phi) + S_{hidden}(\h{\Phi}) + S_{int}(\Phi,\h{\Phi})
~,
\ee
where $\Phi$ are collectively the fields of the visible QFT and $\h{\Phi}$ the fields of the hidden QFT. The interaction term $S_{int}$ can be formally described by a sum of irrelevant interactions of increasing scaling dimension
\be
\label{2_2}
S_{int} = \sum_i \int d^4 x\, \lambda_i \, \vis{O}_i(x) \h{O}_i (x)
~,
\ee
where $\vis{O}_i$ are general operators of the visible QFT and $\h{O}_i$ are general operators of the hidden QFT. $S_{int}$ arises by integrating out massive messenger degrees of freedom of the UV QFT with characteristic mass scale $M$. This scale then defines a natural UV cutoff of the effective description. It is hence a physical scale determining the point in energy where the theory splits into two sectors, weakly interacting with each other at low energies.

\item[$(d)$] If we further assume that the hidden $\h{QFT}$ is a theory with mass gap $m$, at energies $E$ in the range $m\ll E \ll M$ we can employ the description \eqref{2_1} to describe a general process involving both visible and hidden degrees of freedom. For energies $E\ll m \ll M$ on the other hand, it is more natural to integrate out the hidden degrees of freedom and obtain an effective field theory in terms of the visible degrees of freedom only.

\end{itemize}

Our main focus then will be the low-energy $(E\ll M)$ behaviour of observables {\it defined exclusively in terms of elementary or composite fields in the visible QFT}, relevant for observers who have only access to visible QFT fields. In addition, we focus on an effective description of global $U(1)$ symmetries and the possibility that an emergent vector can appear
and couple to the visible theory.

 In \cite{u1} the generating functional of correlation functions (Schwinger functional) for the visible QFT defined as
\be
\label{2_3}
e^{- W(\JJ)} = \int [D\Phi] [D\h{\Phi}] \, e^{-S_{visible}(\Phi,\JJ) - S_{hidden}(\h{\Phi}) - S_{int}(\vis{O}_i, \h{O}_i)}
~.
\ee
was considered.

We use a Euclidean signature convention (that can be rotated to Lorentzian) where $\JJ$ is collective notation that denotes the addition of arbitrary sources in the visible QFT. This path integral is a Wilsonian effective action below the UV cutoff scale $M$. By integrating the hidden sector fields $\h{\Phi}$ we obtain
\be
\label{2_4}
e^{- W(\JJ)} = \int [D\Phi] \, e^{-S_{visible}(\Phi,\JJ) - \WW (\vis{O}_i) }
~,
\ee
where $\WW$ is the generating functional in the hidden QFT,
\be
 e^{-{\cal W}(\h{J})}\equiv \int [D\h{\Phi}] \, e^{- S_{hidden}(\h{\Phi}) - \int \h{O}\h{J}}
~.
\label{2_5}\ee
The first thing to observe is that from the point of view of the hidden QFT, the visible operators $\vis{O}_i$ appearing in the interaction term $S_{int}$ in \eqref{2_1} and \eqref{2_2} are dynamical sources.

The formal series of increasingly irrelevant interactions in \eqref{2_4}, can be reformulated in terms of emergent fields that interact with the visible theory. The effective action of these fields is given by the standard effective action for operators of the hidden theory, \cite{grav,axion,u1}.
 We focus here on the case of $U(1)$ symmetries.
In \cite{u1} it was shown that for each global conserved current in the hidden theory coupled to the visible theory, an emergent U(1) vector appears, mediating interactions in the visible theory.
Our purpose here is to study such interactions in more detail, using holography and contrast them with similar interactions obtained in string theory realizations of the SM.

The coupling of (very weakly-coupled) vectors to the SM is usually called the ``vector portal" of the SM, \cite{ship}. The relevant vector fields are usually referred to as ``dark photons".
The most sensitive coupling in this portal is the kinetic mixing of the dark photon to the SM hypercharge\footnote{Minimal couplings to the SM fields are excluded as typically they do not arise if the vector originates in the hidden sector and there is no mixing of the symmetry with a SM symmetry as is generically the case.  In the opposite case, the minimal couplings to the SM fields are the most important couplings and the hypercharge mixing calculation is a small perturbation.}. It is mostly this coupling we analyze and survey here, as well as other couplings that upon SM quantum corrections generate this mixing term with the hypercharge. We do this by using information from effective field theory, holography and string theory.

In our subsequent analysis we do not analyse a single hidden theory and a single messenger sector, but rather we search for generic results associated with this framework, that do not depend on the details of the theories involved.

\subsection{The brane-world picture\label{s21}}

Our main case of interest is when the hidden $\h{QFT}$ is a (large-N) holographic QFT that has a bulk gravity/closed-string theory dual.
As we shall discuss in more detail further on, the U(1) symmetry relevant here arises in the adjoint (or flavour) sector of QFT$_N$ and will therefore be realized as a local U(1) gauge symmetry in the bulk dual (including possible flavour branes).

To derive the geometrical picture, we start with the QFT picture.
The general action, after integrating out the messengers, can be written as,
\be
S=S_{1}+S_{12}+S_2\sp S_{12}=\int d^4x ~\widehat{J}^{\m} J_{\m}
\label{2_6}\ee
where we assumed a special interaction between the two theories\footnote{As discussed in \cite{u1} this interaction accompanies other interactions between the two theories, including scalar-scalar and higher tensor interactions. The relevant part remaining at low energy involves the interactions of protected operators, the stress tensor giving rise to an emergent dynamical metric, the conserved global vectors giving rise to emergent vector bosons and the instanton density, giving rise to an axion.}.
In (\ref{2_6}), $S_1$ is the action of the holographic theory, $S_2$ is the action of the SM, $\widehat {J}_{\m}$ is the conserved global current of the holographic theory and $J_{\m}$ is a gauge invariant vector operator of the SM.

Applying the holographic correspondence, \cite{mald1}, we can write\footnote{For a current $\widehat{J}^{\m}(x)$ dual to the gauge field $A_M(x,z)$ the asymptotic behaviour will be
$\lim_{z\to 0}A_{\m}(x,z)\approx a_{\mu}(x)+{\cal O}(z^2)$ in a (bulk) axial gauge where $A_z=0$.}
\be
\langle e^{iS_{12}}\rangle=\int_{\lim_{z\to 0}A_{\mu}(x,z)=J_{\m}(x)} {\cal D}A_{M}~e^{iS_{\rm bulk}[A_{M}]}
\label{2_7}\ee
where on the left, the expectation value is taken in the hidden holographic theory $\widehat {QFT}$. $S_{\rm bulk}[A]$ is the bulk gravity action, $z$ is the holographic coordinate, $A_{\m}$ is the bulk field dual to the conserved global current $\widehat{J}_{\m}$ of dimension $\Delta=3$ and the gravitational path integral has boundary conditions for $A_{\m}$ to asymptote to the operator $J_{\m}(x)$ near the AdS boundary. We have also neglected to show in (\ref{2_7}) the other bulk fields.

By inserting a functional $\delta$-function we may rewrite (\ref{2_7}) as
\be
\langle e^{iS_{12}}\rangle=\int_{\lim_{z\to 0}A_{\m}(x,z)= B_{\m}(x)} {\cal D}A_{M}(x,z){\cal D}B_{\m}(x){\cal D}C_{\m}(x)~e^{iS_{\rm bulk}[A]+i\int d^4x C_{\m}(B^{\m}-J^{\m})}
\label{2_8}\ee
If we now integrate $B_{\m}(x)$ first in the path integral transform, we obtain the Legendre transform of the Schwinger functional of the bulk gauge field, which becomes the bulk effective action. This corresponds in holography to switching boundary conditions at the AdS boundary from Dirichlet to Neumann, and where $C_{\m}(x)$ is the expectation value of the operator $\widehat{J}_{\m}$.
We finally obtain,
\be
\langle e^{iS_{12}}\rangle=\int_{\lim_{z\to 0}\partial_zA_{\m}(x,z)=z^2 C_{\m}(x)} {\cal D}A_{\m}(x,z){\cal D}C_{\m}(x)~e^{iS_N[A]-i\int d^4x~ C_{\m}(x)J^{\m}(x)}
\label{2_9}\ee
We may imagine the SM action as coupled at the radial scale $z_0\sim 1/M$ to the bulk action.
Following holographic renormalisation \cite{Bianchi:2001kw, Bianchi:2001de}, we may then rewrite the full bulk+brane action of the emergent gauge field $C_{\m}$ as
 \be
S_{total}=S_{bulk}+S_{brane}
\label{2_10}\ee
\be
S_{bulk}=M_P^3\int d^5x\sqrt{g}\left[{Z\over 4}(F_A)^2+\dots\right]
\label{2_11}\ee
\be
S_{brane}=\int dz\delta(z-z_0)\int d^4x\sqrt{\gamma}\left[{m^2\over 4}\widehat{F}_{\m\n}^2+ {m\chi\over 2}\widehat{F}_{\m\n}F_{Y}^{\m\n}+{1\over 4g_Y^2}F_{Y}^2+\cdots\right]
\label{2_12}\ee
where $\widehat{A}_{\m}(x)\equiv A_{\m}(z_0,x)$ is the induced gauge-field on the brane and $\widehat{F}_{\m\n}$ its field-strength.
The ellipsis in the bulk action stands for other terms involving different fields as well as couplings of $A_{\m}$ to charged bulk fields.
The ellipsis in the brane action stands for the SM model fields. We have added explicitly a possible mixing terms of $\hat A_{\m}$ to the hypercharge gauge boson, and assumed that no SM particle is charged under $\hat A_{\m}$.

The upshot of this discussion is that in the presence of couplings between the holographic theory and the SM, global conserved U(1) currents of the QFT$_N$ can appear as novel U(1) vectors coupled to the SM.
When they originate in the adjoint sector, they should correspond to the closed string U(1)'s (graviphotons) discussed later-on in this paper.

\subsection{A classification of global symmetries}

We now classify the type of U(1) symmetries of the SM and the type of global U(1) symmetries of the messenger and large-N sector.

There are three distinct classes of SM symmetries:

\begin{enumerate}

\item U(1) non-anomalous gauge symmetries: The SM has only the hypercharge Y. In more exotic cases, B-L could appear as such a symmetry. However, in such a case, the B-L gauge boson is massless and the case excluded experimentally\footnote{{Unless the gauge symmetry is broken spontaneously by the Higgs mechanism. However, in orientifold realizations of the SM, although B-L is gauged and non-anomalous in 4d, the relevant gauge boson can be massive due to higher dimensional anomalies that exist in decompactification limits of the theory, \cite{AKR,bianchi,anasta}.}}. Hypercharge traces in the SM satisfy among others $Tr[Y]=0$.

\item U(1) ``anomalous" symmetries that are guaranteed to exist in all bi-fundamental realizations of the SM (see \cite{ADKS}). Such symmetries have massive gauge bosons (whose effective interactions were analyzed in \cite{irges}) and B-L should be in that class. The structure of anomalies was analyzed in \cite{bianchi}. Typically the charge matrices of these symmetries satisfy $Tr[Q]\not= 0$

\item Non-abelian gauge symmetries, namely the SU(2) and SU(3) of the SM.

\end{enumerate}

On the other hand, we must also classify the global symmetries of the large-N sectors: these involve the ``hidden" large-N $\h{QFT}$, as well as the massive messengers.

We may have the following types of symmetries:

\begin{enumerate}

\item[a.] Global symmetries that affect only the messengers.
The particles made out of the messengers are messenger-mesons and messenger-baryons. The lightest are the mesons.
How heavy they are depends on the interplay between their mass scale, $M$, and how strong their interaction is. If their interactions were weak, their masses would be of order ${\cal O}(2M)$. However, a strong binding force will reduce this mass. As we know from holographic conformal theories with large enough supersymmetry (${\cal N}=2$ for flavour, ${\cal N}=4$ for the glue), such a mass becomes ${2M\over \lambda}$ where $\l$ is the t'Hooft coupling, \cite{flavour}.
  It can become arbitrarily small for sufficiently large $\l$.

  Here we assume that the coupling is such that the masses of messenger-mesons are well above the SM scales.
Such symmetries are therefore irrelevant for our discussion, as they will not be present at SM energies.

  A subset of these symmetries may be gauged by the SM gauge fields. To put it alternatively, the messengers may carry SM minimal couplings.

\item[b.] R-like symmetries\footnote{ {R-like symmetries are global symmetries that resemble R-symmetries in supersymmetric theories. For example one can break supersymmetry without breaking an R-symmetry. Or as it happens in N=1 sQCD, the R-symmetry can mix with another U(1) global symmetry to make a non-anomalous R-like symmetry. In string theory R-symmetries are never exact. They can be symmetries of the massless sector, and are subgroups of the orthogonal symmetry of the internal space. They may be present even in the absence of supersymmetry in the massless sector. They are broken by the compactification of $\alpha'$ corrections.}} that act both on the large-N $\h{QFT}$ and the messengers. Such symmetries cannot be gauged under the SM, as in that case, the ``bulk" states would have SM charges, a phenomenologically unattractive situation. However, as such symmetries act only on $\h{QFT}$, after integrating-out the messengers, they can be effectively treated in the low-energy theory as those of case (c) below.

\item[c.] Flavour symmetries of the large-N QFT that do not act on the messengers. This implies the messengers are not charged under such symmetries, but quantum effects of the large-N QFT will introduce higher-order multipole couplings to the messengers.

All of the above can be in principle abelian or non-abelian.

\end{enumerate}

We must now examine the various cases and estimate the leading couplings. We shall consider abelian hidden symmetries and abelian gauge groups of the SM. We must therefore consider the combinations of cases $1c$ and $2c$.

To set up the calculation, it convenient to consider the Schwinger source functional of the large-N $\h{QFT}$+messengers with sources for the global current (we call this source $\widehat{A}_{\m}$), and all the messenger operators that couple to SM fields in the UV.
Such operators are all hidden-color singlets, and transform as SM bi-fundamentals (so as to produce cubic and quartic invariant messenger-SM couplings).

We shall classify them here to obtain an idea of what they can be.
 We denote the messenger fields generically as $\eta_{ai}$, where the index $a=1,2,\cdots N$ is the color index of $\h{QFT}$, while $i$ is a gauge index of a gauge group of the SM. $\eta_{ai}$ can be a vector, a fermion or a scalar. Note that, in order to be able to write the messenger couplings to the
SM, all the SM fields should be bi-fundamentals\footnote{Including adjoints or (anti)symmetric rank-two tensor products.} with respect to the gauge groups of the SM: $\chi_{ij}$. Again, here $\chi_{ij}$ can be a scalar, a fermion or a vector boson.

 Note that there are several ways to write the SM fields as bi-fundamentals.
 The D-brane realizations of the SM produce automatically an (extended) SM model spectrum where all fields are bifundamentals of the (extended) SM gauge group. All different such realizations have been effectively classified in \cite{ADKS}, and they involve the inclusion of at least two more U(1) gauge groups in the SM.
We denote
the SM fermions collectively as $\psi^{ij}$, the scalars as $\phi^{ij}$ and the gauge fields as $A_{\mu}^{ij}$.
Finally, we label the $\h{QFT}$ fields as $\zeta_{ab}$ as they are in the adjoint of the large-N gauge group.

Below, we classify composite, gauge-invariant operators of the messenger sector that behave as bi-fundamentals under the SM gauge group, and are therefore coupled to respective SM fields.

\begin{itemize}

\item Scalar operators: if they are SM singlets (and have the appropriate UV dimension $\Delta\leq 2$, they can couple to the $ |H|^2$ where H is the SM Higgs.
  However, in \cite{1} it was advocated that for reasons of stability, the hidden theory should not have strongly relevant operators. We therefore ignore this possibility in the sequel, some aspects of which were considered in \cite{del}.

The scalar operators charged under the SM gauge group must be bi-fundamentals. We collectively call them $\Phi_{ij}$, where $ij$ are bifundamental color indices of the SM.
Their coupling to the SM must have dimension at least zero (for UV-completeness reasons) and they must couple to the charged scalars of the SM, namely the Higgs, $\phi^{ij}\Phi_{ij}$. Therefore the dimension of $\Phi_{ij}$ must be 3 or lower.
$\Phi_{ij}$ cannot couple to SM composites again for dimensional reasons except to $A_{\m}^{ik}A^{\mu,jk}$ in order to make gauge-invariant gauge couplings to scalars.

\item Spin 1/2, fermion operators (in the near-free field case, we should think of them as composites of a scalar and a fermion messenger). Their dimension should be $\Delta \geq 3/2$ for unitarity reasons.
  Their dimension is somewhat lower than 5/2 at weak coupling and a bit above 3/2 at strong coupling.
   We denote them collectively with $\Psi_{ij}$.
  They can couple to the SM fermions $\psi^{ij}$.

 \item  Spin-1 bosonic operators coupled to vectors of the standard model. For renormalizability, such couplings must be gauge couplings. In that case, the vectors are conserved currents coupling to the gauge fields of the SM.
   We denote them with ${{V}}^{\m}_{ij}$.

\end{itemize}

Therefore, in this language, the generic couplings of the messenger operators to the SM are written as

\be
S_{m}=\int d^4x\left[ \Phi_{ij}\phi^{ij}+\bar\Psi_{ij}\psi^{ij}+{{V}}^{\m}_{ij}A_{\m}^{ij}+\cdots\right]
\label{2_13}\ee
where the dots indicate possible quartic terms present for gauge invariance purposes.
In the formula above, $\phi^{ij},\psi^{ij},A_{\m}^{ij}$ are scalar, fermion and vector elementary SM fields, carrying their bifundamental SM indices.
On the other hand, $\Phi_{ij},\Psi_{ij},{{V}}^{\m}_{ij}$ are messenger composite operators that are invariant under the large-N gauge group. They are of the form $\Phi_{ij}\sim \sum_a\eta^{\dagger}_{ai}\eta_{aj}$ where $\eta_{ai}$ are the messenger bi-fundamentals.

Consider now the Schwinger functional $W$ of the hidden/large-N+messenger theory with a source $\widehat{A}_{\m}$ for the U(1) global current as well as $J^{\Phi}_{ij},J^{\Psi}_{ij}, J^{{V}}_{\mu ij}$ for the messenger operators
$\Phi^{ij},\Psi^{ij},{{V}}_{\m}^{ij}$.
\be
e^{W(A_{\m},J^{\Phi}_{ij},J^{\Psi}_{ij}, J^{{V}}_{\mu ij})}\equiv\langle e^{\int [\widehat{A}^{\m}J_{\m}+J^{\Phi}_{ij}\Phi^{ij}+\bar J^{\Psi}_{ij}\Psi^{ij}+ J^{{V}\mu}_{ij}V_{\m}^{ij}]d^4x}\rangle
\label{2_14}\ee
Adding now the couplings to Standard model fields, $\phi_{ij},\psi_{ij},A^{\m}_{ij}$ from (\ref{2_13})
we obtain
\be
\langle e^{\int [\widehat{A}^{\m}J_{\m}+(J^{\Phi}_{ij}+\phi_{ij})\Phi^{ij}+(\bar J^{\Psi}_{ij}+\bar\psi_{ij})\Psi^{ij}+ (J^{{V}\mu}_{ij}+A^{\m}_{ij}){{V}}_{\m}^{ij}]d^4x}\rangle=e^{W(\widehat{A}_{\m},J^{\Phi}_{ij}+\phi_{ij},J^{\Psi}_{ij}+\psi_{ij}, J^{{V}\mu}_{ij}+A^{\m}_{ij})}
\label{2_15}\ee
where the expectation value above is taken in the vacuum of the large-N+messenger theory.

When we integrate out the messengers, at low energy, this is equivalent to setting the messenger sources above to zero.
Therefore, the relevant coupling of the U(1) current to the standard model is controlled by
\be
W(\widehat{A}_{\m},\phi^{ij},\psi^{ij}, A_{\m}^{ij})+S_{SM}(\phi^{ij},\psi^{ij},A_{\m}^{ij})
\label{2_16} \ee
In the formula above, $\phi^{ij},\psi^{ij},A_{\m}^{ij}$ are the SM scalars, fermions and vectors while $\widehat{A}_{\m}$ is the hidden U(1) source.

Our goal, therefore, is to study the form of the Schwinger functional in (\ref{2_14}) so that we learn about the induced couplings of $\widehat{A}_{\m}$ to the SM sector.

\subsection{Case 1c: Hypercharge and symmetries of the large-N $\h{QFT}$}

This case corresponds in string theory orientifolds, to the mixing between a U(1) current from the closed string sector and another from the open string sector (where the SM is realized).

The first observation is that none of the operators $\Phi_{ij}$, $\Psi_{ij}$, $V^{\m}_{ij}$ is minimally coupled to (charged under) $\widehat{A}_{\m}$. The same therefore applies to $\phi_{ij},\psi_{ij}, A_{ij}^{\m}$.

A second observation is that if there is bifundamental (massive) matter between the $\h{QFT}$ theory and the SM, that is charged under both the $U(1)_{\widehat{A}}$ and $Y$ then a one-loop computation can in principle produce a mixing between $F_{\widehat{A}}$ and $F_Y$. However, such matter does not exist for this type of bulk symmetries, and therefore no mixing term exists in $W$.

We now consider the allowed low-dimension terms in  $W(\widehat{A}_{\m},\Phi_{ij},\Psi_{ij}, {{V}}^{\m}_{ij})$ that mix $\widehat{A}_{\m}$ to the SM fields, without minimal couplings.
{ They start at dimension 6},
\be
W_{6}(\widehat{A}_{\m},\Phi_{ij},\Psi_{ij}, {{V}}^{\m}_{ij})\sim {1\over N}{1\over M^2}\pa_{\m}\Phi \pa_{\n}\Phi^* F_{\widehat{A}}^{\m\n}+{1\over N^{3\over 2}M^2}\bar\Psi\gamma_{\m\n}\Phi\Psi~F^{\m\n}_{\widehat{A}}+
\label{2_17}\ee
$$+
{1\over N^{3\over 2}M^2}F_{\m\n}^{\widehat{A}}F^{V,\m\n}\Phi\Phi^*+{1\over N^{2}M^4}F_{\m\n}^{\widehat{A}}F^{V,\m\n}\bar \Psi\Phi \Psi+\cdots
$$
where the ellipsis indicates higher-dimension couplings\footnote{As these operators mirror the SM ones, there is no gauge invariant coupling $\bar\Psi\gamma_{\m\n}F^{\m\n}_{\widehat{A}}\Psi$ because of chirality constraints. Equivalently there is no gauge invariant $\bar\psi\gamma^{\m\n}\psi$ coupling in the SM. A similar statement applies to $\bar \Psi \Psi$.}
In (\ref{2_17}) above, traces over SM indices have been suppressed and the mass scale controlling the terms in the Schwinger functional is the messenger scale $M$. Moreover, we have inserted the appropriate powers of $N$ based on standard adjoint and fundamental large-N counting, \cite{1}.

There is no direct mixing term of the type $F_{\m\n}^{\widehat{A}}F^{{V},\m\n}$ as the messengers are neutral under the bulk U(1) symmetry.
The magnitudes in $1/N$ of the amplitudes in (\ref{2_17})
are easily estimated by remembering that each ``glueball" operator of the hidden theory carries a factor $N^{-1}$ while hidden mesonic operators carry a factor of $N^{-{1\over 2}}$.

Therefore, from (\ref{2_16}), the low-dimension terms in  $W(\widehat{A}_{\m},\phi_{ij},\psi_{ij}, A^{\m}_{ij})$ that mix the emergent photon $\widehat{A}_{\m}$ to the fields of the visible theory, start at dimension 6, with the following terms
\be
W_{6}(\widehat{A}_{\m},\phi_{ij},\psi_{ij}, A^{\m}_{ij})\sim {1\over NM^2}F_{\widehat{A}}^{\m\n}D_{\m}H^{\dagger} D_{\n}H+{1\over N^{3\over 2}M^2}F^{\m\n}_{\widehat{A}}\left[\bar\psi\gamma_{\m\n}H\psi+c.c.\right]
\label{2_18}\ee
$$+
{1\over N^{3\over 2}M^2}F_{\m\n}^{\widehat{A}}F^{Y,\m\n}H^{\dagger}H+{1\over N^{ 2}M^4}F_{\m\n}^{\widehat{A}}F^{Y,\m\n}\left[\bar \psi H\psi+c.c.\right]\cdots
$$
where $H$ is the Higgs doublet, $\psi$ collectively denotes the standard model fermions and $F^{Y,\m\n}$ is the hypercharge field strength.

There are dimensionless couplings in all the above terms that we assume to be of order one. We have set them to 1, as we are mainly interested in the dependence on the mass scale $M$ and the number of colors $N$ of the hidden theory.

It should be stressed that these leading estimates are ``generic" in the sense that they are allowed by known symmetries. As we shall see in the second part of this paper, some of these terms will appear only at subleading orders in string theory.
(\ref{2_18}) includes all gauge-invariant, lowest dimension, antisymmetric tensors of the SM coupled to $F_{\widehat{A}}^{\m\n}$.

The first two terms in (\ref{2_18}) can be rewritten as $\widehat{A}_{\m}$ coupling to a topological current\footnote{We call it topological because it is exactly conserved off-shell and its charge is cutoff dependent and may arise only at the boundary of space. } $J^T_{\m}$
\be
\int d^4x \widehat{A}_{\m}J^{T,\m}\sp J^{T}_{\n}=\pa^{\m}\left[{1\over M^2}(\pa_{\m}H^\dagger \pa_{\n}H-\pa_{\n}H^\dagger \pa_{\n}H)+{1\over M^2}\left(\bar\psi\gamma_{\m\n}H\psi+cc\right)\right]
\label{2_19}\ee
\be
\pa^{\m}J^T_{\m}=0
\label{2_20}\ee

These terms can mediate two-point couplings between $F_{\widehat{A}}$ and $F_Y$ due to the quantum effects of the SM fields. They are of the form
\be
{1\over N}\int d^4x d^4y~F^{\widehat{A}}_{\m\n}(x)A^Y_{\rho}(y)\left[{1\over M^2}\langle (\pa^{\mu}H^\dagger\pa^{\n}H)(x) (H^\dagger\pa^{\rho}H - \partial^\rho H^{\dagger} H)(y)\rangle+\right.
\label{2_21}\ee
$$\left.+{1\over \sqrt{N}M^2}
\sum_{Y}\left[Q_Y\langle \bar \psi\gamma^{\m\n}H\psi(x)\bar \psi\gamma^{\rho}\psi(y)\rangle+cc.\right]\right]=
$$
$$=
{1\over N}\int d^4p~F^{\widehat{A}}_{\m\n}(p)A^Y_{\rho}(-p)\left[{1\over M^2}\langle(\pa^{\mu}H^\dagger\pa^{\n}H)(H^\dagger\pa^{\rho}H{-}\partial^\rho H^{\dagger} H)\rangle(p)+\right.
$$
$$
+\left.
\sum_{Y}{Q_Y\over \sqrt{N}M^2}\langle \left(\bar \psi\gamma^{\m\n}H\psi +cc\right) \, \bar\psi\gamma^{\rho}\psi\rangle(p)\right]
$$
where in the last term the sum runs over all fermions charged under the hypercharge, and we have rewritten the terms in momentum space.

The two-point functions are calculated in appendix \ref{mixing}.
The Higgs contribution has as leading term
\be
\sim {\Lambda^2\over N~M^2} F_{\m\n}^{\widehat{A}}F^{\m\n}_Y\sim {1\over N} F_{\m\n}^{\widehat{A}}F^{\m\n}_Y
\label{2_22}\ee
with $\Lambda\simeq M$.

The fermionic dipole contribution
is smaller and proportional to
\be
Tr\left[Q_Y{m_{Y}\langle H\rangle\over N^{3\over 2}\,M^2}\right]\log{\Lambda ^2\over M^2}
\label{2_23}\ee
as shown in appendix \ref{mixing}. This is triply suppressed compared to the Higgs contribution in (\ref{2_22}). There is an extra factor of $N^{-{1\over 2}}\ll 1$, a factor ${m_{Y}\over M}\ll 1$ and a factor ${\langle H\rangle\over M}\ll 1$.

The contribution of the third term in (\ref{2_18}) is proportional to
{
\be
{\langle H^{\dagger}H\rangle\over N^{3\over 2}M^2}\sim {\langle H\rangle^2\over N^{3\over 2}M^2}\
 \label{2_29}
 \ee
 and is therefore smaller than  (\ref{2_22})  by a power of $\sqrt{N}$ as well as by the ratio $\langle H\rangle^2/M^2\ll 1$.
To obtain this we assumed a tuning of the UV theory so that the Higgs vev is small compared to the messenger scale.
}

Finally the contribution of the fourth term in (\ref{2_18}) is proportional to
\be
Tr\left[{m\langle H\rangle~\Lambda^2\over N^{ 2}\,M^4}\right]\sim Tr\left[{m\langle H\rangle\over N^{ 2}\,M^2}\right]
\label{2_26}\ee

Higher order terms may have additional suppressions.
The algorithm is as follows
\begin{enumerate}

\item Higher derivatives acting on the terms in (\ref{2_18}) give similar contributions as they contain additional powers of $\Lambda/M$ that is of order one.

\item Terms containing additional SM fields have additional suppressing powers on N: an $N^{-{1\over 2}}$ per additional field. They may also give contributions that have an additional suppression due to multiplication by SM coupling constants.

 \end{enumerate}

In a sense, both bosonic and fermionic contributions in the SM are due to the Higgs. The first is direct while the second is due to the fermion masses, that are generated by the Higgs-Englert-Brout mechanism.

The size of the mixing is small as $N\gg 1$, but may also be constrained by experiments as the limits on kinetic mixing are strong. The emergent vector $\hat A_{\m}$ here is expected to be light, so it will mediate a fifth force even though none of the SM particles are directly charged under it. They will acquire an effective charge because of the mixing.

Standard model quantum effects also correct the coupling of the vector $\aa_{\m}$, but this effect is even smaller. It has the structure
\be
{\delta g_{hidden}^2 \over g_{hidden}^2}\sim {1\over N^2}{m_{SM}^2\over M^2}
\label{2_24}
\ee
where $m_{SM}$ stands for a SM mass scale, typically a mass scale of one of the SM particles.
 Therefore the correction has an extra suppression (beyond the large N suppression) due to the hierarchy, $m_{SM}\ll M$. It is calculated in the appendix \ref{ggc}.

The generic QFT contributions will be compared to similar effects in string theory contributions in the second part of this paper. The emergent vectors discussed here will be similar to what we call graviphotons in the case of string theory, later on.

\subsection{Case 1b: Hypercharge and global symmetries that act on the messengers}

The main difference in this case, compared to the previous subsection, is that
a direct mixing term $F_{\m\n}\hat F^{\m\n}$ can appear as the messengers are charged under the global symmetry.
This coupling comes suppressed by a single power of $N$.

The rest of the relevant couplings are as discussed in the previous section, and all statements made there apply to these cases also.

Such emergent U(1)s are similar to what we called dark photons in string theory, later on.

\subsection{Case 2c: SM Anomalous U(1)'s and exact symmetries of the large-N QFT\label{s25}}

Realizing the SM in terms of bifundamental fields only of the associated gauge group, is equivalent to embedding the SM in a string theory orientifold where the SM fields are localized on a stack of intersecting D-branes, \cite{AKT,Ib,AD}. The general embedding of the SM gauge group was classified in \cite{ADKS} where it was shown that any such realization includes at least two anomalous U(1)'s\footnote{Typically extra right-handed neutrino states render B-L anomalous. It is possible, however, that in special cases, B-L is not anomalous and has a massless gauge boson. Obviously, this case is phenomenologically excluded unless B-L breaks due to symmetry breaking.}
which are massive. The interaction of such U(1)'s with the symmetry breaking pattern of the SM is interesting, \cite{irges}, as the Higgses are charged under these massive anomalous U(1)'s. Their masses are a one-loop effect, and this is why they are probably the lightest stringy states, \cite{AKR,muon}. They can be almost as light as the EW symmetry breaking scale, but can also be quite heavier.

Such U(1)'s, have non-zero $Tr[Q]$, but can also have non-zero mixed traces with the hypercharge. This is the generic case, as shown in \cite{bianchi}. The associated hypercharge anomalies are cancelled by generalized Chern-Simons terms.

A non-zero trace $Tr[Q_a]$ for an anomalous U(1)$_a$, is still not enough to allow perturbative mixings with the hidden sector conserved currents.
Therefore, the effective action in this case is similar to (\ref{2_18}), with
associated conclusions that follow.
Note that in this case, the kinetic mixing of an emergent massless vector boson $\widehat{A}$ with the massive $B_\mu$ of an anomalous U(1)$_a$, does not produce SM minimal couplings for the rotated massless combinations and therefore the constraints on this mixing, unlike the previous one of case 1c, are largely innocuous.

We assume that there is a massive gauge field $B_\mu$ coupled to an anomalous U(1)$_a$ extension of the SM. The effective action will be
\begin{align}
{\cal L} &= \frac{1}{4g^2} F^2_B + \frac{1}{2}(\partial\beta + {\cal M} B)^2 + {\cal C}_{ij} \beta Tr[G_i\wedge G_j]
\label{2_25}\\
& + E_{YBB} Y\wedge B\wedge F_B + E_{YBY} Y\wedge B\wedge F_Y \nn
\end{align}
where $Y_\mu$ is the gauge field coupled to hypercharge, $\beta$ is an axion that it is eaten by the anomalous U(1)$_a$ and renders it massive \`a la St\"uckelberg. $G^{\mu\nu}_i$ collectively denote the field strengths of the abelian or non-abelian gauge fields of the SM. The last two terms are the generalized Chern-Simons couplings (GSC). We also assume that all SM matter fields (fermions and the Higgs) are charged under the anomalous U(1)$_a$.

We proceed to study the mixing between the hidden gauge field $\widehat{A}$ and an anomalous U(1)$_a$ and the differences between the anomalous $B$ and non-anomalous $Y$ cases.

In both cases, the leading contribution to the mixing is coming by the one-loop diagram with a circulating Higgs in the loop, and it is suppressed by $1/N$ due to the first coupling in \eqref{2_18}. The anomalous behaviour of the U(1)$_a$ appears at three loops. The relevant diagrams are also suppressed by factors of $1/N$. However, they are further suppressed by several SM couplings.

Therefore, we do not expect any significant difference in the mixing of $\widehat{A}$ with non-anomalous $Y$ or anomalous $B$ gauge fields of the SM. More details are provided in the appendix \ref{anomalousU1}.

\subsection{The effects of mixing in the holographic case\label{s26}}

We shall now discuss the effective interactions and mixing, when the dark photon emerges from a holographic hidden sector. The details are treated in the appendix \ref{gau12}.

We consider the bulk plus brane action,
 \be
S_{total}=S_{bulk}+S_{brane}
\label{t1}\ee
\be
S_{bulk}=M_P^3\int d^5x\sqrt{g}\left[{Z\over 4} F_{\hat A,mn}F_{\hat A}^{mn}+ {\cal J}^{m}\hat A_m\dots\right]
\label{t2}\ee
\be
S_{brane}=\int dz\delta(z-z_0)\int d^4x\sqrt{\gamma}\left[{m_{\hat A}^2\over 4}\widehat{F}_{\m\n}\widehat{F}^{\m\n}+{m\chi\over 2}\widehat{F}_{\m\n}F_{Y}^{\m\n}+{1\over 4g_Y^2}F_{Y}^2+Y^{\m}J_{\m}+\cdots\right]
\label{t3}\ee
where $m,n$ are five-dimensional indices and $\m,\n$ are four-dimensional indices. $\hat A_{m}$ is the bulk gauge field, dual to a global conserved current of the hidden holographic theory. ${\cal J}_{m}$ is the bulk current coupled to $\hat A_{m}$ and obtaining contributions from fields dual to operators that are charged under the global symmetry.
$\widehat{A}_{\m}(x)\equiv A_{\m}(z_0,x)$ is the induced gauge-field on the brane and $\widehat{F}_{\m\n}$ its field-strength.
$Y^{\m}$ is the SM hypercharge gauge field, $F_Y$ its four-dimensional field strength and $J_{\m}$ is the gauge invariant SM hypercharge current.

We have allowed for a possible non-zero coupling constant, $m_{\hat A}$,  for $\widehat A_{\m}$ localized on the brane.
$M_P$ is the five dimensional Planck mass.
$\widehat A_{m}$ is dimensionless, while the hypercharge $Y_{\m}$ has the canonical mass dimension one. Therefore $m_{\hat A},M_P$ are mass scales while $\chi$ and $g_Y$ are dimensionless. $Z$ is a function of various scalar fields, some of them having a non-trivial radial profile in the fifth (holographic) dimension. For all practical purposes, we can therefore consider $Z$ as a function of $z$.

The ellipsis in the bulk action stands for other terms involving different fields.
The ellipsis in the brane action stands for the SM model fields. We have added explicitly a possible mixing terms of $\widehat A_{\m}$ to the hypercharge gauge boson and assumed that no SM particle is charged under $\widehat A_{\m}$.

We would now like to calculate the induced interaction on the hypercharge current $J_{\m}$ and the bulk (hidden) current ${\cal J}_{m}$ generalizing the calculation in \cite{h}. This is performed in appendix \ref{gau12} with the following result,

\be
S_{int}
={1\over 2}\int {d^4p\over (2\pi)^4}J_{\mu}(-p)G_{44}(p)J^{\m}(p)+{1\over 2}\int dz\int {d^4p\over (2\pi)^4}J_{\m}(-p)G_{45}(z,p){\cal J}^{\m}(z,p)+
\label{t4}\ee
$$
+{1\over 2}\int dz\int dz'{d^4p\over (2\pi)^4} {\cal J}_m(z,-p)G^{mn}_{55}(z,z',p){\cal J}_{n}(z',p)
$$
with

\be
G_{44}(p)=-{g_Y^2\over p^2}{{1\over m_{\hat A}^2p^2}+{G_5(z_0,z_0,p)\over M_P^3 Z_0}\over {1\over m_{\hat A}^2p^2}+{(1-\chi^2 g_Y^2)G_5(z_0,z_0,p)\over M_P^3Z_0}}
\label{t5}\ee
\be
G_{45}(z,p)=\left({e^{3\tilde A(z)}~G_5(z,z_0,p)\over Z_0}-{e^{2\tilde A(z)}\over Z(z)}G_5(z_0,z,p)\right) {m_{\hat A}\chi g_Y^2\over 1+{m_{\hat A}^2(1-\chi^2 g_Y^2)G_5(z_0,z_0,p)p^2\over M_P^3 Z_0}}
\label{t6}\ee

\be
G_{55}^{zz}(z,z',p)={M_P^3 e^{5\tilde A(z')}\over Z(z')}{1\over p^2}\sp G_{55}^{z\m}(z,z',p)=0
\label{t7}\ee
\be
G_{55}^{\m\n}(z,z',p)={M_P^3 e^{5\tilde A(z')}\over Z(z')}{G_{5}(z,z',p)\over 1+{m_{\hat A}^2(1-\chi^2 g_Y^2) \over M_P^3Z_0 }p^2 G_5(z,z_0,p)}\left(\eta^{\m\n}-{p^{\m}p^{\n}\over p^2}\right)
\label{t8}\ee

We have chosen the Lorentz gauge for the hypercharge interaction and the five-dimensional Coulomb gauge for the bulk U(1).
$G_5(z,z',p)$ is the scalar Green's function in the nontrivial bulk metric, (\ref{n8}) and $Z$ kernel,
\be
\left( \pa_z^2+\left(A'+{Z'\over Z}\right)\pa_z-p^2\right)G_5(x,z;x',z')=\delta(z-z')\eqp
\label{t9}\ee

Unitarity implies that
\be
\chi^2 g_Y^2<1
\label{t10}\ee
which is satisfied in the large N limit as $\chi\sim {\cal O}(N^{-1})$.

The effect of the induced kinetic term due to the SM corrections and the mixing term for the current-current interactions in (\ref{n26}), \cite{dgp}, have the following consequences.
\begin{itemize}

\item The hypercharge interaction in (\ref{t5}) is modified. However, in the far UV or far IR it is unchanged to leading order.
  At intermediate distances, there is a new DGP-like resonance mediating the hypercharge interaction with strength proportional to $\chi^2$.
This resonance however is heavy. Taking into account the estimates
\be
M_P^3\sim {\cal O}(N^2)\sp Z_0,m_{\hat A}\sim {\cal O}(1)\sp \chi\sim {\cal O}(N^{-1})
\label{t11}\ee
we obtain that the mass scales as ${\cal O}(N^2)$.

\item The interaction of hypercharge via (\ref{t6}) with the bulk global charge is present due to the non-trivial mixing. It is doubly weak:  once because of the smallness of $\chi$ and second because of the large-N suppression of bulk interactions.

\item The bulk gauge interaction in (\ref{t8}) is modified both at short and long distances. This was analysed in detail in section 6 of \cite{u1}. There it was shown that the emergent vector mediates a five-dimensional interaction at intermediate distances while the interaction is four-dimensional at short and long distances. Moreover there is an effective  mass generated from bulk effects.
    Because of the brane mixing that appears here, similar remarks apply to the interaction between bulk charges and hypercharges as well as between hypercharges only.

\item Overall, the fact that the dark photon arises in a holographic hidden sector provides additional suppression to the effects of its mixing to hypercharge.
\end{itemize}

\section{Dark photons in string theory}
\label{stringsetup}

In models with open and unoriented strings \cite{Bianchi:1990yu}-\cite{Bachas}, very much as in other string setups, the SM is usually accompanied by extra fermions, scalars or gauge bosons. In particular, in the intersecting D-brane context, it has been shown, that at least four stacks in the oriented case or three in the unoriented case, whereby $SU(2)_W = Sp(2)$, are necessary to embed the SM fermions with their correct charge assignments and the necessary Higgs scalars, \cite{AKT,Ib,rev,AD}.
As a result, the hyper-charge $Y$ is realized as a non-anomalous linear combination of four (three) $U(1)$'s \cite{ADKS}\footnote{U(1)s that are free of four-dimensional anomalies may be massive, \cite{AKR,anasta}, due to higher-dimensional anomalies in various decompactification limits. Moreover, there could be mixed anomalous charge traces between non-anomalous U(1)'s like the hypercharge and anomalous U(1)'s that are cancelled by generalized Chern-Simons terms, \cite{bianchi}.} .

The remaining $U(1)$'s are typically anomalous and massive. Even $U(1)_{B{-}L}$, though non anomalous, is typically broken at the string scale by coupling to closed-string axions \cite{Ibanez:2006da}-\cite{Addazi:2015yna}. Yet, the string scale $M_s$ can be a priori relatively light, \eg~$M_s\approx \text{TeV}$ or at an `intermediate' scale $M_s\approx 10^{9\div 12} \text{ GeV}$ and anyway much lower than the Planck scale $M_P = M_s {\cal V}/g_s$, where ${\cal V}$ denotes the (large) internal volume transverse to the volume of the cycles wrapped by the SM branes. In summary, extra abelian gauge fields living at the same intersecting/magnetized D-branes as the SM are quite natural in this setting. Their phenomenology has been studied extensively in the past \cite{irges,bianchi}, \cite{Coriano:2007fw}-\cite{Fucito:2008ai} and we shall not dwell on it any further.
Instead, we would like to identify another two classes of (non)abelian vector bosons that can also appear in this setup: closed-string vectors and `flavour' or hidden brane vectors.

The former comes from the closed string sector and corresponds to massless vector fluctuations of the metric, when the compactification manifold has continuous isometries, or to massless vector fluctuations of the $(p{+}1)$-form potentials, when the compactification manifold support harmonic $p$-forms.

The second class of extra gauge bosons comes from the open string sector and they are associated to extra (intersecting and/or magnetized) D-branes that either do not intersect with the SM branes, therefore they are `hidden' to lowest order, or wrap large cycles, so much so that the gauge theory they support is extremely weakly coupled. In either case, their mixing with SM gauge bosons is very small in a sense to be made more precise momentarily.
We shall refer to vectors arising in the closed string sector as {\em graviphotons}, and those from hidden branes as {\em dark brane photons}.

In this section, we evaluate couplings between the gravi/dark $U(1)$'s with the SM sector resulting from string scattering amplitudes and then compare them with the analogous couplings derived in the Effective Field Theory (EFT) of the holographic setup.
Specifically, we evaluate a) the kinetic mixing of these $U(1)$'s with SM gauge fields and b) other effective interactions with SM matter fields that can give an effective kinetic mixing once SM quantum corrections are included.
Such terms have been classified and studied in EFT, in section \ref{holosetup}.
{We perform our computations in flat four-dimensional space-time with internal tori, orbifolds or CY's at points in their moduli space where a CFT is available. However, we include closed-string fluxes both in the NSNS and RR sector that represent the first step towards a computation in a warped space such as AdS or alike.}

In this procedure, we neglect numerical factors and we provide the dependance of the couplings on the various scales of the two setups. In particular, in the string setup, we explicitly evaluate the powers of the string scale, the dependence on the volumes of the internal space, the distance of the D-branes etc and on the holographic setup, the dependance on $N$, the mass of the messengers etc. The details of the computations are provided in the appendix \ref{Mixing_string}.

\vskip 1cm

\subsection{Mixing between visible gauge bosons and closed-string `graviphotons'}
\label{mixing_closed_string}

\vskip 1cm

We define as `graviphotons' in this paper, all four-dimensional vectors that emerge in the closed string sector of type II string theory and its orientifolds.

We start with the couplings and mixings of the `graviphotons' with the SM vector bosons. We are mostly interested in chiral vacua with ${\cal N}=1$ SUSY in $D=4$. After a brief reminder of the situation in ten dimensions and in toroidal compactifications, we focus mostly on RR gravi-photons that appear in CY (orbifolds) or flux compactifications.
 This is because NSNS gravi-photons are only present when the compactification manifold has some topologically non-trivial 1-cycle such as $T^6$, $K3\times T^2$ or shift-orbifolds with lower or no supersymmetry (e.g., Scherk-Schwarz models). For completeness, in appendix \ref{NSNSgraviphotons} we study amplitudes involving $NSNS$ gravi-photons, when present in the massless closed-string spectrum, and show that they
interact the same way as RR gravi-photons, both in the absence and in the presence of bulk RR and NSNS fluxes.

In this section we set the stage of our analysis. We start with toroidal compactifications of orientifolds of Type II superstrings and study the mixing of the open-string vectors $A_\mu$, arising from the gauge fields $A_M$, with the closed string vectors $G_{\mu i}$ and $C_{\mu i}$, arising from the metric $G_{MN}$ and the RR 2-form $C_{MN}$. The Type I Lagrangian in $D=10$ contains terms of the form
\be
{\cal L}_{GVV} = G^{MN} G^{KL} Tr(F_{MK} F_{NL}) \quad , \quad {\cal L}_{CVV} = \der^{L}C^{MN} Tr(A_M F_{NL})
\label{3_1}\ee
In lower dimensions, setting $\mu,\nu =1, ... D$, $i=D+1,..., 10$, we have the following decompositions
\be
G_{MN} \rightarrow \{G_{\mu\nu}, G_{\mu i}, G_{ij}\}~,~~~~~
C_{MN} \rightarrow \{C_{\mu\nu}, C_{\mu i}, C_{ij}\} ~,~~~~~ A_{M} \rightarrow
\{A_{\mu}, \varphi_{i}\}
\label{3_2}\ee
Keeping only terms with $G_{\mu\nu}=\eta_{\mu\nu}$ and $G_{\mu i}$ or $C_{\mu i}$,
we obtain the following mixings
\be
{\cal L}_{GVV} = F_{G,i}^{\mu\nu} Tr(\varphi^i F_{\mu\nu}) \quad , \quad {\cal L}_{CVV} = F_{C,i}^{\mu\nu} Tr(\varphi^i F_{\mu\nu})
\label{3_3}\ee
where
\be
F^{\mu\nu}_{[G/C],i} = \der^\mu [G/C]^\nu_i - \der^\nu [G/C]^\mu_i \quad, \quad
F^{\mu\nu} = \der^\mu A^\nu - \der^\nu A^\mu
\label{3_4}\ee
Notice the similarity of the couplings of RR and of NSNS graviphotons. In the following, we extract these kinetic mixings from string amplitudes.

As mentioned earlier, we use vertex operators in four-dimensional flat space-time with internal tori, orbifolds or CY's with CFT description on the world-sheet.

\begin{figure}[t]
\centering
i.~~\includegraphics[height=0.30\textwidth]{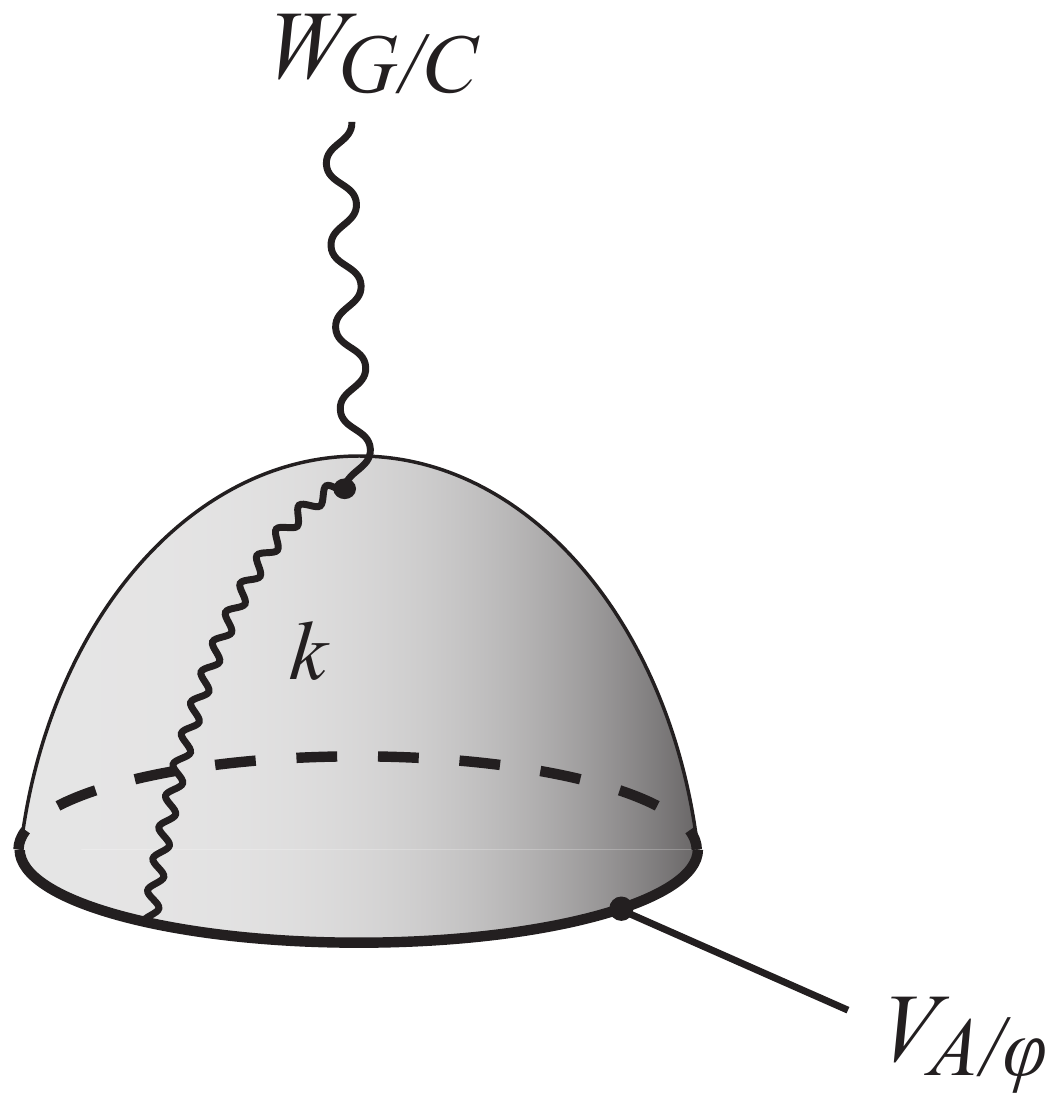}~~~~~~~~~~
ii.~~\includegraphics[height=0.30\textwidth]{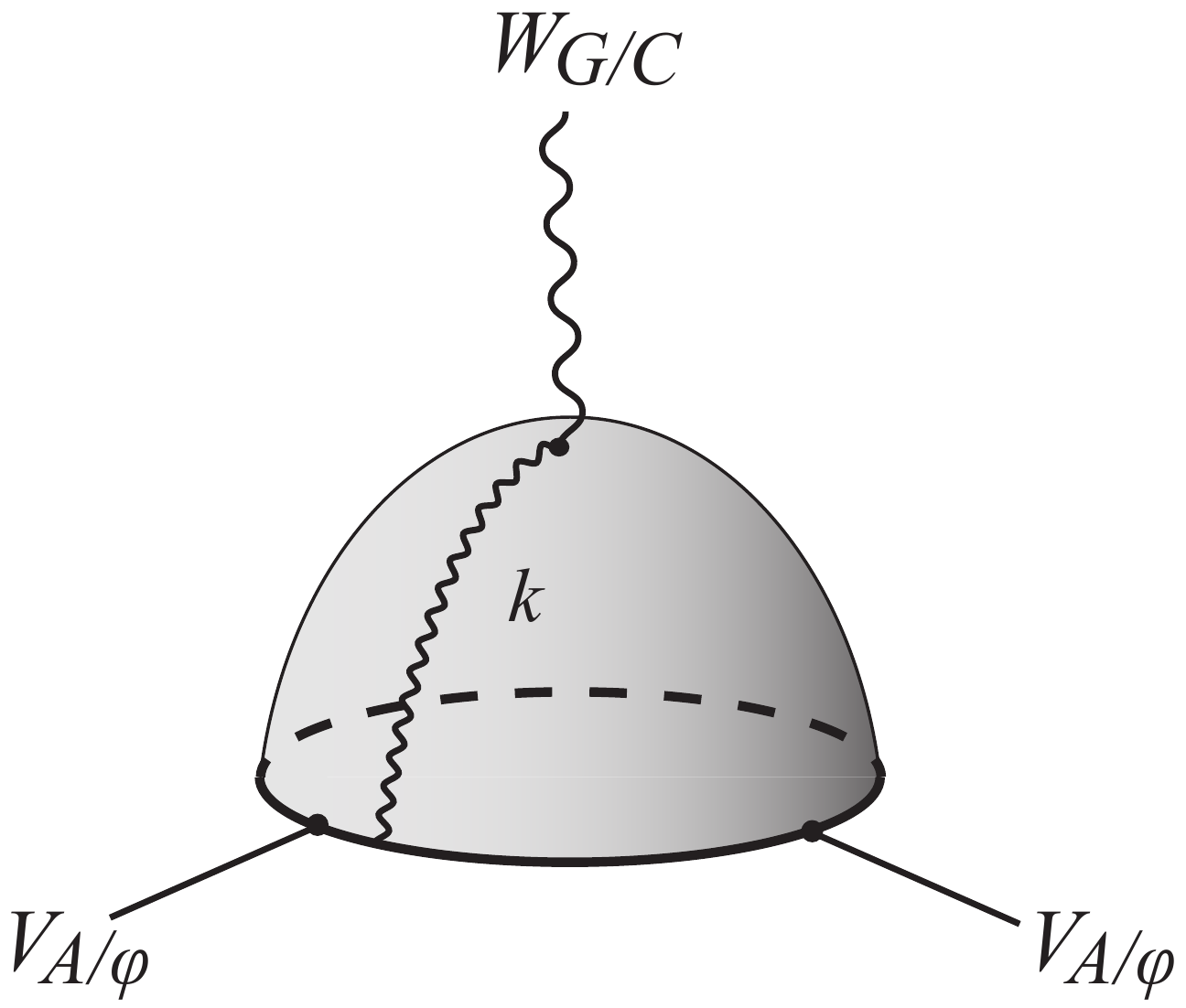}
\caption[]{The disk amplitude with the insertion of a closed-string $W_{G/C}$ in the bulk and (i.)
one or (ii.) two open-strings $V_{A/\phi}$ (${\cal N}=4$). The $k$ line denotes the twist.}
\label{WVtwist}
\end{figure}

The simplest amplitude to consider for the kinetic mixing between RR graviphotons and the SM gauge field is a disc with the insertion of only two vertex operators (VO's), as it is shown in fig \ref{WVtwist}. However, it vanishes, regardless of supersymmetry (as shown in Appendix \ref{Mixing_string}).

The next relevant string amplitude, that generates the coupling \eqref{3_3} in toroidal compactifications with ${\cal N}=4$ supersymmetry in $D=4$, is a disk amplitude, with one closed-string vertex $W_{RR}$ inserted in the bulk and two open-string vertices $V_{\phi/A}$ inserted on the boundary (figure \ref{WVtwist}.ii).
The amplitude is computed in the appendix \ref{Mixing_string}, and one finds the interaction term
\be
\D S= \int d^4x~g_s\, \frac{\ell_s}{\sqrt{{\mathcal{V}_6}}}~F_{RR,i}^{\mu\nu} Tr(\varphi^i F_{\mu\nu})
\label{3_5}\ee
which is suppressed by
\be
\sqrt{G_4}= {1 \over M_P } =g_s \frac{\ell_s^4}{\sqrt{V_{6}}}=g_s \frac{\ell_s}{\sqrt{{\mathcal{V}_6}}}
\label{3_6}\ee
where $G_4$ and $M_P$ are the Newton constant and the Plank mass in four dimensions, $g_s$ is the string coupling while $V_{6}$ is the compactification volume and ${\mathcal{V}_6}=V_6/\ell_s^6$ is the dimensionless compactification volume in units of the fundamental string scale.

In compactifications with lower supersymmetry, the situation changes.
\begin{itemize}
\item In models with ${\cal N}=2$ supersymmetry in $D=4$, the lowest-order mixing terms in the effective action are\footnote{A similar result obtains for NSNS gravi-photon which is associated to the `untwisted' direction $u$. See appendix \ref{NSNSgraviphotons} for details.}
\be
\D S=g_s \frac{\ell_s }{\sqrt{ {{\mathcal{V}_6}} }} \int d^4x~F_{{{RR}},u}^{\mu\nu} Tr(\varphi^u F_{\mu\nu})
\label{3_7}\ee
where $u=1,2$ are the `untwisted' directions that (usually) span a 2-torus $T^2$.

\item In models with ${\cal N}=1$ supersymmetry in $D=4$, the lowest-order mixing terms in the effective action are
\be
\D S=g_s \frac{\ell_s }{\sqrt{ {{\mathcal{V}_6}} }} \int d^4x~ F_{RR,h_{21}^-}^{\mu\nu} Tr(\varphi_{A} F_{\mu\nu})
\label{3_8}\ee
where $h_{21}^-$ denotes harmonic (2,1) forms that are odd under the un-oriented projection $\Omega$. $\varphi_A \equiv \text{Re} \, \phi_A$ denotes the real component of a complex scalar in a chiral multiplet, transforming in the adjoint of the gauge group. In principle, $\varphi_{A}$ is the internal component of a gauge-field in $D=10$ (or rather in $D=p+1$ for D$p$-branes with $p<9$) and therefore it can obtain a VEV and/or a flux ({\it e.g.} magnetised and/or intersecting branes or even T-branes \cite{Angelantonj:2000hi}-\cite{Collinucci:2014qfa}).
\end{itemize}

\subsubsection{Interaction terms in the absence of closed-string ``bulk'' fluxes}
\label{interactions_closed_string}

\begin{figure}[t]
\centering
\includegraphics[height=0.30\textwidth]{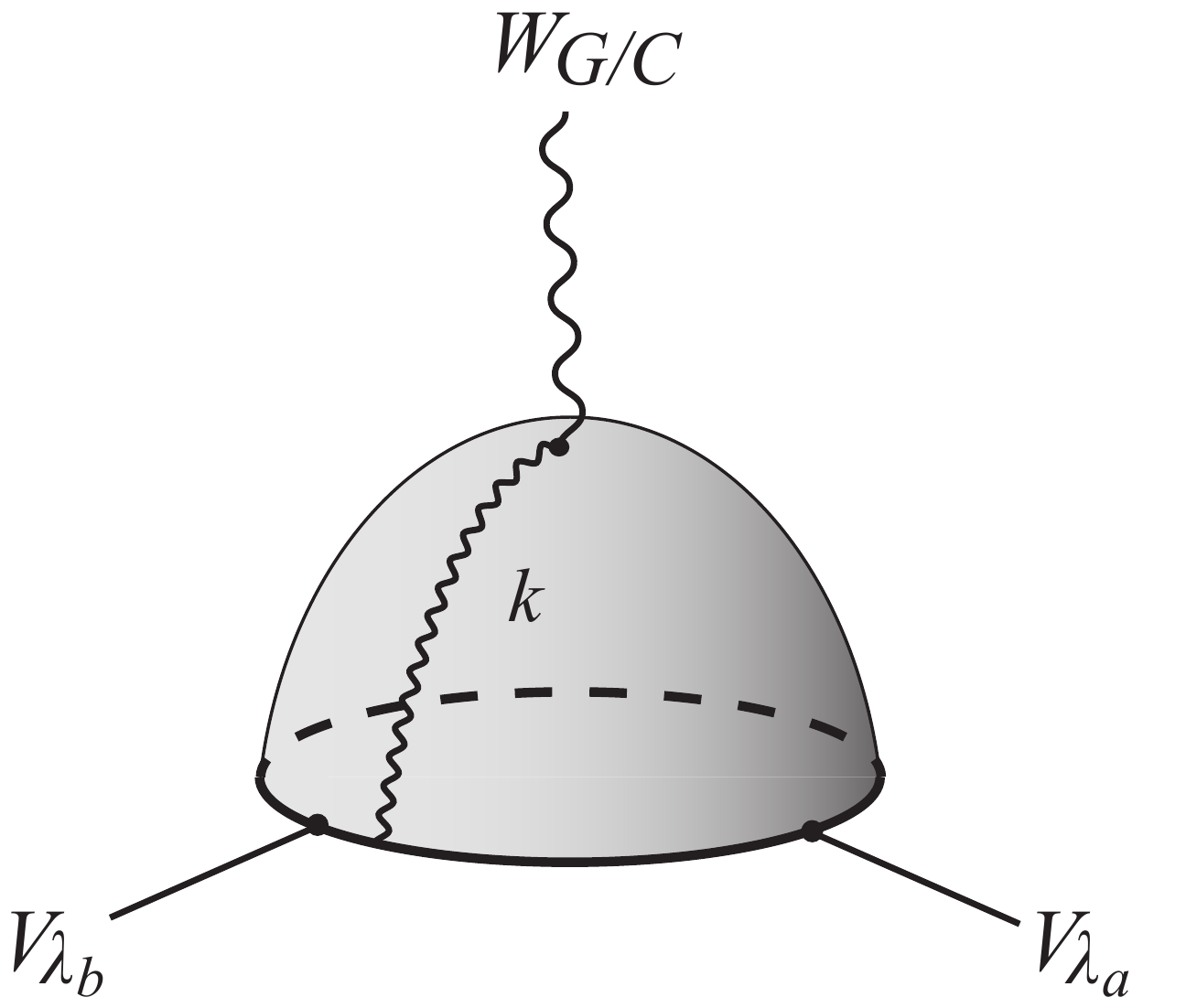}\\
\caption[]{The disk amplitude that contains the dipole coupling.}
\label{AphiAphidisk2}
\end{figure}

After the kinetic mixings, we investigate the interaction terms between the dark photon and the SM fields that mimic the couplings in \eqref{2_18}.

Starting from the couplings involving the Higgses $F^{\mu\nu}_{\hat{A}} D_\mu H^\dagger D_\nu H$, we have to consider disk amplitudes with a RR graviphoton in the bulk and with the insertion of two Higgses on the boundary. To reproduce the sub-leading coupling $F^{\mu\nu}_{\hat{A}} F^Y_{\mu\nu} H^\dagger H$ shown in \eqref{2_18}, we have to insert a SM gauge field too.

In the absence of bulk (closed-string) fluxes, the relevant amplitudes vanish at tree level (disk), and more details are provided in Section \ref{no_fluxes_computations}. Yet, the same amplitudes receive a non-vanishing contribution at one loop (annulus/cylinder $A$) in (twisted) sectors with $\langle \partial X^i \rangle_A\neq 0$ that is suppressed by an additional power of the string coupling $g_s$. Notwithstanding possible cancellations between powers of the momenta in the numerator with poles (denominator), the effective couplings are also suppressed by powers of $\ell_s^2 k_H{\cdot}k_{H^\dagger}$, which are however compatible with EFT suppression as we shall see later on.

\paragraph{Dipole couplings.}
The situation is different for interaction terms involving SM fermions. Dipole couplings are present at the two-derivative order in the low-energy expansion of the string theory effective action, \cite{book}, and can appear in particular in D-brane realizations of the SM, \cite{muon}.
They are of the form,
\be
{\cal L}_{F\chi\chi} = F^{\mu\nu}_{RR} {\rm tr}(\bar\chi \gamma_{\mu\nu} \chi)
\label{3_9}\ee
They may accompany the kinetic mixing terms $F^{\mu\nu}_{RR} {\rm tr}(\phi F_{\mu\nu})$ {\eqref{3_3}}.
Indeed, in toroidal compactifications, there are as many RR graviphotons $C_{\mu,i}$ as scalars $\phi_i$ in vector multiplets. The relevant coupling reads
\be
{\cal L}_{F\lambda\lambda} = F^{\mu\nu}_{RR, i} [\Gamma^i_{ab}{\rm tr} (\lambda^a \gamma_{\mu\nu} \lambda^b) +{\rm h.c.}]
\label{3_10}\ee
where $\lambda^a$ denote the gaugini (\eg ~$a=1,..,4$ in $D=4$).
One can extract the above coupling from a disk amplitude with one RR vertex in the bulk and two open-string fermion vertices on the boundary (see fig \ref{AphiAphidisk2}). Up to numerical factors, the interaction term in the effective action is the following
\be
\D S = g_s \frac{\ell_s}{\sqrt{ {{\mathcal{V}_6}} }} \int d^4x \l_{\a, a} F^{\a \b, a b}_{RR} \l_{\b, b}
\label{3_11}\ee
The overall constants are of order $M_P^{-1}$, this is in agreement with the normalization of $F^{\mu\nu}_{RR} {\rm tr}(\phi F_{\mu\nu})$ obtained in \eqref{3_5}.

In models with reduced supersymmetry, the situation is subtler and richer.

\begin{itemize}
\item In models with ${\cal N}=2$ supersymmetry, RR photons are in vector multiplets and they can have non-minimal dipole couplings only to gaugini. Indeed, by explicit computation of the relevant disk amplitudes, one can check that
\be
{\cal A}_{{\cal F}\lambda\lambda} = g_s \frac{\ell_s}{\sqrt{ {{\mathcal{V}_6}} }}
~ F_{RR}^{\mu \nu} {\rm tr}
(\lambda^r \sigma_{\mu\nu} \lambda^s) \varepsilon_{rs} \qquad {\rm while} \qquad {\cal A}_{{\cal F}\zeta\zeta} = 0
\label{3_12}\ee
where $\lambda$ denote open-string gaugini while $\zeta$ denote (open-string) hyperini.

The easiest way to derive the above selection rule is based on the $U(1)$ R-symmetry of ${\cal N}=2$ supersymmetry. Vertex operators for gaugini carry R-charge $r=+1/2$ while hyperini have $r=-1/2$. Since VO's for RR photons with negative helicity \ie $\,(\alpha\beta)$ involve spectral flow of $(c,c)$ deformations, they carry R-charge ($-1/2,-1/2$) and can only combine with gaugini. Positive helicity RR photons with R-charge ($+1/2,+1/2$), couple to anti-gaugini and not to hyperini, due to Lorentz invariance constraints.

\item In models with ${\cal N}=1$ supersymmetry, RR photons are in vector multiplets, associated to $h_{2,1}^-$ complex structure deformations that are odd under $\Omega$, and they can have non-minimal dipole couplings to gaugini combined with matter fermions in the adjoint
\be
{\cal A}_{{\cal F}\lambda\lambda} = g_s \frac{\ell_s}{\sqrt{ {{\mathcal{V}_6}} }} ~ F^{\mu \nu}_{RR,({-}{1\over 2},{-}{1\over 2})} {\rm tr}
(\lambda_{{+}{3\over 2}} \sigma_{\mu\nu} \chi_{{-}{1\over 2}})
\label{3_13}\ee
where the subscripts denote the $U(1)$ R-charge (that can be identified with the $U(1)$ R-charge of the underlying ${\cal N}=2$ SCA on the world-sheet) of the VO's in the canonical super-ghost picture.
\end{itemize}

The above dipole couplings (\ref{3_11}-\ref{3_13}) perfectly match with the kinetic mixing term $ F^{\mu\nu}_{RR} {\rm tr} (F_{\mu\nu} \phi)$ with $F$ the super-partner of $\lambda$ and $\phi$ the one of $\chi$. In fact both terms arise from the unique manifestly ${\cal N}=1$ supersymmetric gauge-invariant term
\be
{\cal W}^\alpha_{RR} {\rm tr} (W_\alpha \Phi)
\label{3_14}\ee
where $\Phi$ is a chiral multiplet in the adjoint.
Similar dipole couplings exist for `hidden' brane photons with messenger fermions at tree level (disk $D$) or with SM fermions at one loop in models with lower or no supersymmetry.

For chiral fermions, as in the SM, gauge invariance of the dipole couplings require the insertion of a Higgs field
the interaction term in the effective action is the following
\be
\D S = g^{3/2}_s \frac{\ell^2_s}{\sqrt{ {{\mathcal{V}_6}} }} \int d^4x H \psi^{(1)}_{\a} F^{\a \b}_{RR} \psi^{(2)}_{\b}
\label{3_14a}\ee
The corresponding amplitude
\be
{\cal A}_{{\cal F}H\psi\psi} = g^{3/2}_s \frac{\ell^2_s}{\sqrt{ {{\mathcal{V}_6}} }} F^{\mu\nu}_{RR} H \psi_1\sigma_{\mu\nu} \psi_2
\label{3_14b}\ee
is suppressed wrt (\ref{3_13}) by an additional power of $\sqrt{g_s}$ for the open-string vertex of the Higgs as well as a power of $\ell_s$ by dimensional analysis. It may also receive further suppression in $\ell_s$ due to possible higher powers of the momenta resulting from the numerator.

\paragraph{Other couplings with fermion bilinears.}

With the further insertion of a SM gauge boson vertex on the disk, we can also compute the coupling $F^{\mu \nu} \hat{F}_{\mu \nu} H \bar{\psi} \psi$ in \eqref{2_18} and obtain the interaction term:
\be
\Delta S = \int d^4x \, g_s^{5/2} \frac{\ell_s^4}{\sqrt{ {{\mathcal{V}_6}} }} \,
F^{RR}_{\mu \nu} F_A^{\mu \nu} H \psi_\alpha^{(1)} \psi^{^{(2)}\alpha}
\label{3_15}
\ee
The details are provided in Appendix \ref{no_fluxes_computations}.

\subsubsection{Mixing in the presence of closed-string (bulk) fluxes}
\label{mixing_closed_string_fluxes}

In view of the later comparison with holographic settings, we now consider the effect of bulk fluxes on the mixing and the couplings with the SM. For clarity, we henceforth work in a T-dual framework with D3-branes. We work at linear order in the fluxes and perform all the computations in flat target space-time with internal tori, orbifolds or solvable SCFT describing CY's.

The VO's for the fluxes in ten-dimensions are the followings
\begin{align}
W_\textup{RR,flux}^{(-\frac{1}{2},-\frac{1}{2})}&= C_{RR}e^{-\frac{\varphi(z)}{2}} e^{-\frac{\tilde{\varphi}(\bar{z})}{2}} \calF_{AB} S^A(z) \tilde{S}^B(\bar{z})e^{i K_2 {\cdot} X_L (z)} e^{i \tilde{K}_2 {\cdot} X_R (\bar{z})} \label{3_16}\\
W_\textup{NSNS,flux}^{(-1,0)}&= C_G \mathcal{H}_{PQR} \left[ e^{-\varphi} \Psi^P
(\tilde{X}^Q \bar{\partial} \tilde{X}^R{+}\tilde{\Y}^Q \, \tilde{\Psi}^R){+}
(L{\leftrightarrow}R) \right] e^{i K_2 {\cdot} X_L} e^{i \tilde{K}_2 {\cdot} X_R}
\label{3_17}
\end{align}
The normalizations $C_{RR}$ and $C_{G}$ are shown in Appendix \ref{Mixing_string}. In $D=4$, the RR and NSNS fluxes decompose as
\begin{align}
{\cal F}_{[AB]} &= {\cal F}_{MNP}\Gamma^{MNP}_{[AB]} \rightarrow
{\cal F}_{ijk}\varepsilon_{\a\b} \Gamma^{ijk}_{(rs)} + ...
={\cal F}_{(rs)} \varepsilon_{\a\b} + ...
\label{3_18}\\
\mathcal{H}_{PQR} & \rightarrow
\mathcal{H}_{ijk} +\dots \qquad, \qquad \mathcal{H}_{(ab)}= \Gamma^{ijk}_{(ab)} \mathcal{H}_{ijk}
\label{3_19}
\end{align}
By `...' above, we denote all terms that contain fluxes with Lorentz indices, like ${\cal F}_{\m ij}$ etc, and that are irrelevant for our present analysis.

In a complex basis
\be
{\cal F}_{ijk} \rightarrow {\cal F}_{(3,0)} \oplus {\cal F}_{(2,1)} \oplus {\cal F}_{(1,2)} \oplus {\cal F}_{(0,3)}
\label{3_20}\ee

\begin{figure}[t]
\centering
\includegraphics[width=0.4\textwidth]{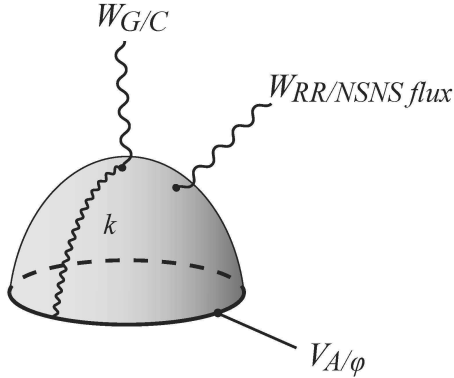}
\caption[]{The disk amplitude with the insertion of a closed-string $W_{G/C}$ in the bulk and an open-string $V_{A/\phi}$ on the boundary. The insertion of an extra $W_{RR/NSNS~flux}$ VO is considered.
}
\label{WWVdisk}
\end{figure}

\paragraph{Fermion masses from fluxes.}

We incidentally notice that starting from the
$\Lambda^t \Gamma^{ijk} \Lambda {\cal F}_{ijk}$
term in $D{\,=\,}10$, where $\Lambda$ denote the Majorana-Weyl gaugini, a flux ${\cal F}_{(0,3)}$ is known to generate a susy-breaking mass for the gaugini $\lambda$ in $D{\,=\,}4$. The amplitude associated, is similar to the dipole coupling, however, in this case we are interested in the components of the ten-dimensional RR flux shown in \eqref{3_18} that yield,
\be
\calA= g_s \frac{\ell_s}{\sqrt{ {{\mathcal{V}_6}} }} \, \tilde{\lambda}_{a,\a} \lambda^\a_b \calF^{(ab)}
= \frac{\calF^{(ab)}}{M_P}\, \tilde{\lambda}_{a,\a} \lambda^\a_b
\label{3_21}\ee
where $M_P$ is the four-dimensional Plank mass. Here $\calF^{(ab)}$ should be interpreted as a background field instead of a dynamical field, therefore the amplitude above should be considered as a `2-{point}' amplitude, and $\calF^{(ab)}$ has mass dimension 2. The corresponding term in the effective action is then
\be
\D S=\int d^4x~\frac{\calF^{ab} }{M_P}~ \tilde{\lambda}_{a,\a} \lambda^\a_b
\label{3_22}\ee

On the other hand, ${\cal F}_{(2,1)}$ combined with the NSNS flux ${\cal H}_{(2,1)}$ and the dilaton ${\cal T}$ can generate masses for vector-like pairs $\chi, \tilde{\chi}$ of brane fermions. This can be done in a supersymmetric way, if the fluxes and the dilaton give rise to an imaginary (anti)-self dual primitive (2,1)-form, ${\cal G}_{(2,1)} = {\cal F}_{(2,1)} +{\cal T} {\cal H}_{(2,1)} = i * {\cal G}_{(1,2)}$ with ${\cal J}_{(1,1)} \wedge {\cal G}_{(2,1)} = 0$.

\paragraph{Mixing SM/brane-photon / RR gravi-photon.}
The amplitudes corresponding to the kinetic mixings are shown in fig \ref{WWVdisk} and computed in Appendix \ref{string_setup_mixing_RR_fluxes}. Up to numerical factors, the amplitudes yield the following effective interaction terms:
\begin{align}
\Delta S_\textup{RR flux}&= \int d^4 x\,
g_s^{3/2} \frac{\ell_s^2}{ {{\mathcal{V}_6}} }\,
 Tr(\gamma_k F^{\mu\nu}) F_{RR,\mu\nu}^{(ab)} {\cal F}_{(ab)}
\label{3_23}\\
\Delta S_\textup{NSNS flux}&= \int d^4 x\,
g_s^{3/2} \frac{\ell_s^2}{ {{\mathcal{V}_6}} }\,
 Tr(\gamma_k F^{\mu\nu}) F_{RR,\mu\nu}^{(ab)} {\cal H}_{(ab)}
\label{3_24}
\end{align}
where $F^{(ab)}_{RR,\mu\nu}$ is the relevant component of the ten-dimensional field strength $F_{RR,MN}$.
{Notice here the pairing of the internal indices $a,b$ in the `untwisted' RR gravi-photon vertex, with the ones in the `untwisted' RR/NSNS fluxes. For `twisted' or more general RR gravi-photons, the relevant pairing is dictated by world-sheet selection rules on the string amplitudes.}

\subsubsection{Interaction terms in the presence of closed-string (bulk) fluxes}
\label{interactions_closed_string_fluxes}

With computations similar to the ones we have performed in the absence of fluxes, we consider the effect of fluxes in the following.

\paragraph{Dipole couplings to SM fermions.}

In comparison to the previous case, the amplitudes that should yield the couplings with SM fermions are subleading. The reason is the presence of a fermionic VO in the picture $+1/2$ in the amplitudes:
\begin{align}
{\cal A}_{\bar{\psi} \hat{F} \psi}&{=} \big\langle c\tilde{c} W^{(-\frac{1}{2},-\frac{1}{2})}_{RR, flux}(i,{-}i) \int W^{(-\frac{1}{2},-\frac{1}{2})}_{RR}(z, {\bar z}) d^2 z\, V^{(-\frac{1}{2})}_{\tilde{\psi}}(x_1)\, c V^{(+\frac{1}{2})}_\psi(x_2) \, \big\rangle \label{3_25}\\
{\cal A}_{\hat{F} F \bar{\psi} \psi}&{=} \big\langle c\tilde{c} W^{(-\frac{1}{2},-\frac{1}{2})}_{RR, flux}(i,{-}i) \int W^{(-\frac{1}{2},-\frac{1}{2})}_{RR}(z, {\bar z}) d^2 z\, V^{(-\frac{1}{2})}_{\tilde{\psi}}(x_1)\, V^{(+\frac{1}{2})}_\psi(x_2) \, c V^{(0)}_A(x_3) \big\rangle \label{3_26}
\end{align}
The VO $V^{(+1/2)}_\psi$ contains the term $k^\mu_\psi \Psi_\mu$\footnote{Here $\Psi_\mu$ denotes a world-sheet fermion.}. This momentum cannot be combined to form a field strength, because $\hat{F}$ ($F$) is already present in the VO $W^{(-1/2,-1/2)}_{RR}$ ($V^{(0)}_A $). Therefore the interaction terms coming from these amplitudes are sub-leading in comparison to couplings in \eqref{2_18}.

Pretty much as in the absence of closed-string fluxes, for chiral fermions, like in the SM, gauge invariance requires the inclusion of a Higgs field leading to a coupling of the form
\be
{\cal F}_3 F^{\mu\nu}_{\hat{A}} H \psi_1\sigma_{\mu\nu} \psi_2
\ee
The relevant amplitude is quite laborious to compute but should receive a non-zero contribution already at disk level with a suppression of $\sqrt{g_s}$ due to the extra open-string insertion (Higgs) and possibly a further suppression by powers of $\ell^2_s k_ik_j$ where $k$'s are the momenta involved in the process.

\paragraph{Couplings to Higgses or scalar messengers.}
As for the interaction terms involving the Higgses, we find the following terms
\begin{align}
\D S &= \int d^4x~ g_s^2 \frac{\ell_s^4}{ {{\mathcal{V}_6}} } \calF_3 F_{RR}^{\mu \nu} \, \partial_\mu \tilde{\phi}\, \partial_\nu \phi
\label{3_27}\\
\D S &= \int d^4x~ g_s^{5/2} \frac{\ell_s^6}{ {{\mathcal{V}_6}} }\,
\partial_\rho\tilde{\phi}{\cdot} \partial^\rho\phi \,
\calF_3 F_{RR,\mu \nu} F^{\mu \nu}
\label{3_28}
\end{align}
The former may appear at tree level, depending on the `overlap' of the flux $\calF_3$ with the internal components of the VO's for the RR field and for the two Higgses, that can be replaced by messenger scalars. The latter may only arise at one loop, if the two scalars are SM Higgses or at disk level if the two scalars are messengers.

\subsection{Mixing between visible and dark open-string vectors}
\label{mixing_open_string}

In the following, we evaluate the mixing of visible gauge bosons, denoted by $A_{ab}$ with open-string vector bosons $A_{ij}$ that arise from hidden or flavour branes.

\paragraph{Mixing at tree level.}

\begin{figure}[t]
\centering
\includegraphics[width=0.45\textwidth]{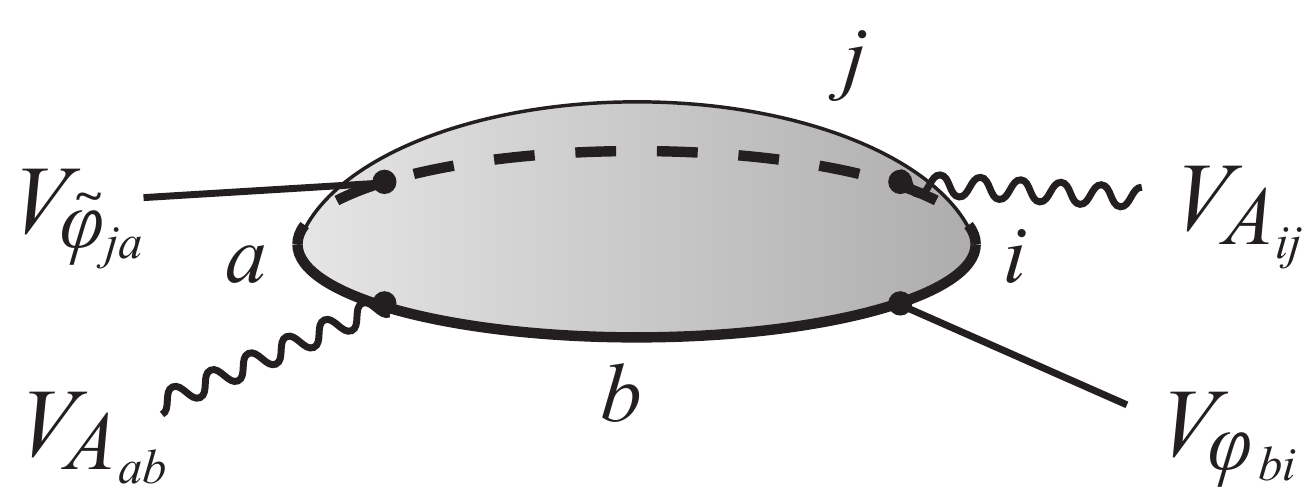}\\
\caption[]{The disk amplitude which gives the coupling between $A_{ab}$ and $A_{ij}$ when the scalar messengers $\phi_{bi}$, $\tilde{\phi}_{ja}$ acquire VEVs.}
\label{AphiAphidisk}
\end{figure}

Mixing terms can appear at tree-level (disk) if there are scalar `messengers' denoted by $\phi_{bi}$, $\tilde{\phi}_{ja}$ which obtain VEV's\footnote{{In weakly-coupled string theory, scalar messengers are typically massive with positive mass squared . Therefore, they  cannot acquire VEV's. Yet one cannot exclude the possibility of some Higgs mechanism involving them, in cases where the mass-squared is negative. This appears in several cases where supersymmetry is broken.}}. The relevant amplitude is (fig \ref{AphiAphidisk})
\be
\big\langle V^{(0)}_{ab} V^{(-1)}_{\phi_{bi}} V^{(0)}_{ij} V^{(-1)}_{\tilde{\phi}_{ja}} \big\rangle_{disk}
= g_s ~ Tr[{T}_{ab} {T}_{\phi_{bi}} {T}_{ij} {T}_{\tilde{\phi}_{ja}}] ~a_1\cdot a_3 ~\phi\cdot \tilde \phi
\label{3_29}\ee
where $V^{(0)/(-1)}_i$ are the VO's of $A_{ab}$, $A_{ij}$, ${\phi_{bi}}$, ${\tilde{\phi}_{ja}}$ fields in $0$ and $-1$ picture. Also $T_i$'s are the Chan-Paton factors and $a_1$ and $a_3$ the polarizations of $A_{ab}$, $A_{ij}$. More details are given in the appendix \ref{string_setup_mixing_scalars}.

\paragraph{Mass mixing at one loop.}

Mass and kinetic mixings between vector bosons can be also generated at one loop (annulus) via messenger\footnote{Here by messenger we mean all the states, including massive open strings \cite{Anastasopoulos:2011hj}-\cite{Anastasopoulos:2016yjs}, carrying charge with respect to both visible and hidden/dark photons. Clearly the contribution of individual very massive states will be suppressed in the loop but the high degeneracy can compensate in models with low string tension.} loops
\be
\big\langle V^{(0)}_{ab} V^{(0)}_{ij} \big\rangle_{ann} \sim Tr(\gamma^{(a)}_K Q_{a}) Tr(\gamma^{(i)}_K Q_{i})
\sim (B^r_i N_i) (B^s_a N_a) \kappa_{r,s}
\label{3_30}\ee
where $\kappa_{rs}$ denotes the kinetic mixing matrix for closed-string moduli and $B^r_i$, $B^s_a$ denote the bulk-boundary coupling of the closed-string sectors $r,s$ to the open-string sectors $i$, $a$. Indeed, the relevant factorisation channel is via closed-string axions that couple to both visible $a$ and dark $i$ sectors via mixed disk amplitudes
\be
\big\langle V^{(0)}_{A,i} W^u_{ax} \big\rangle = {\cal M}^s_i = M_s Tr(\gamma_{(i)}^s Q_i) \quad , \quad \big\langle V^{(0)}_{A,a} W^s_{ax} \big\rangle = \widetilde{\cal M}^s_a = M_s Tr(\gamma_{(a)}^s Q_a)
\label{3_31}\ee
where $M_s$ denotes the string scale, that can be hierarchically smaller than the Planck scale. Summing over the various sectors $s$, we obtain a mass-mixing matrix
\be
{\cal M}^2_{ia} = \sum_{s,r} {\cal M}^s_i \widetilde{\cal M}^r_a \kappa_{r,s}
\label{3_32}\ee
Notice that the kinetic mixing matrix for the axions $\kappa_{rs}$ may decompose into several mutually orthogonal blocks. In orbifold models, for instance, `fractional' D-branes located at different fixed points couple to different axions so the above amplitudes can be zero. This is the case also for `regular' D-branes of different kinds, {\it e.g.} `colour' D3-branes and `flavour' D7-branes, since the first couple to the axio-dilaton $S$ while the latter couple to the $T$ modulus of the `divisor' they wrap, so much so that the overlap is zero.

\begin{figure}[h]
\centering
\includegraphics[width=0.9\textwidth]{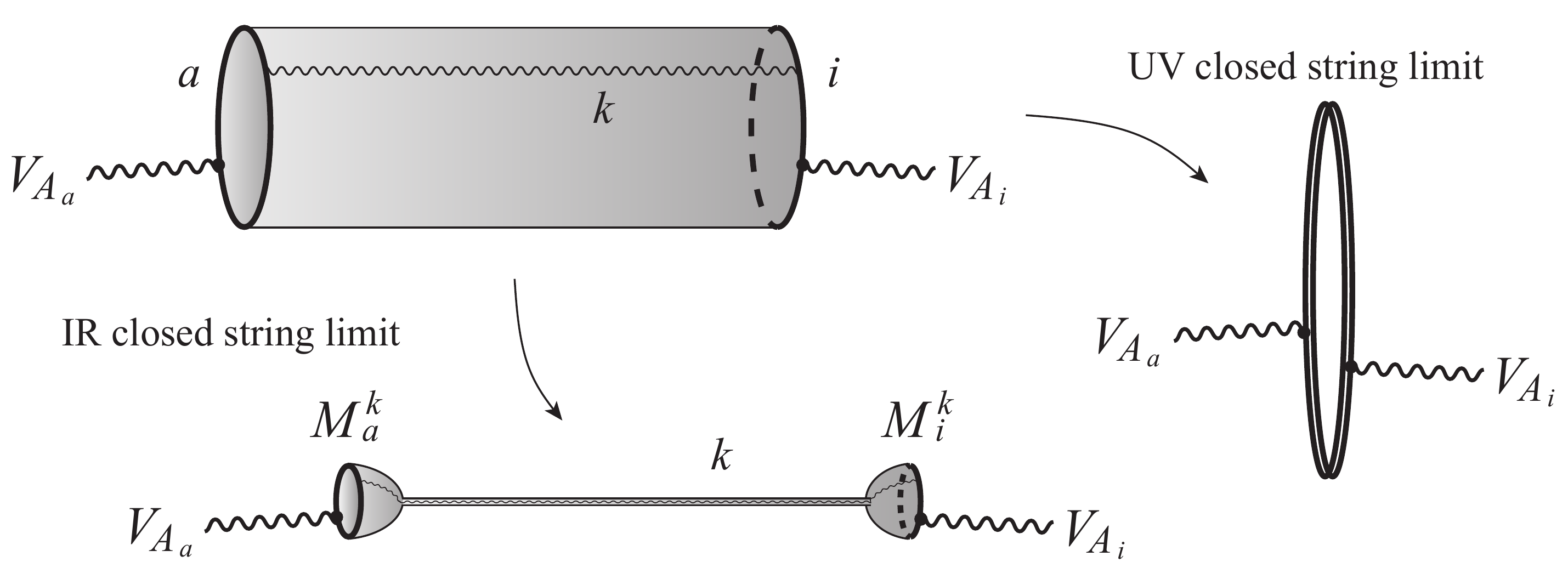}\\
\caption[]{The one-loop amplitude and the factorisation at the long tube limit where axions propagate in-between.}
\label{AA1loop}
\end{figure}

\paragraph{Kinetic mixing at one loop.}
In addition to mass-mixing, we obtain also kinetic mixing of the vis/dark photons at one loop. The former comes from the massless axion poles in the closed-string transverse channel \cite{AKR} that by factorisation produces the above result, the latter from (possibly light) massive poles in the closed-string channel \cite{AbelGoodsell}. Two-point functions are subtle in string theory since they cannot be viewed as {\it bona fide} scattering amplitudes. Yet, very much as in the computation of threshold corrections, one can either relax momentum conservation \cite{Minahan}-\cite{Klapuetc} or insert a soft closed-string modulus field (dilaton or alike). Anyway, the result can be un-ambiguously found and in the helicity basis reads \cite{MBConsoli,Bianchi:2006nf}
\be
{\cal A}^{\rm 1-loop}_{a,i} = g_s \ell_s^4 \,{\rm tr} (Q_a) {\rm tr} (Q_i) F_a^{\pm}(k_1) F_i^{\pm}(k_2) \int_0^\infty T dT ( {\cal E}_{\cal N} \pm i {\cal C}_{{\cal N}}) \int_0^1 d{{z}} \, e^{ \ap k_1{\cdot}k_2 \mathcal{G}({{z}})}
\label{3_33}
\ee
where $F^{\pm}(k)$ denote complex (anti)self-dual field strengths, $\mathcal{G}({{z}})$ is the bosonic propagator on the annulus, ${\cal N}$ denotes the number of SUSY preserved in the relevant `messenger' sector\footnote{ ${\cal N} = 4$ sectors do not contribute {\it i.e.} ${\cal E}_{{\cal N} = 4} = 0$
and ${\cal C}_{{\cal N} =4} =0$. For ${\cal N} = 2$ sectors ${\cal C}_{{\cal N} =2}=0$ due to uncancelled fermionic zero-modes.}. The functions ${\cal E}_{\cal N}$ and ${\cal C}_{{\cal N}}$ are defined in Appendix \ref{string_setup_mixing_scalars}.

In the IR limit, $T \to \infty$, the contributions from ${\cal E}_{\cal N}$ and ${\cal C}_{{\cal N}}$ are dominated by the lightest particles circulating in the loop. In our case, they are the lightest messengers, whose masses depend on the branes separation, $M = \Delta x/\alpha'$. Therefore ${\cal E}_{\cal N}$ and ${\cal C}_{{\cal N}}$ reduce to
\be
{\cal E}_{\cal N},{\cal C}_{{\cal N}} \sim \frac{e^{-\ap \pi M^2 T}}{(\alpha' T)^2}
\label{3_34}\ee
Regarding $T$ as a Schwinger parameter, the amplitude can be seen as an integral over a four-dimensional loop momentum, whereby ${{z}}$ assumes the role of the Feynman parameter. The integral over the loop momentum is logarithmically divergent and can be regulated, for instance, using a cut-off regularization. The leading contribution of the messengers at low energy, is then
\be
\Delta S \sim \int d^4 x \,g_s \, {\rm tr} (Q_a) {\rm tr} (Q_i)
F^{\mu \nu}_a F_{i,\mu \nu}\log \frac{\Lambda^2}{M^2}
\label{3_35}\ee
which is in line with field theory expectations in this limit.

\subsubsection{Interaction terms between visible and dark open-string vectors}
\label{interactions_open_string}

Interactions terms involving SM fermions are subleading in comparison to the couplings in \eqref{2_18}. The reason is the same as for the fermionic interaction terms in the presence of fluxes in Section \ref{interactions_closed_string_fluxes}, \ie , one of the fermionic VO is in the picture $+1/2$ that yields an extra momentum factor.

\paragraph{Couplings of Higgses with dark brane photons.}
To obtain the effective coupling $D_\mu H D_\nu H F^{\mu \nu}$ we consider the 3-{point} amplitude on the annulus with two scalars in one boundary and a `dark photon' in the other boundary. As shown in Appendix \ref{string_setup_mixing_scalars} we obtain the effective interaction term
\be
\Delta S \sim \frac{g_s^{3/2}}{M^2}  {\rm tr} (Q_a) {\rm tr} (T_i T_j) \int d^4 x \, F_{a}^{\mu \nu} \der_{[\mu} \varphi^i \der_{\nu]} \widetilde{\varphi}_i
\label{3_36}
\ee

The next coupling involving Higgses, can be obtained by inserting another SM gauge boson, thus obtaining a 4-{point} amplitude. The calculations are quite involved. However, one can estimate the scales and the power of the string coupling:
\be
\Delta S \sim \frac{g_s^2}{M^4} \int d^4 x \, F_{a}^{\mu \nu} F^{SM}_{\mu \nu} \der_{\rho} \varphi^i \der^{\rho} \widetilde{\varphi}_i
\label{3_37}
\ee

\section{Comparison between the QFT and string theory pictures}
\label{comparison}

In this section, we compare the kinetic mixings and couplings obtained in section \ref{holosetup} using effective field theory and in section \ref{stringsetup} using string theory.
The rational for doing this is the string-theory/gauge theory correspondence.

The reason that such a comparison is interesting, is because the AdS/CFT correspondence indicates that weakly-coupled string theories are dual to strongly coupled, large N (and holographic) QFTs.
Therefore, we expect to learn about strongly coupled QFTs from weakly coupled string theory and about about strongly coupled string theory from weakly coupled QFTs.

The vectors emerging from global symmetries of the hidden QFT, are similar to graviphotons and other bulk U(1)'s like the RR U(1)'s of closed string theory.
Vectors emerging from flavour sectors, like the messenger sector, are qualitatively similar to  U(1)'s appearing on D-branes distinct from the SM D-brane stack in orientifold constructions that we called dark brane photons.
This holographic map, allows a correspondence between calculations in a strongly coupled large $N$-theory and related calculations in string theory.

Of course, the string theory calculations are done in string theory around flat space, while the string theory dual of the field theory framework in (\ref{i1}) is expected to by in a non-trivial asymptotically AdS gravitational background. However, as we know from the holographic correspondence, the leading dependence of couplings in the effective gravitational theory, especially on the string coupling,  is expected to be similar in the two frameworks.

\subsection{Closed sector\label{css}}

\begin{table}[h!]
\begin{tabular}{llllll}
&~{\bf EFT coupling} & &
{\bf string theory diagram}
&~~~& {\bf string theory}
 \\
~~~&~ & &
&~~~& {\bf effective coupling}
 \\
 1.&~$\displaystyle{\Lambda^2 \over NM^2} F^{\mu \nu} \hat F_{\mu \nu}$&&
\raisebox{-7mm}{\includegraphics[width=0.25\textwidth]{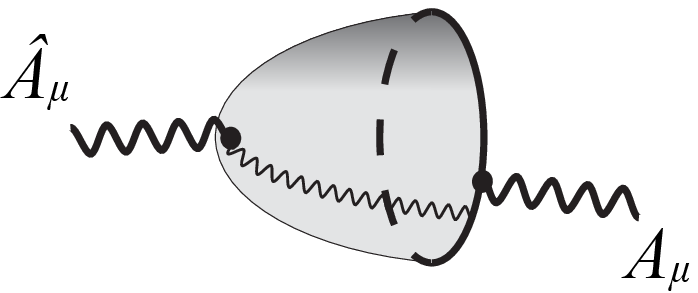}}
&~~~& $\displaystyle{0\cdot g_s+{\cal O}(g_s^2)}$
\\
2.&~$\displaystyle{\Lambda^2 \over NM^2} F^{\mu \nu} \hat F_{\mu \nu}$ &&
\raisebox{-7mm}{\includegraphics[width=0.25\textwidth]{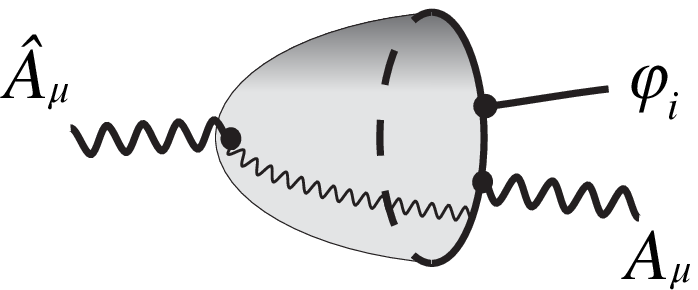}}
&~~~& $\displaystyle{g_s \frac{\ell_s}{\sqrt{ {{\mathcal{V}_6}} }} \,
\varphi^i \, \hat{F}_i^{\mu\nu} F_{\mu\nu} }$
\\
3.&~$\displaystyle{1\over NM^2} D_\mu H^\dagger D_\nu H \hat F^{\mu \nu}$ &&
\raisebox{-7mm}{\includegraphics[width=0.25\textwidth]{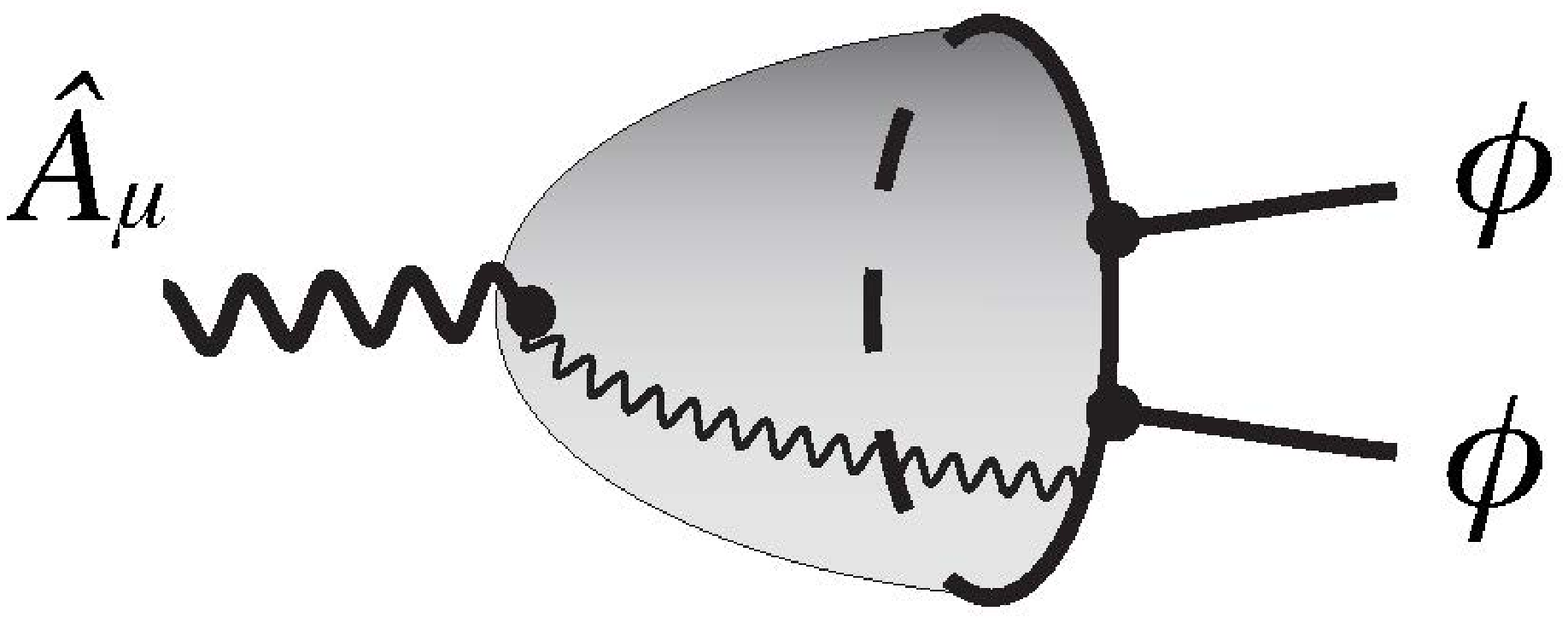}}
&~~~& $\displaystyle{0\cdot g_s+{\cal O}(g_s^2)}$
 \\
4.&~$\displaystyle{1\over N^{\frac{3}{2}} M^2} \bar \psi \gamma_{\mu \nu} H \psi\hat F^{\mu \nu}$ &&
\raisebox{-7mm}{\includegraphics[width=0.25\textwidth]{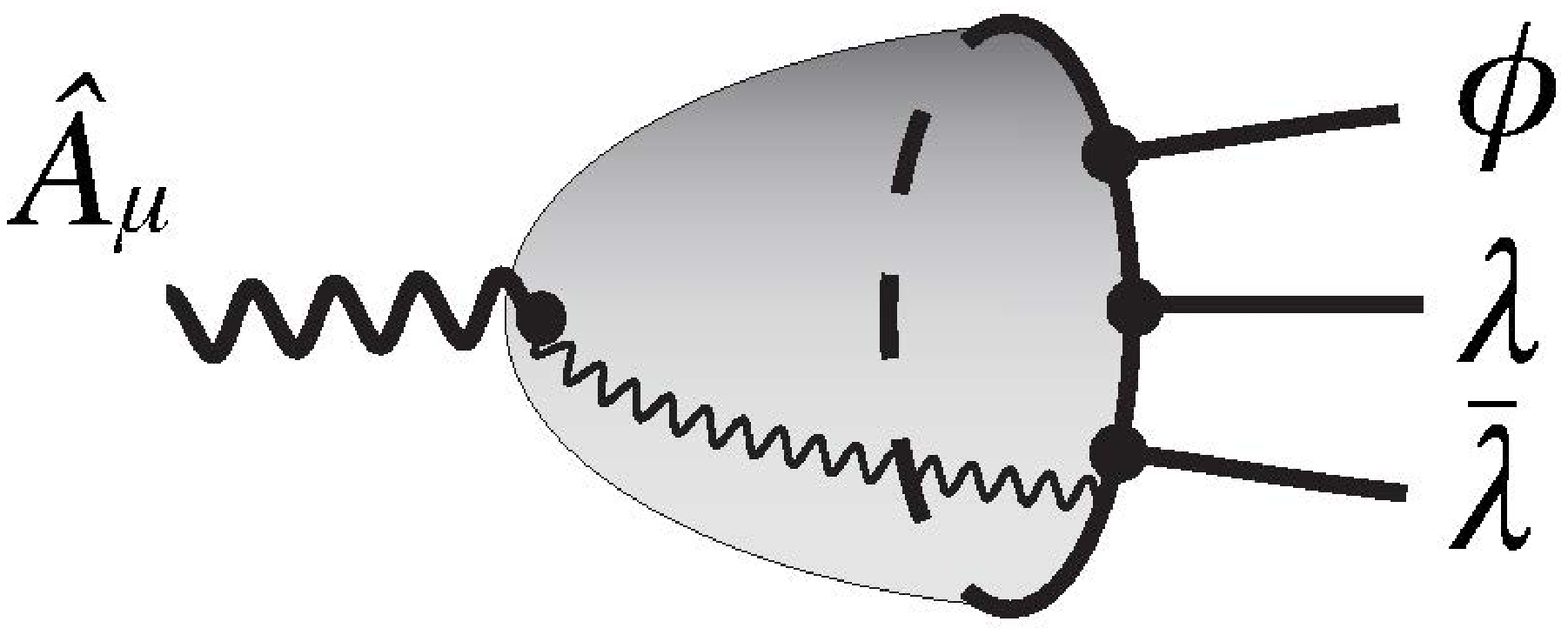}}
&~~~& $\displaystyle{g_s^{\frac{3}{2}} \frac{\ell_s^2}{\sqrt{ {{\mathcal{V}_6}} }} v_\a \hat F^{\a\b}\varphi v_\b}$
\\
5.&~$\displaystyle{1\over N^{3\over 2}M^2} F^{\mu \nu}\hat F_{\mu \nu}~H^\dagger H$ &&
\raisebox{-7mm}{\includegraphics[width=0.25\textwidth]{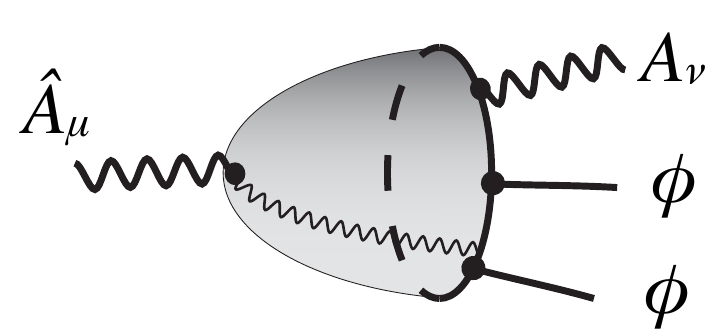}}
&~~~& $\displaystyle{0\cdot g_s^{3\over 2}+{\cal O}(g_s^{5\over 2})}$
\\
6.&~$\displaystyle{1\over N^{2}M^4} F^{\mu \nu}\hat F_{\mu \nu} \bar \psi H \psi$ &&
\raisebox{-7mm}{\includegraphics[width=0.25\textwidth]{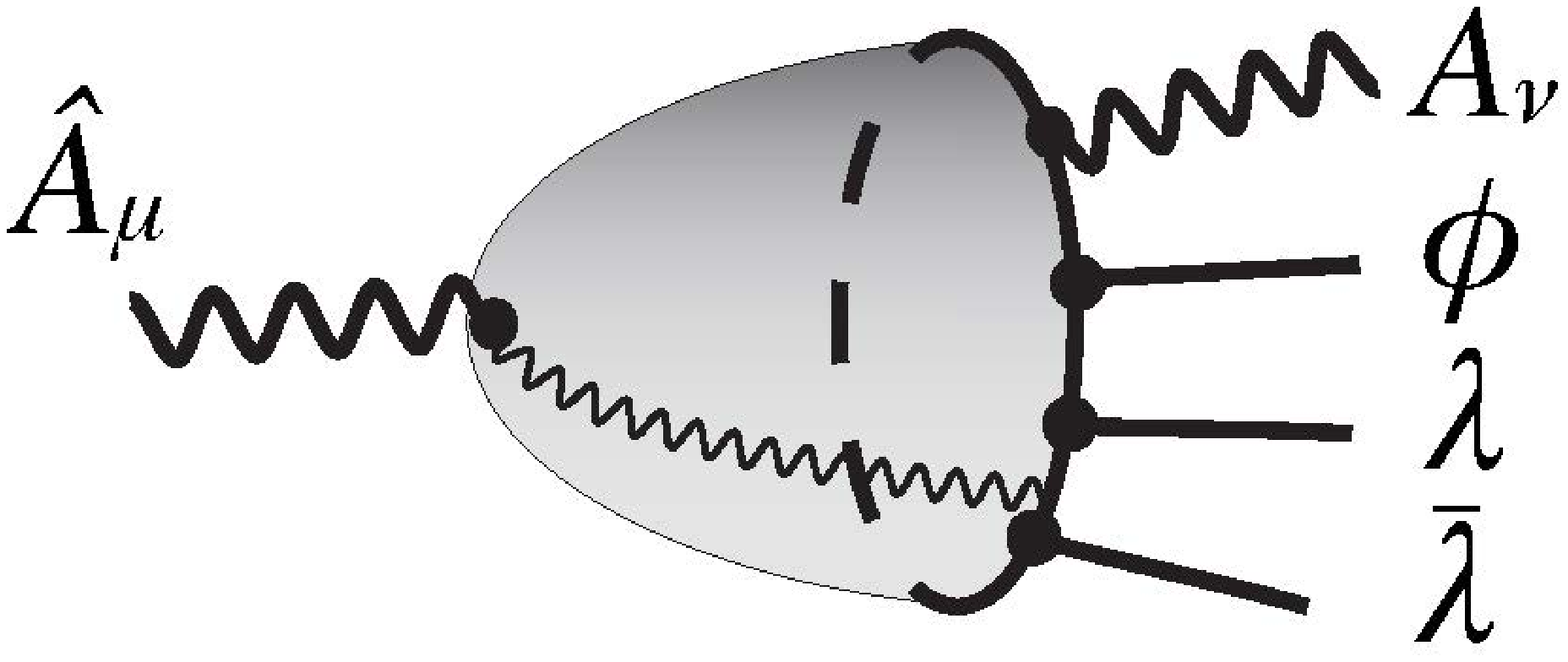}}
&~~~& $\displaystyle{g_s^2 \frac{\ell_s^4 }{\sqrt{ {{\mathcal{V}_6}} }} u_\a v^\a \varphi F\hat F }$
\\
\end{tabular}
\caption{Comparison of kinetic mixing (1., 2.) and the couplings (3., 4., 5. and 6.) between the QFT results (on the left side) and the string results (on the right side) in the closed sector in the absence of fluxes. $u$ and $v$ in the right-hand column are Dirac spinor wave-functions defined in appendix \ref{Mixing_string}. The mass dimensions of the spinors $u,v$ are 1/2 while the scalar $\varphi$ has dimension one. The EFT couplings 1, 3, 5 are vanishing at lowest order (disk $D$) in $g_s$, they can be generated at one-loop order (annulus $A$) in (twisted) sectors with $\langle \partial X^i \rangle_A\neq 0$. }
\label{tab_closed_noflux}
\end{table}

We start with the closed sector without fluxes. As shown in table \ref{tab_closed_noflux}, we have three non-vanishing couplings/mixings: \eqref{3_11}, \eqref{3_15} and \eqref{3_5} respectively to leading order in perturbation theory.
{Although we mostly focussed on RR graviphotons, {\it mutatis mutandis} similar results are obtained for NSNS graviphotons, as shown in appendix \ref{NSNSgraviphotons}.}

Comparing \eqref{3_11} and \eqref{3_15} with the effective field theory couplings \eqref{2_18}, we find that the powers of $g_s$ match when the couplings are non-zero and we are led to define two scales
\be
M_s= \frac{1}{\ell_s} \quad , \quad M_{KK}= \frac{ {{\mathcal{V}_6}} ^{-1/6}}{\ell_s}
\label{4_1}
\ee
For our purposes we can take these scales to be of the same order and of the order of the messenger scale $M$,
\be
M_s\sim M_{KK}\sim M
\label{4_2}\ee

Notice that the loop order of the Feynman diagrams in Appendix \ref{mixing} does not match the loop order in $1/N \sim g_s$. This is due fact that we do not assume any specific relation between the Standard Model couplings and the coupling constant or the number of `colors' of the hidden sector. Yet we assume that the messenger scale $M$ be of the same order as the string and KK scale $M_s\sim M_{KK}\sim M$ as indicated.

If we consider the dipole couplings (line 4 of table \ref{tab_closed_noflux}, corresponding to an operator of dimension $\Delta =6$), we find they match:
\be
g^{3/2}_s \frac{\ell^2_s}{\sqrt{ {{\mathcal{V}_6}} }} H \bar\psi\gamma_{\m\n}\psi F^{\m\n}_A=
\frac{g^{3/2}_s M_{KK}^3}{M_s^3} \frac{1}{M_{s}^2} H {\bar\psi\gamma_{\m\n}\psi} F^{\m\n}_A \longleftrightarrow {1\over N^{3/2} M^2} H {\bar\psi\gamma_{\m\n}\psi}F^{\m\n}_A
\label{4_3}
\ee
Using the identification (\ref{4_2}), we find agreement for the next fermionic coupling (line 6 of table \ref{tab_closed_noflux}, corresponding to an operator of dimension $\Delta =8$)
\be
g_s^{2} \frac{\ell_s^4}{\sqrt{ {{\mathcal{V}_6}} }} F_{\m\n}^AF^{Y,\m\n}H\bar \psi\psi=
\frac{g_s^{2} M_{KK}}{M_s} \frac{1}{M_{KK}^4} F_{\m\n}^AF^{Y,\m\n}H \bar \psi\psi \longleftrightarrow
{1\over N^{2}M^4} F_{\m\n}^AF^{Y,\m\n}H \bar \psi\psi
\label{4_4}
\ee
For the kinetic mixing (line 6 of table \ref{tab_closed_noflux}, corresponding to an operator of dimension $\Delta =5$), comparison between \eqref{3_8} and \eqref{2_22} yields
\be
g_s \frac{\ell_s}{\sqrt{ {{\mathcal{V}_6}} }} \tr[\phi^i F^{\m\n}_Y] F_{i,\m\n}^{A} =
\frac{g_s M_{KK}^4}{M_s^4} \tr\left[\frac{\phi^i}{M_{KK}} F^{\m\n}_Y\right] F_{i,\m\n}^{A} \longleftrightarrow
{\Lambda^2\over N~M^2} F_{\m\n}^{A}F^{\m\n}_Y
\label{4_5}
\ee
Taking into account (\ref{4_2}) and that in the EFT computation $\Lambda\sim M$, the two results match.

However, one should keep in mind that the scalar $ \phi $ belongs to the adjoint of the visible gauge groups. Therefore, it would seem that we cannot strictly identify it with the SM Higgs. However, in D-brane realizations of the SM, the Higgs doublet is a state originating in a string stretching between the SU(2) brane stack and a U(1) brane stack, and in this sense, it is in the adjoint of the SM brane stack. In general, in a D-brane SM realization, all SM particles are in the adjoint of the non-simple SM gauge group, \cite{ADKS}.

That being said, the SM quantum numbers, and the D-brane embedding of the SM preclude on group theoretical grounds such a coupling. Therefore, no kinetic mixing can appear here. If, however, the SM stack of branes contains other scalars beyond the (two) Higgses, then such a coupling may be possible.
A typical SM stack is composed of a stack of 3 branes realizing SU(3), a stack of 2 branes realizing SU(2), and two U(1) branes, \cite{ADKS}.

As the hypercharge is a specific linear combination of all four U(1)'s the scalar in order to appear in a coupling of the form in (\ref{4_5}) must be in the adjoint of SU(3), SU(2) or the two U(1)'s. If it obtains a vev it will break SU(3) or give mass to the photon. Therefore, the only possible adjoint Higgses that can obtain vevs are those that are associated to either of the two extra U(1)'s that are neutral under the SM. In any case, these are particles that are beyond the SM, and their phenomenological discussion goes beyond the scope of this paper.

We conclude that for the minimal SM spectrum, such mixing terms cannot be generated.

\subsubsection{Closed sector in the presence of bulk fluxes}

Comparing the kinetic mixing, that corresponds to operators of dimension $\Delta=4$, in the presence of RR flux \eqref{3_23} and the one obtained in section \ref{holosetup} \eqref{2_22} we find
\be
g_s^{3/2} \, \frac{\ell_s^2}{ {{\mathcal{V}_6}} }\,
\calF_3 F_{\m\n}^{A}F^{\m\n}_Y=
{g_s^{3/ 2}\calF_3\over M_2^2}
F_{\m\n}^{A}F^{\m\n}_Y=
\frac{1}{N^{3/2}}
{\calF_3\over M_s^2}F_{\m\n}^{A}F^{\m\n}_Y \longleftrightarrow
{\Lambda^2\over N~M^2} F_{\m\n}^{A}F^{\m\n}_Y ~~~~~
\label{4_6}
\ee
\begin{table}[b!]
\begin{tabular}{llllll}
&~{\bf EFT coupling} &&
{\bf string theory diagram}
&& {\bf string theory}
 \\
& &&
&~& {\bf effective coupling}\\
1.&~$\displaystyle{\Lambda^2 \over NM^2} F^{\mu \nu} \hat F_{\mu \nu}$ &&
\raisebox{-7mm}{\includegraphics[width=0.25\textwidth]{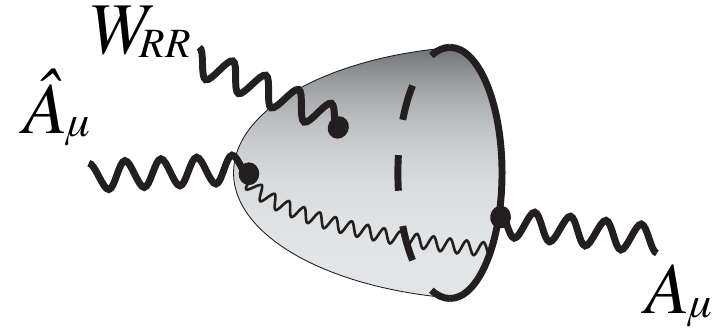}}
&~& $\displaystyle{g_s^{3/2} \frac{\ell_s^2}{ {{\mathcal{V}_6}} }\,
{\cal F}_{(ab)}\,F^{\mu\nu} \hat{F}_{\mu\nu}^{(ab)}}$
\\
2.&~$\displaystyle{1\over NM^2} D_\mu HD_\nu \bar H \hat F^{\mu \nu}$ &&
\raisebox{-7mm}{\includegraphics[width=0.25\textwidth]{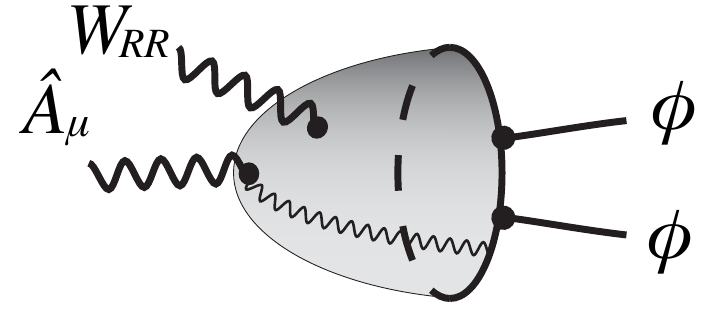}}
&~& $\displaystyle{g_s^2 ~{ \ell_s^4 \over {{\mathcal{V}_6}} } {\cal F}^{ab} k_\phi{\cdot} \hat F_{ab} {\cdot} k_{\bar \phi} ~\phi \bar \phi}$
\\
3.&~$\displaystyle{1\over N^{\frac{3}{2}} M^2} \bar \psi \gamma_{\mu \nu} H \psi\hat F^{\mu \nu}$ &&
\raisebox{-7mm}{\includegraphics[width=0.25\textwidth]{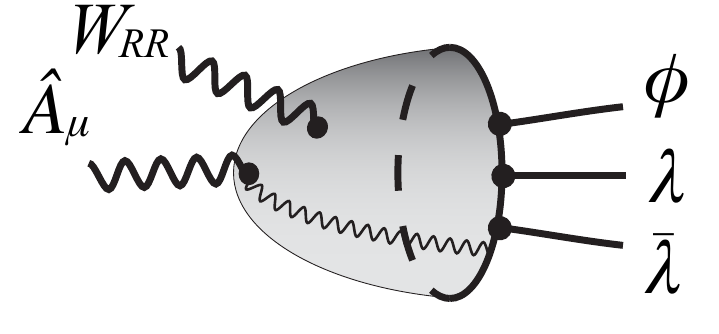}}
&~& $\displaystyle{0\cdot g^{3\over 2}+{\cal O}(g_s^{\frac{5}{2}})}$ (subl. in $\pa$)
\\
4.&~$\displaystyle{1\over N^{3\over 2}M^2} F^{\mu \nu}\hat F_{\mu \nu}H^\dagger H$ &&
\raisebox{-7mm}{\includegraphics[width=0.25\textwidth]{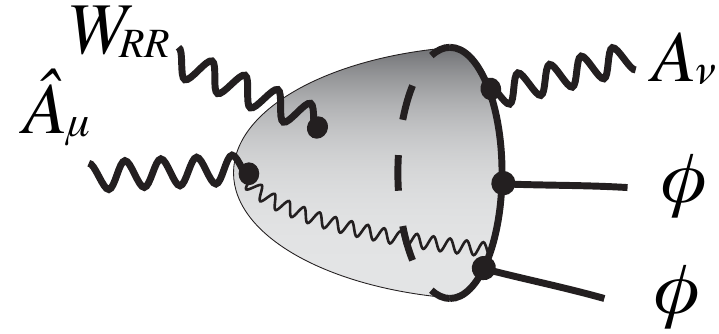}}
&~& $\displaystyle{g_s^{5/2} \frac{\ell_s^6}{ {{\mathcal{V}_6}} }\,{\cal F}^{ab} F_{\mu \nu} \hat F_{ab}^{\mu \nu} k_\phi {\cdot} k_{\bar \phi} ~\phi \bar \phi}$
 \\
5.&~$\displaystyle{1\over N^{2}M^4} F^{\mu \nu}\hat F_{\mu \nu} \bar \psi H \psi$ &&
\raisebox{-7mm}{\includegraphics[width=0.25\textwidth]{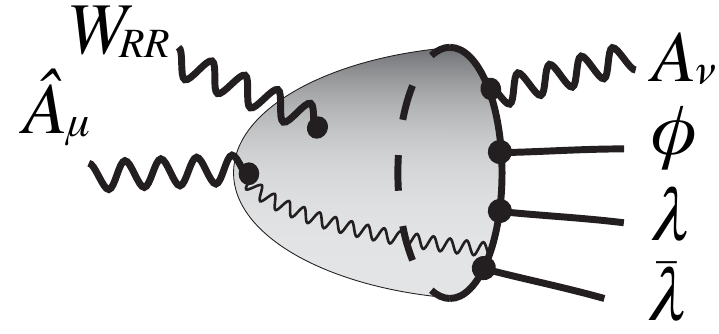}}
&~& $\displaystyle{0\cdot g_s^2+{\cal O}(g_s^3)}$(subl. in $\pa$)
\\
\end{tabular}
\caption{Comparison of the kinetic mixing (1.) and the couplings (2., 3., 4. and 5.) between the QFT results (on the left side) and the string results (on the right side) in the closed sector in the presence of fluxes. The amplitudes 3. and 5. yield higher-derivative terms. The EFT couplings 3, 5 are therefore generated at higher-derivative order (one more derivative at least) in the presence of fluxes for NON-chiral fermions or require insertions of Higgs fields for gauge invariance in the case of chiral fermions. Higher-derivative couplings, however, can generate leading corrections to the mixing, once SM quantum effects are included. }
\label{tab_closed_withflux}
\end{table}
where in the third term above we have used \eqref{4_1} and \eqref{4_2} while $M_2$ is
\be
M_2= M_s \sqrt{ {{\mathcal{V}_6}} } \gg M_s
\label{4_7}
\ee
where we made the reasonable assumption that the flux ${\cal F}_3$ is independent on $g_s$. The kinetic mixing above in the string setup is sub-leading with respect to the one in the holographic setup, due to an extra power of $\sqrt{g_s}\sim \sqrt{N}$.

As argued in section \ref{interactions_closed_string}, fermionic couplings to the graviphoton are subleading in the IR, ie. they have extra derivatives compared to the EFT couplings. We therefore turn our attention on the couplings involving open-string scalar fields.

Starting from the leading coupling, that corresponds to an operator of dimension $\Delta=6$, we have to consider \eqref{3_27} and \eqref{2_18}
\begin{align}
&g_s^2 \frac{\ell_s^4}{ {{\mathcal{V}_6}} } \calF_3 Tr[D_{\m}H D_{\n}H^{\dagger} ]F_{A}^{\m\n}
=
\frac{g_s^2}{M_s^2}\frac{\calF_3}{M_2^2} Tr[D_{\m}H D_{\n}H^{\dagger} ]F_{A}^{\m\n}=\label{4_8}\\
& ~~~~~ \frac{1}{N^2 M_{KK}^2}\frac{\calF_3}{M_s^2} Tr[D_{\m}H D_{\n}H^{\dagger} ]F_{A}^{\m\n}
 \longleftrightarrow
{1\over NM^2}Tr[D_{\m}H D_{\n}H^{\dagger} ]F_{A}^{\m\n} \nonumber
\end{align}
where in the third term we have used \eqref{4_1} and \eqref{4_3} once again.

{Comparing string theory and field theory results, we find two apparent differences. One is the presence of the ratio $\frac{\calF_3}{M_s^2}$ as an overall factor. The normalization of ${\calF_3}$ and similar vev's were assumed to be of order ${\cal O}(1)$ in the EFT calculation. Therefore, taken this vev to be ${\cal O}(1)$ there is no discrepancy as far as this term is concerned.
The second apparent difference is an extra power of the string coupling $g_s\sim {1\over N}$. This makes the string theory result sub-leading as compared to the EFT result on the right.
This can be understood as follows.
{}From the string theory side, this is due to the fact that ${\calF_3}$ is a flux in the RR sector and as such it has a modified scaling with the dilaton that explains the different power of the string coupling or $N$, \cite{book}.
A similar phenomenon is known in quantum field theory when the relevant coupling is a $\theta$ angle, \cite{witten,diss}. }

For the next couplings, corresponding to operators of dimension $\Delta = 6$ or $\Delta = 8$, we have to consider \eqref{3_28} and \eqref{2_18}
\begin{align}
&g_s^{5/2} \frac{\ell_s^4}{ {{\mathcal{V}_6}} } [\ell_s^2 k_H {\cdot} k_{H^{\dagger}}]\,{\cal F}_3 Tr[H H^{\dagger} ]F_{A}^{\m\n} F_{Y,\m\n}=
\label{4_9}\\
&~~~~~ \frac{\calF_3}{N^{5/2} M_{KK}^4} \left[\frac{k_H {\cdot} k_{H^{\dagger}}}{M_s^2}\right] Tr[H H^{\dagger} ]F_{A}^{\m\n} F_{Y,\m\n}
 \longleftrightarrow
{1\over {N^{3/2}M^2}}Tr[H H^{\dagger} ]F_{A}^{\m\n} F_{Y,\m\n} \nonumber
\end{align}
As mentioned in \eqref{3_28}, the scalar product $k_H {\cdot} k_{H^{\dagger}}$, that shifts the dimension from $\Delta = 6$ to $\Delta = 8$, may be absent if it is absorbed by poles arising from the world-sheet integration. If this is the case, the coupling is suppressed by $g_s$. On the other hand, if additional powers of the momenta remain, the combination $\calF_3/M_s^2$ will appear, as in \eqref{4_6}, and the coupling is doubly suppressed by extra powers of $g_s\sim {1\over N}$ and $M_{KK}\sim M$.

 As before, our results here are similar to those obtained for NSNS graviphotons.

\subsection{The open sector}

\begin{table}[h!]
\begin{tabular}{llllll}
~~~&~{\bf EFT coupling} & &
{\bf string theory diagram}
&~~~& {\bf string theory}
 \\
~~~&~ & ~~&
&~~~& {\bf effective coupling}
 \\
1.~~~&$\displaystyle{\Lambda^2 \over NM^2} F^{\mu \nu} \hat F_{\mu \nu}$ ~~~&&
\raisebox{-7mm}{\includegraphics[width=0.25\textwidth]{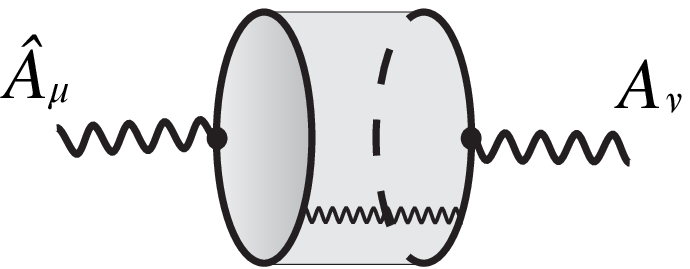}}
&~~~& $\displaystyle{g_s \hat{F}^{\mu \nu} F_{\mu \nu}\log \frac{\Lambda^2}{M^2}}$
\\
2.~~~&~$\displaystyle{1\over NM^2} D_\mu HD_\nu \bar H \hat F^{\mu \nu}$ ~~~&&
\raisebox{-7mm}{\includegraphics[width=0.25\textwidth]{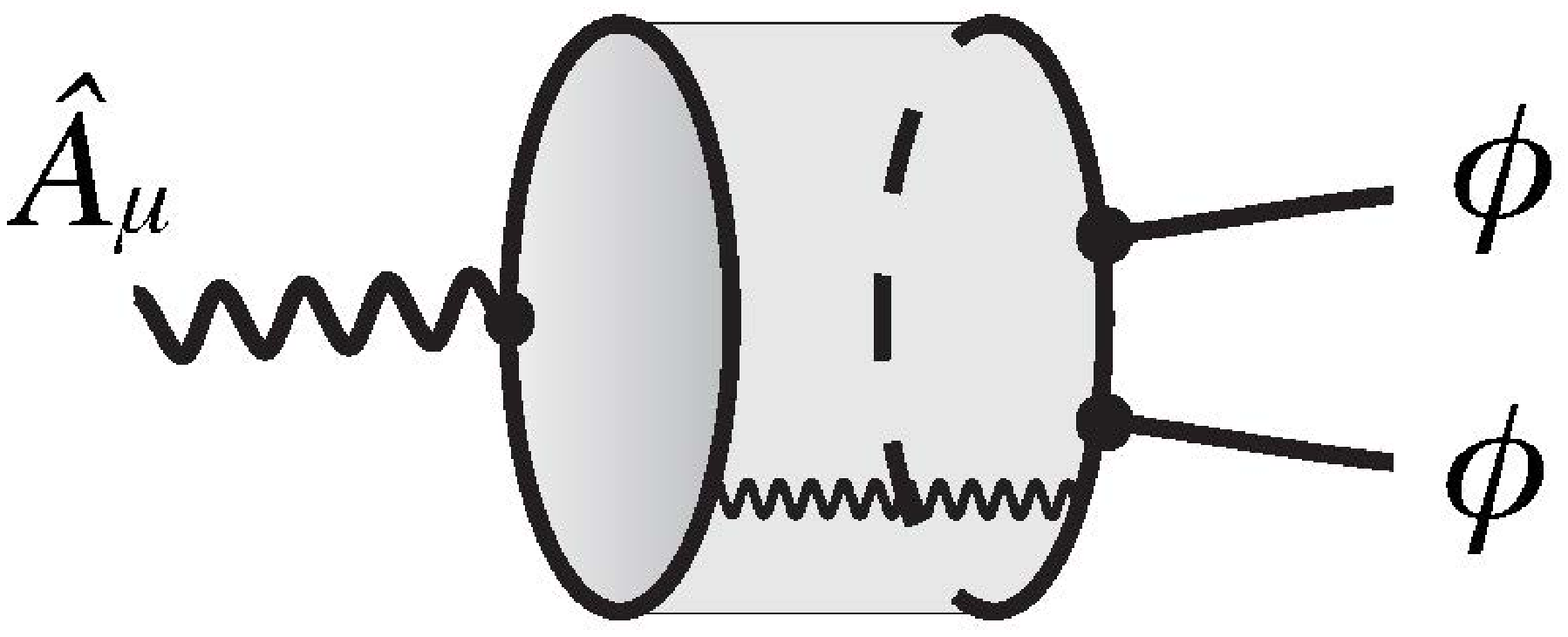}}
&~~~& $\displaystyle{{g_s^{3/2}\over M^{2}}k_\phi{\cdot} \hat F {\cdot} k_{\bar \phi} ~\phi~ \bar \phi}$
\\
3.~~~&~$\displaystyle{1\over N^{\frac{3}{2}} M^2} \bar \psi \gamma_{\mu \nu} H \psi\hat F^{\mu \nu}$ &&
\raisebox{-7mm}{\includegraphics[width=0.25\textwidth]{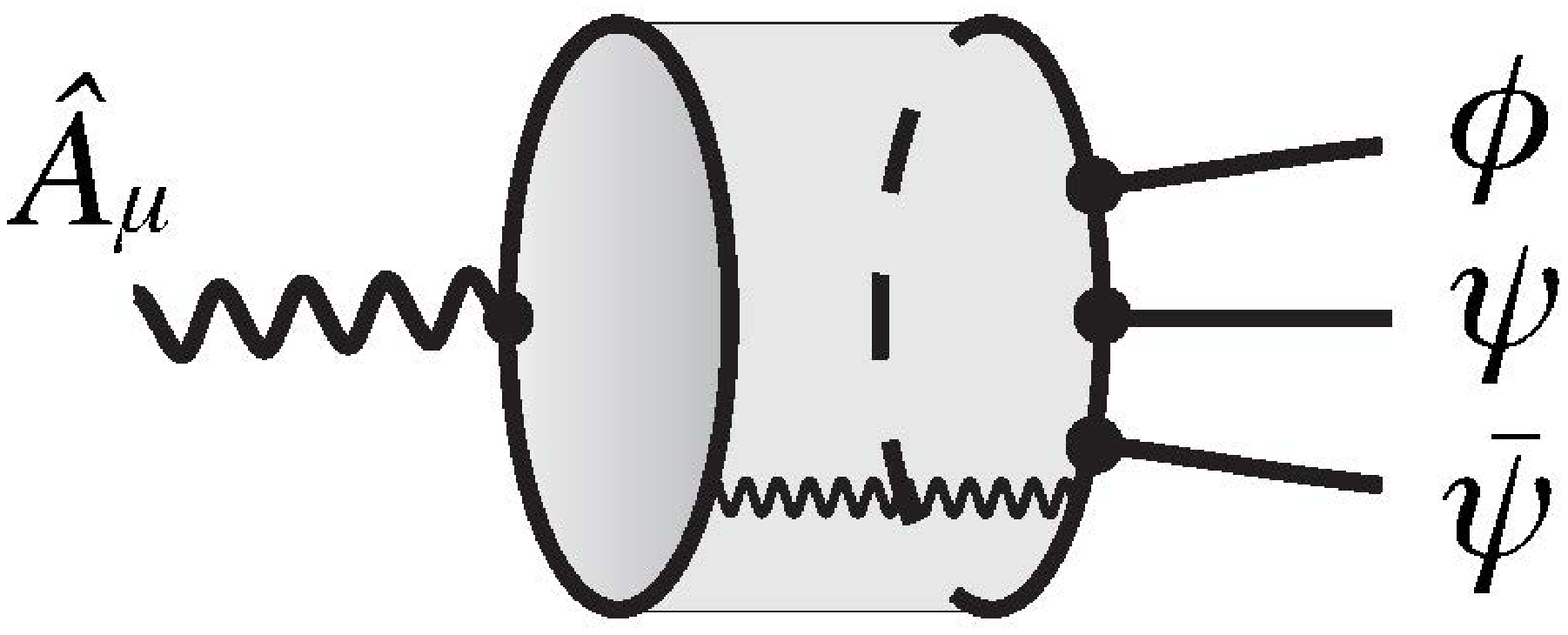}}
&~~~& $\displaystyle{0\cdot g_s+{\cal O}(g_s^{2})}$(subl. in $\pa$)
\\
4.~~~&~$\displaystyle{1\over N^{3\over 2}M^2} F^{\mu \nu}\hat F_{\mu \nu}~H^\dagger H$ ~&&
\raisebox{-7mm}{\includegraphics[width=0.25\textwidth]{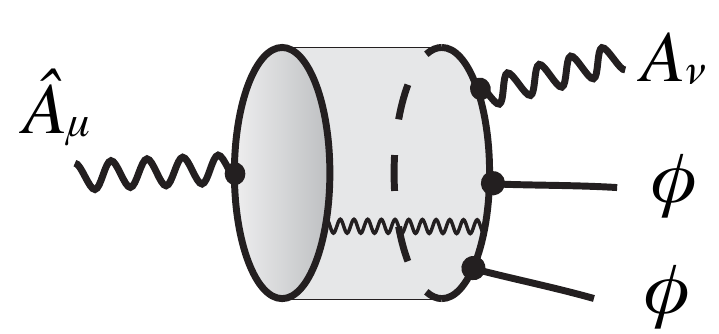}}
&~~~& $\displaystyle{{g_s^{2}\over M^{4}} ~F \hat{F} ~\phi ~\bar \phi \, k_{\phi}{\cdot}k_{\bar \phi} }$ %
\\
5.~~~&~$\displaystyle{1\over N^{2}M^4} F^{\mu \nu}\hat F_{\mu \nu} \bar \psi H \psi$ ~&&
\raisebox{-7mm}{\includegraphics[width=0.25\textwidth]{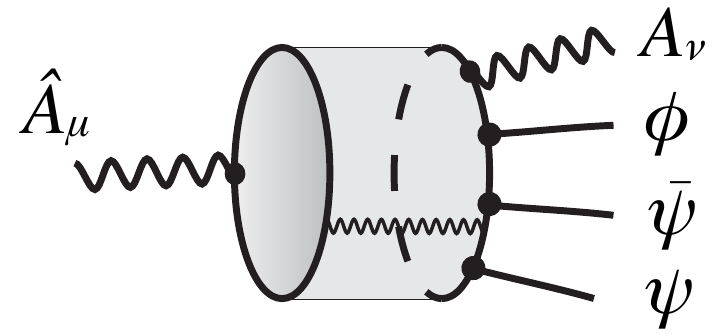}}
&~~~& $\displaystyle{{\cal O}(g_s^{\frac{5}{2}})}$(subl. in $\pa$)
\\
\end{tabular}
\caption{Comparison of the kinetic mixing (1.) and the couplings (2., 3., 4. and 5.) between the QFT results (on the left side) and the string results (on the right side) in the open sector. The amplitudes 3. and 5. yield higher-derivative terms. Higher-derivative couplings, however, can generate leading corrections to the mixing, once SM quantum effects are included.}
\label{tab_open}
\end{table}

 We consider the mass mixing term, that corresponds to a dimension $\Delta = 4$ operator of the form\footnote{The full gauge invariant amplitude for scalar Compton scattering reads
${\cal A} = g_s \eta_{\mu\nu} \left(a^\mu_Y - {a_Y p\over k_Yp} k_Y^\mu\right)
\left(a^\nu_A - {a_A p\over k_Ap} k_A^\nu\right) \phi(p){\cdot}\tilde \phi(\tilde{p})$.}
\be
g_s \, Y^\mu A_{\mu} ~\phi\cdot \tilde \phi
\label{4_10}
\ee
Since, as already observed around Eq \eqref{3_29}, these scalars are messengers, they are assumed to be very massive and it is unlikely they will acquire VEVs. The term above should therefore be interpreted as a interaction term rather than a mass term.

Next we consider the gauge kinetic mixing, that corresponds to a dimension $\Delta = 4$ operator, and set $M= \Delta x/\ap$. On the left side, we find
\be
g_s \, \log \frac{\Lambda^2}{M^2} \, F_{\m\n}^{A}F^{\m\n}_Y \longleftrightarrow
{\Lambda^2\over N~M^2} F_{\m\n}^{A}F^{\m\n}_Y
\label{4_11}
\ee
Taking into account that $g_s\sim {1\over N}$ and $\Lambda\sim M$ in the EFT calculation, the two results are broadly compatible. The extra log, will also appear in the EFT result if higher loop corrections are included.

We finally consider couplings corresponding to dimension $\Delta=6$ operators involving scalars.
\begin{align}
\frac{g_s^{3/2}}{M^2} Tr[D_{\m}H D_{\n}H^{\dagger} ]F_{A}^{\m\n} & \longleftrightarrow
{1\over NM^2}Tr[D_{\m}H D_{\n}H^{\dagger} ]F_{A}^{\m\n}
\label{4_12}\\
{g_s^2 \over M^4}F_{\m\n}^A F^{Y,\m\n}HH^{\dagger}k_H{\cdot}k_{H^\dagger} &\longleftrightarrow
{1\over N^{3\over 2}M^2}F_{\m\n}^AF^{Y,\m\n}HH^{\dagger}
\label{4_13}
\end{align}
Both interaction terms above are subleading in powers of $g_s$. The fermionic couplings are sub-leading being higher derivative terms.

The comparison presented above is nicely summarized in table \ref{tin}. Further comments on the comparison are advanced in the introductory section  \ref{out}.

\vskip 1cm

\section*{Aknowledgments}
\addcontentsline{toc}{section}{Aknowledgements\label{ack}}
\vskip 1cm

We would like to thank I. Antoniadis, A. Arvanitaki, P. Betzios, E. Dudas, D. Luest, F. Marchesano, V. Niarchos, F. Nitti, O. Papadoulaki, N. Tsamis, L. Witkowski, and R. Woodard for discussions.
{We also thank the referee, whose questions, suggestions and criticism helped ameliorate the presentation of this paper.}

\noindent This work was supported in part by the Advanced ERC grant SM-grav, No 669288.
P.A. and D.C. were supported by FWF Austrian Science Fund via the SAP P30531-N27.

\noindent

\newpage
\appendix

\begin{appendix}
\renewcommand{\theequation}{\thesection.\arabic{equation}}
\addcontentsline{toc}{section}{Appendices}
\section*{APPENDIX}

\section{Mixing computations in Effective Field Theory\label{mixing}}

\begin{figure}[h]
\centering
\includegraphics[width=0.45\textwidth]{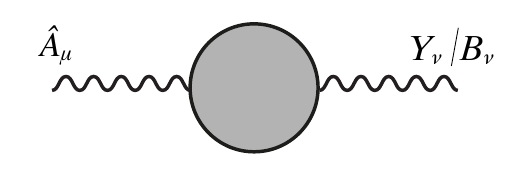}\\
\caption[]{The general structure of the amplitude that provides the mixing between the hidden gauge field $\hat A$ and non-anomalous $Y$/anomalous $B$ gauge fields of the SM in the un-broken phase, and $Y_\n / B_\n$ should be replaced by $A_\n / Z_\n/ Z'_\n$ in the broken phase.}
\label{AtoYB}
\end{figure}

In this appendix, we compute the emergence of mixing terms of the type $F_{\m\n}^{\widehat{A}} F^{\m\n}_Y$ due to the SM quantum effects (as shown schematically in fig \ref{AtoYB}), from the generic effective action of the hidden theory in (\ref{2_18}).

We denote by $\hat A^\mu$ the gauge field originating in the hidden sector, emergent from a U(1) global symmetry, along the lines of \cite{u1}. On the SM side, we separate cases for the unbroken/broken phases. More precisely, in the unbroken phase, the SM gauge fields are $Y^\m/B^\m$ the hypercharge or an anomalous U(1) respectively, and in the broken phase, $A_\n / Z_\n/ Z'_\n$ are respectively the photon, the $Z$ and the $Z$-prime, see fig \ref{AtoYB}.

As we already mentioned in the main text, we do not assume any specific relation between the Standard Model couplings and the coupling constant or the number of `colors' of the hidden sector.

In sections \ref{SMup} and \ref{SMbp}, we analyze the kinetic mixing in the absence of anomalous $U(1)$'s while in \ref{anomalousU1} we discuss the mixing between the emergent gauge field with anomalous $U(1)$'s.

\subsection{The unbroken phase of the SM \label{SMup}}

\begin{figure}[h]
\centering
\includegraphics[width=0.65\textwidth]{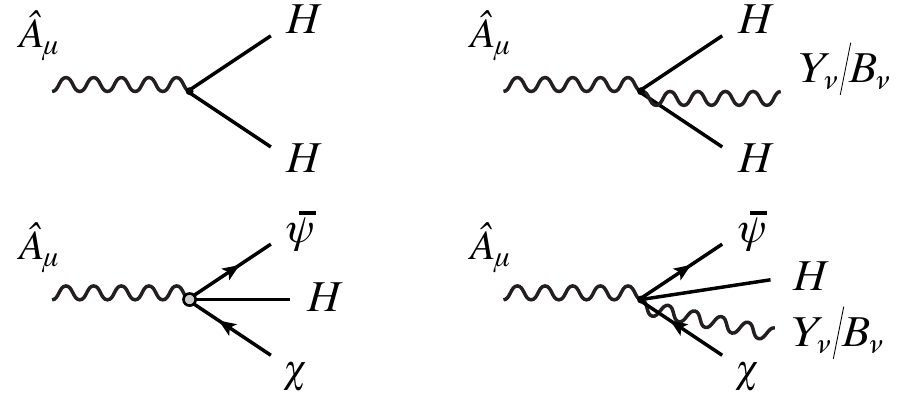}\\
\caption[]{The couplings of the hidden gauge field $\hat A$ to SM fermions and the Higgs respectively. The small grey blob includes the $\g^{\m\n}$ of the dipole coupling between $\hat A$ and SM fermions $\psi$, the Higgs.}
\label{AcouplingstoSM}
\end{figure}

We start from the induced couplings of the emergent photon $\widehat{A}_{\mu}$ to the SM fields which were presented in \eqref{2_18}
\begin{align}
W_{6}(\widehat{A}_{\m}&,\phi_{ij},\psi_{ij}, A_{\m}^{ij}) \sim
{1\over NM^2}Tr[D_{\m}H D_{\n}H^{\dagger} ]F_{\widehat{A}}^{\m\n}+
{1\over N^{3\over 2}M^2}F^{\m\n}_{\widehat{A}}\left[\bar\psi\gamma_{\m\n}H\psi+c.c.\right]+\nn\\
&+{1\over N^{3\over 2}M^2}F_{\m\n}^{\widehat{A}}F^{Y,\m\n}HH^{\dagger}+
{1\over N^{ 2}M^4}F_{\m\n}^{\widehat{A}}F^{Y,\m\n}\left[\bar \psi H\psi+c.c.\right]\cdots
\label{a_1}
\end{align}
Schematically, we present the basic structure of the (simplest) couplings in fig \ref{AcouplingstoSM}, without explicitly presenting informations that can be easily extracted from \eqref{a_1}. Terms with the same number of fields but more derivatives give similar results, because the cutoff scale $\Lambda$ is of the same order as the messenger scale $M$ that controls the size of these terms.

We do not show all the details of the couplings in fig \ref{AcouplingstoSM}, like the momentum dependence and Lorentz structure, but just the structure (fields involved). We also do the same in fig \ref{AcouplingstoSMbroken}.

The first coupling in \eqref{a_1} can be further manipulated. Neglecting the SU(2) covariant derivative we can write
\be
Tr[D_{[\m}H D_{\n]}H^{\dagger} ]=(\pa_{\m}H\cdot \pa_{\n}H^*-({\mu\leftrightarrow \nu}))+iQ_Y^H(A_{\mu}^{Y}\pa_{\n}|H^2|-A_{\nu}^{Y}\pa_{\m}|H^2|)
\label{a_2}\ee
and since by integration by parts
\be
\int d^4 x F_{\widehat{A}}^{\m\n} \Big(A_{\mu}^{Y}\pa_{\n}|H^2|-A_{\nu}^{Y}\pa_{\m}|H^2|\Big)=-\int d^4 x F_{\widehat{A}}^{\m\n} F^Y_{\m\n}|H^2|
\label{a_3}\ee
{this last term is similar to the third term of \eqref{a_1} but is larger by an extra power of $\sqrt{N}$. We will not discuss it separately, but when we calculate the contribution of the third term, we shall remember that there is one at order $1/N$.}

Next, we have the couplings of the hypercharge gauge boson to SM operators
\be
W_Y=\int d^4x~ A^{Y}_{\m}
\left( iQ_Y^H Tr[H\pa_{\mu}H^{\dagger}-H^{\dagger} \partial_{\mu} H]
+ Tr[Q_Y^{\psi}\bar \psi \gamma^{\m}\psi]\right)
\label{a_4}\ee

Our aim is to compute the kinetic mixing of the hidden and visible U(1)'s via the couplings \eqref{a_1} and \eqref{a_4}. {The coefficient of the mixing term $F_{\widehat{A}}^{\m\n} F^Y_{\m\n}$ is a weighted sum, to leading order, of the following vevs below.}
\begin{itemize}
\item Vev $V_1$
\be
V_1\equiv \langle Tr[\pa_{\m}H \pa_{\n}H^{\dagger} ](x)~
Tr[H\pa_{\rho}H^{\dagger}-H^{\dagger} \partial_{\rho} H](y)\rangle
\label{a_5}\ee

\item Vev $V_2$
\be
V_2\equiv \langle Tr[\pa_{\m}H \pa_{\n}H^{\dagger} ](x)~
Tr[Q_Y^{\psi}\bar \psi \gamma_{\rho}\psi](y)\rangle
\label{a_6}\ee

\item Vev $V_3$
\be
V_{3}=\langle Tr[\bar{\psi} \gamma^{\mu \nu} \psi H](x) ~
Tr[Q_Y^{\psi}\bar \psi \gamma_{\rho}\psi](y)\rangle
\label{a_6aaaa}\ee

\item Vev $V_4$
\be
V_{4}=\langle Tr[\bar{\psi} \gamma^{\mu \nu} \psi H](x) ~
Tr[H\pa_{\rho}H^{\dagger}-H^{\dagger} \partial_{\rho} H](y)\rangle
\label{a_8}\ee

\item Vev $V_5$
\be
V_5=\langle
F^{Y,\m\n}HH^{\dagger}(x)~
Tr[H D_{\rho}H^{\dagger}-H^{\dagger} D_{\rho} H](y)\rangle
\label{a_9}\ee

\item Vev $V_6$
\be
V_6=\langle
F^{Y,\m\n}HH^{\dagger}(x) ~ Tr[Q_Y^{\psi}\bar \psi \gamma_{\rho}\psi](y)\rangle
\label{a_10}\ee

\item Vevs which include the last coupling in \eqref{a_1} will be highly suppressed by several powers of $N$ and being at least three-loop contributions. Therefore, we neglect them here.

\end{itemize}
Next, we compute the vevs above.

\subsubsection{The $V_1$ term: one-loop contribution.}

\begin{figure}[h]
\centering
\includegraphics[width=0.40\textwidth]{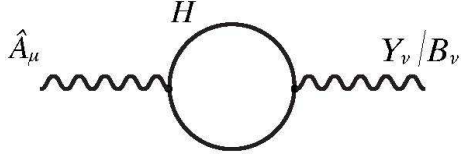}
\caption[]{The one-loop contribution to $V_1$ vev.}
\label{AcouplingstoSMV1}
\end{figure}

The leading contribution to $V_1$ is coming at one-loop (fig \ref{AcouplingstoSMV1}) and it reads
\bea
M_1&\equiv& {Q_Y^H\over NM^2}\!\int \!d^4x \,d^4y~F^{\widehat{A}}_{\m\n}(x)A^Y_{\rho}(y)
\left\langle (\pa^{\mu}H\pa^{\n}H^{\dagger})(x) ~(H\pa^{\rho}H^{\dagger}-H^{\dagger} \partial^\rho H)(y)\right\rangle
\label{a_11}\\
&{\,=\,}&{-\,}
{Q_Y^H\over NM^2}\!\int \!d^4x \,d^4yF^{\widehat{A}}_{\m\n}(x)A^Y_{\rho}(y)
\left[\pa^{\m}_{x}G(x{-}y)\pa^{\n}_x\pa^{\rho}_xG(x{-}y){-}\pa^{\m}_x\pa^{\rho}_xG(x{-}y)\pa^{\n}_xG(x{-}y)\right]\nn
\eea
{where $G(x-y)$ is a massive scalar (Euclidean) propagator in flat space-time,}\footnote{Henceforth we drop the factor of $(2\pi)^4$ in the denominator.}
\be
G(x)=\int {d^4k\over (2\pi)^4} {e^{ik\cdot x}\over k^2+m^2}
\label{a_12}\ee
and in the last line we converted all derivatives to $x$ derivatives.
Transforming to momentum space
\be
M_1=-{Q_Y^H\over NM^2}\int d^4p~d^4k~F^{\widehat{A}}_{\m\n}(k{+}p)~A^Y_{\rho}({-}k{-}p)\left(k^{\m}p^{\n}p^{\rho}-k^{\m}k^{\rho}p^{\n}\right)G(p)G(k)
\label{a_13}\ee
Changing variables to $q=k+p$ we can rewrite as
\be
M_1={Q_Y^H\over NM^2}\int d^4p~d^4q~F^{\widehat{A}}_{\m\n}(q)~A^Y_{\rho}({-}q)\left(q^{\m}p^{\n}q^{\rho}{\,-\,}2q^{\m}p^{\n}p^{\rho}\right)G(p)G(p{\,-\,}q)
\label{a_14}\ee
where in the last line we dropped terms symmetric under the exchange of $\mu\leftrightarrow \n$.

We now introduce the Feynman parametrisation and keeping only the loop integral for simplicity (which we denote by $\tilde M_1$) we have
\be
\tilde M_1^{\m\n\rho}
=\int d^4p \int_0^1 dx~ {q^{\m}p^{\n}q^{\rho}-2q^{\m}p^{\n}p^{\rho}\over [(p-xq)^2+Q(x)^2]^2} \label{a_15}
\ee
where $Q(x)^2\equiv x(1-x)q^2+m^2$. Shifting the integration variable $p-xq\to p$ we obtain
\be
\tilde M_1^{\m\n\rho}
=-{1\over 2}q^{\m}\delta^{\n\rho}\int d^4p \int_0^1 dx~ {p^2\over [p^2+Q(x)^2]^2}
\label{a_16}
\ee
where we have dropped terms that are symmetric under the $(\m\leftrightarrow \n)$ interchange as well as terms linear in $p$ that will give vanishing contributions upon integration.

Direct integration with a momentum cutoff $\Lambda$ gives
\begin{align}
\tilde M_1^{\m\n\rho}
&=-{1\over 4}q^{\m}\delta^{\n\rho}\Omega_3\int_0^1 dx\left[\Lambda^2+Q(x)^2-{Q(x)^4\over \Lambda^2+Q(x)^2}-2Q(x)^2\log{\Lambda^2+Q^2(x)\over Q(x)^2}\right]\nn\\
&=
-{1\over 4}q^{\m}\delta^{\n\rho}\Omega_3\Lambda^2+{\cal O}(p^3)
\label{a_17}
\end{align}
so that
\be
M_1={Q_Y^H\over NM^2}\int d^4q~F^{\widehat{A}}_{\m\n}(q)~A_{\rho}(-q)\tilde M_1^{\m\n\rho}
={\Omega_3\over 8}{Q_Y^H\over N}{\Lambda^2\over M^2}\int d^4q~F_{\widehat{A}}^{\m\n}(q)~F^Y_{\m\nu}(-q)+\cdots
\label{a_18}
\ee

\begin{figure}[t]
\centering
\includegraphics[width=0.40\textwidth]{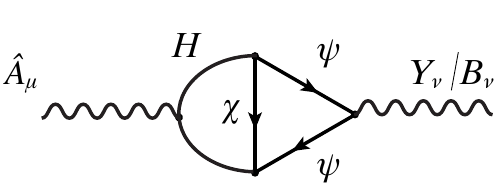}
\caption[]{The two-loop contribution to $V_2$ vev.}
\label{AcouplingstoSMV2}
\end{figure}

The correction $M_1$, as most of the other corrections shown below, is quadratically divergent. One may expect that the corrections to the kinetic terms be logarithmically divergent, however, due to the fact that the effective couplings between the hidden sector and the Standard Model are non-renormalizable, we find quadratically divergent corrections.

\subsubsection{The $V_2$ term: two-loop contribution.}

The leading contribution to $V_2$ is coming at two loops (fig \ref{AcouplingstoSMV2}) and it reads
\be
M_2\equiv {Q_Y^{\psi}\over NM^2}\int d^4x ~d^4y~F^{\widehat{A}}_{\m\n}(x)A^Y_{\rho}(y)~
\langle (\pa^{\mu}H\pa^{\n}H^{\dagger})(x)~(\bar\psi \gamma^{\rho}\psi)(y)\rangle
\label{a_19}\ee
where $\psi$ are chiral left-handed fermions while $\chi$ are right-handed fermions and ${\cal Y}_{H\psi\chi}$ denotes the Yukawa coupling between $H \psi \chi$. In momentum space we obtain
\begin{align}
&M_2= {Q_Y^{\psi} \bar{\cal Y}_{H\psi\chi} {\cal Y}_{H\psi\chi} \over NM^2} \frac{1}{2(2\pi)^2}
\int d^4 p \, F^{\widehat{A}}_{\m\n}(p)A^Y_{\rho}(p) I^{\mu \nu \rho}(p)
\label{a_20}\\
&I^{\mu \nu \rho}(p)= \int \frac{d^4\ell_1}{(2\pi)^4} \frac{d^4\ell_2}{(2\pi)^4}
\frac{(p^\mu \ell_1^\nu-p^\nu \ell_1^\mu)Tr[\lslash_2 \gamma^\rho \lslash_2 (\lslash_1-\lslash_2)]}{(\ell_1^2+m^2)^2(\ell_2^2)^2 (\ell_1-\ell_2)^2}
\label{a_21}
\end{align}
$I^{\mu \nu \rho}(p)$ is not the full result but we simplified it in the following ways.
We kept external momentum $p$ in the right place to obtain the field strength of $A$ and we have neglected $p$ in the rest of the amplitude so that we obtain the leading IR Contribution. The $\gamma_5$ terms, which should be present since $\psi$ and $\chi$ are chiral fermions, are negligible since they provide higher order terms in the external momentum $p$ and therefore have been dropped. Next, we evaluate the trace and we define the scalar integral $I$
\be
I^{\mu \nu \rho}(p) = (p^\mu \eta^{\nu \rho} - p^\nu \eta^{\mu \rho}) I
\quad, \quad
I= \int \frac{d^4\ell_1}{(2\pi)^4} \frac{d^4\ell_2}{(2\pi)^4}
\frac{2 (\ell_1{\cdot} \ell_2)^2-\ell_1^2 \ell_2^2-(\ell_1{\cdot} \ell_2) \ell_2^2}{(\ell_1^2+m^2)^2(\ell_2^2)^2 (\ell_1-\ell_2)^2}
\label{a_22}
\ee

We show below that $I$ is quadratically divergent, therefore to obtain the leading term we can neglect $m^2$ in the denominator. Introducing the variables
$D_1= \ell_1^2$, $D_2= \ell_2^2$ and $D_3= (\ell_1-\ell_2)^2$, $I$ can be written as
\be
I= \int \frac{d^4\ell_1}{(2\pi)^4} \frac{d^4\ell_2}{(2\pi)^4}
\left(\frac{1}{2} \frac{1}{D_2^2 D_3}
-\frac{1}{D_1 D_2 D_3}
-\frac{3}{2}\frac{1}{D_1 D_2^2}
+\frac{1}{2}\frac{1}{ D_1^2 D_3}
-\frac{3}{2}\frac{1}{D_1^2 D_2}
+\frac{D_3}{D_1^2 D_2^2}\right)
\label{a_23}
\ee
In all the integrals above we have a permutation symmetry between $D_1$, $D_2$ and $D_3$, which can be shown using redefinition of the loop momenta. To simplify the integral we can also use the identity
\be
\int \frac{d^4\ell_1}{(2\pi)^4} \frac{d^4\ell_2}{(2\pi)^4}
\frac{D_3}{D_1^2 D_2^2} =
\int \frac{d^4\ell_1}{(2\pi)^4} \frac{d^4\ell_2}{(2\pi)^4}
\frac{\ell_1^2{+}\ell_2^2{-}2 \ell_1 {\cdot}\ell_2 }{(\ell_1^2)^2 (\ell_2^2)^2}
= \int \frac{d^4\ell_1}{(2\pi)^4} \frac{d^4\ell_2}{(2\pi)^4} \frac{2}{D_1 D_2^2}
\label{a_24}
\ee
where $\ell_1 {\cdot}\ell_2$ vanishes due to parity symmetry. The simplified integral is then
\be
I=-\int \frac{d^4\ell_1}{(2\pi)^4} \frac{d^4\ell_2}{(2\pi)^4}
\frac{1}{\ell_1^2 \, \ell_2^2 \, (\ell_1-\ell_2)^2}
\label{a_25}
\ee
To calculate it we use the following expansion
\be
\frac{1}{(\ell_1-\ell_2)^2}=\frac{1}{(\ell_1^2+\ell_2^2)-2 \ell_1 {\cdot} \ell_2 }
=\frac{1}{\ell_1^2+\ell_2^2}\frac{1}{1-\frac{2 \ell_1 {\cdot} \ell_2 }{\ell_1^2+\ell_2^2}}
=\sum_{n=0}^\infty \frac{2^n (\ell_1{\cdot}\ell_2)^n}{(\ell_1^2+\ell_2^2)^{n+1}}
\label{a_26}
\ee
Separating the angular and radial part we obtain
\be
I=-\sum_{n=0}^\infty 2^{n-2}\int \frac{d \ell_1}{(2\pi)^4} \frac{d\ell_2}{(2\pi)^4}
 \frac{\ell_1^{1+n} \ell_2^{1+n}}{(\ell_1^2+\ell_2^2)^{n+1}} \int d \Omega_1 \, d \Omega_2 \, (n_1{\cdot}n_2)^n
\label{a_27}
\ee
The angular integral can be computed introducing a new set of angular coordinates such that $n_1{\cdot} n_2 = \cos \theta$ and parametrizing $n_1$ in hyperspherical coordinates\footnote{with $\phi\in(0,2\pi)$, $\psi\in(0,\pi)$ and $\theta\in(0,\pi)$.}
\begin{align}
n_1&= (
\cos \phi \sin \psi \sin \theta,
\sin \phi \sin \psi \sin \theta,
\cos \psi \sin \theta,
\cos \theta)
\label{a_28}\\
\int d \Omega_1 \, d \Omega_2 \, \cos^n \theta&= (8 \pi^2)(2 \pi) \int_0^\pi d\psi\, \sin \psi\, \int_0^\pi \sin^2 \theta \, \cos^n \theta
= \frac{(-)^{n+1} 2^{n+6} \pi^5}{(n+2)!\, \Gamma(\frac{1-n}{2})\Gamma(\frac{-1-n}{2})} \nn
\end{align}
The integral over $\ell_1$ and $\ell_2$ can be interpreted as an integral over the first quadrant of the plane $(\ell_1,\ell_2)$
\be
\int d \ell_1\, d\ell_2
\frac{\ell_1^{1+n} \ell_2^{1+n}}{(\ell_1^2+\ell_2^2)^{n+1}} = \int_0^{\pi/2} d\varphi \,(\sin \varphi \cos \varphi)^{n+1}\int_0^\Lambda L\,dL =
\frac{\Lambda^2}{2^n} \sqrt{\pi} \frac{\Gamma(\frac{2+n}{2})}{\Gamma(\frac{3+n}{2})}
\label{a_29}
\ee
where $\ell_1= L \cos \varphi$ and $\ell_2= L \sin \varphi$. Therefore the integral is quadratically divergent, however we have to check that the series is convergent at finite cutoff. The sum over $n$ can be computed exactly, obtaining
\be
I=-\frac{\log 2}{(2 \pi)^4 } \Lambda^2
\label{a_30} \ee
{Due to the non-renormalizable couplings between the hidden sector and the Standard model,
we find as before that $M_2$ depends  quadratically on the cutoff}
\begin{align}
M_2 & \sim Q_Y^{\psi} \bar{\cal Y}_{H\psi\chi} {\cal Y}_{H\psi\chi} {\Lambda^2 \over N M^2}
\int d^4 p \, F^{\widehat{A}}_{\m\n}(p)F^{Y,\mu \nu}(p)+ O (m^2,p) \nn \\
& \sim Q_Y^{\psi} \bar{\cal Y}_{H\psi\chi} {\cal Y}_{H\psi\chi} {1\over N}
\int d^4 p \, F^{\widehat{A}}_{\m\n}(p)F^{Y,\mu \nu}(p)+ O (m^2,p)
\label{a_31}
\end{align}
where in the last line we have eventually taken $\Lambda\simeq M$.

\subsubsection{The $V_3$ and $V_4$ terms: two-loop contribution.}

\begin{figure}[t]
\centering
~~~~~~~~ i. \includegraphics[width=0.35\textwidth]{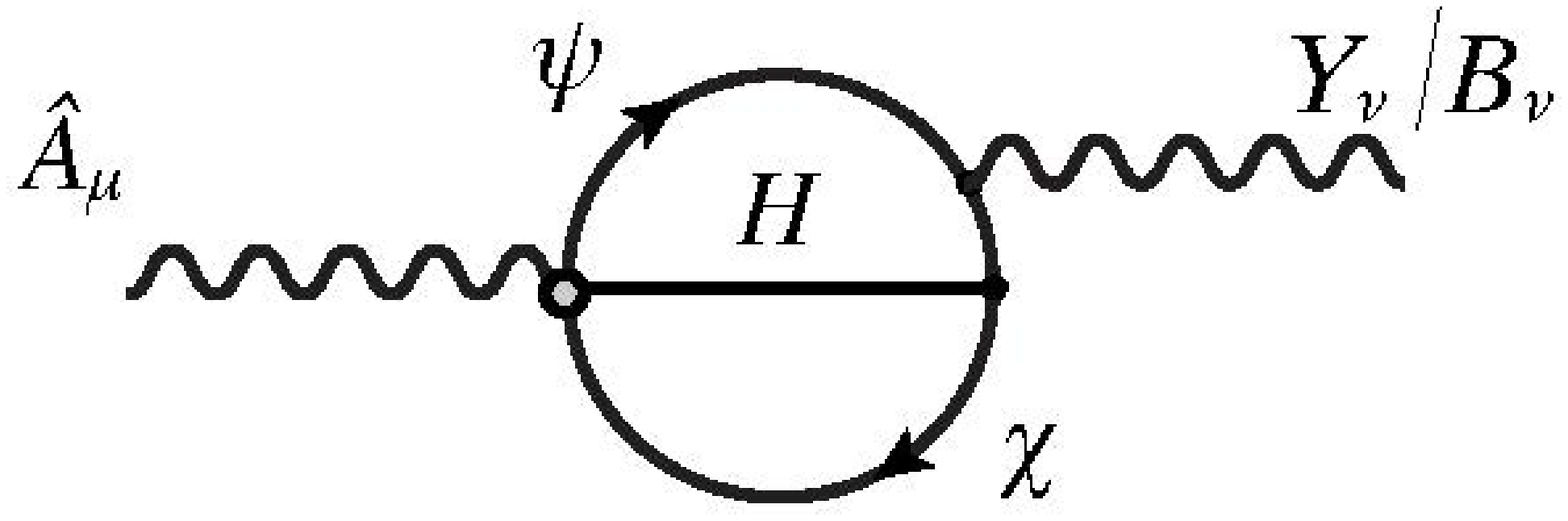}~~~~~~~~~~
ii.\includegraphics[width=0.35\textwidth]{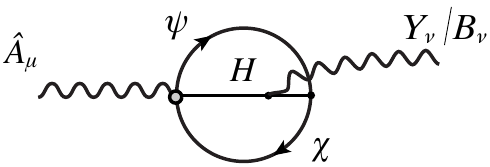} ~~~~~~~~~~
\caption[]{The lowest, two-loop contribution to $V_3$ and $V_4$ vevs are shown in fig i. and ii. above.}
\label{AcouplingstoSMV2ab}
\end{figure}

In this section we estimate the $V_3$, $V_4$ vevs. The relevant diagrams are similar, and they are presented in fig \ref{AcouplingstoSMV2ab}, where a Yukawa coupling is involved.
The coupling with the hidden gauge boson in the left side carries a suppression $N^{-\frac{3}{2}}$. Comparing the $N$ suppression of $V_{3}$ and $V_{4}$ with $V_{2}$, they have an extra suppression $\sqrt{N}$.

We start from $V_3$
\be
M_3\equiv \frac{ {\cal Y}_{H\psi\chi} }{M^2 N^{\frac{3}{2}}} \int d^4x ~d^4y~F^{\widehat{A}}_{\m\n}(x)A^Y_{\rho}(y)~
\langle Tr[\bar{\psi} \gamma^{\mu \nu} \psi H](x) ~
Tr[Q_Y^{\psi}\bar \psi \gamma^{\rho}\psi](y)\rangle
\label{a_19a}\ee
The contractions lead to the following 2-loop integral
\begin{align}
M_3\equiv &\frac{ {\cal Y}_{H\psi\chi} }{M^2 N^{\frac{3}{2}}} \int d^4 p~
F^{\widehat{A}}_{\m\n}(p)A^Y_{\rho}(p) ~I^{\mu \nu \rho}
\label{a_20a}\\
I^{\mu \nu \rho}= & \int {d^4 \ell_1 \over (2 \pi)^4} {d^4 \ell_2 \over (2 \pi)^4} ~
{Tr [\gamma^{\mu \nu} \slashed{\ell_1}(\slashed{\ell_1}+\slashed{p})\gamma^\rho \slashed{\ell_2} ]
\over
\ell_1^2 (\ell_1+p)^2 \ell_2^2 [(\ell_1-\ell_2+p)^2+m^2]}
\label{a_21a}
\end{align}
We follow the same steps done for $M_2$, however this time we keep $m$ non zero and we expand integral only in terms of the external momentum $p$. After the calculation of the trace in the numerator, the expansion in terms of $p$ and the reduction to a scalar integral, we find
\be
I^{\mu \nu \rho}= 2 m^2 (p^\nu \eta^{\mu \rho}-p^\mu \eta^{\nu \rho}) \int {d^4 \ell_1 \over (2 \pi)^4} {d^4 \ell_2 \over (2 \pi)^4} \left({1 \over \ell_1^2 \ell_2^2 [(\ell_1-\ell_2)^2+m^2]^2 }+{1 \over (\ell_2^1)^2 \ell_2^2 [(\ell_1-\ell_2)^2+m^2] }\right)
\label{a_22a} \ee
We show only the calculation of the first integral, the second one gives a similar result. The integral can be computed using the identity
\be
\frac{1}{[(\ell_1-\ell_2)^2+m^2]^2}=
\sum_{n=0}^\infty \frac{2^n (n+1) (\ell_1{\cdot}\ell_2)^n}{(\ell_1^2+\ell_2^2+m^2)^{n+2}}
\label{a_23a}
\ee
Following the same steps done for $M_2$ we obtain
\be
\int {d^4 \ell_1 \over (2 \pi)^4} {d^4 \ell_2 \over (2 \pi)^4} {1 \over D_1 D_2 D_3^2 } = \sum_{n=0}^\infty \frac{2}{(4 \pi) ^4 (n{+}1)^2}
 {_2F_1}\left(
\begin{array}{c}
2 (n{+}1),2 (n{+}1) \\
2 n{+}3
 \end{array} \!\!, {-}\frac{\Lambda ^2}{m^2}\right) \left(\frac{\Lambda ^2}{m^2}\right)^{4 (n+1)}
\label{a_24a} \ee
The above hypergeometric function can be transformed to expose the presence of logarithmic divergences. Then after some manipulations and in the approximation of large $\Lambda$ we obtain
\be
\int {d^4 \ell_1 \over (2 \pi)^4} {d^4 \ell_2 \over (2 \pi)^4} {1 \over \ell_1^2 \ell_2^2 [(\ell_1-\ell_2)^2+m^2]^2 } \sim \log \frac{\Lambda^2}{m^2}
\label{a_25a} \ee
where we are neglecting the overall $O(1)$ coefficient. A similar computation can be done for the second integral in \eqref{a_22a} obtaining again a logarithmic divergence, therefore the contribution $M_2$ is logarithmically divergent, up to an $O(1)$ factor,
\be
M_3\sim \frac{ {\cal Y}_{H\psi\chi} m^2}{M^2 N^{\frac{3}{2}}} \log \frac{\Lambda^2}{m^2}
\int d^4 p \, F^{\widehat{A}}_{\m\n}(p)F^{Y,\mu \nu}(p)
\label{a_26a}\ee

We now focus on the contribution $V_4$
\be
M_4\equiv \frac{Q_Y^H {\cal Y}_{H\psi\chi} }{M^2 N^{\frac{3}{2}}}
\int d^4x ~d^4y~F^{\widehat{A}}_{\m\n}(x)A^Y_{\rho}(y)~
\langle Tr[\bar{\psi} \gamma^{\mu \nu} \psi H](x) ~
Tr[H\pa_{\rho}H^{\dagger}-H^{\dagger} \partial_{\rho} H](y)\rangle
\label{a_27a}\ee
Computing the contractions we obtain
\begin{align}
M_4= &\frac{Q_Y^H {\cal Y}_{H\psi\chi} }{M^2 N^{\frac{3}{2}}}
\int d^4 p~ F^{\widehat{A}}_{\m\n}(p)A^Y_{\rho}(p) ~I^{\mu \nu \rho}
\label{a_28a}\\
I^{\mu \nu \rho}= & \int {d^4 \ell_1 \over (2 \pi)^4} {d^4 \ell_2 \over (2 \pi)^4} ~
{(2 \ell_2-p)^\rho Tr [\gamma^{\mu \nu} \slashed{\ell_1}(\slashed{\ell_1}-\slashed{\ell_2}+\slashed{p})]
\over
\ell_1^2 (\ell_2^2+m^2) (\ell_1-\ell_2)^2 [(\ell_2-p)^2+m^2]}
\label{a_29a}
\end{align}
This time we can expand w.r.t. $p$ and also take $m=0$. The calculations lead to the integral
\be
I^{\mu \nu \rho}=2 (p^\mu \eta^{\nu \rho}-p^\nu \eta^{\mu \rho}) \int \frac{d^4\ell_1}{(2\pi)^4} \frac{d^4\ell_2}{(2\pi)^4}
\frac{1}{\ell_1^2 \, \ell_2^2 \, (\ell_1-\ell_2)^2} \sim \frac{ \Lambda^2}{(2 \pi)^4} (p^\mu \eta^{\nu \rho}-p^\nu \eta^{\mu \rho})
\label{a_30a} \ee
where we have used that, up to an overall factor, the integral is the same of \eqref{a_25}. Therefore we have found that $M_4$ is quadratically divergent
\be
M_4\sim \frac{Q_Y^H {\cal Y}_{H\psi\chi} }{N^{\frac{3}{2}}}
{\Lambda^2 \over M^2} \int d^4 p~
F^{\widehat{A}}_{\m\n}(p)F_Y^{\mu \nu}(p)
\sim \frac{Q_Y^H {\cal Y}_{H\psi\chi} }{N^{\frac{3}{2}}}
\int d^4 p~
F^{\widehat{A}}_{\m\n}(p)F_Y^{\mu \nu}(p)
\label{a_39a} \ee

\subsubsection{The $V_5$ terms: two-loop contribution.}

\begin{figure}[h]
\centering
i. \includegraphics[width=0.35\textwidth]{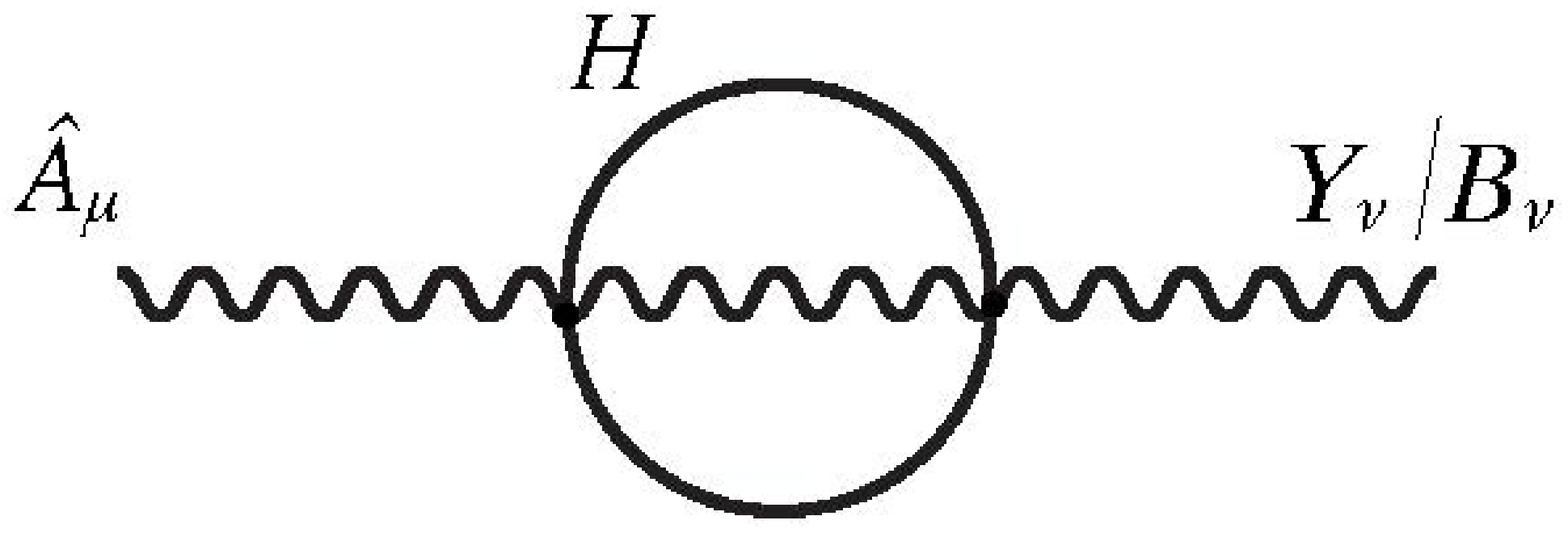}~~~~~~~~~~
ii.\includegraphics[width=0.35\textwidth]{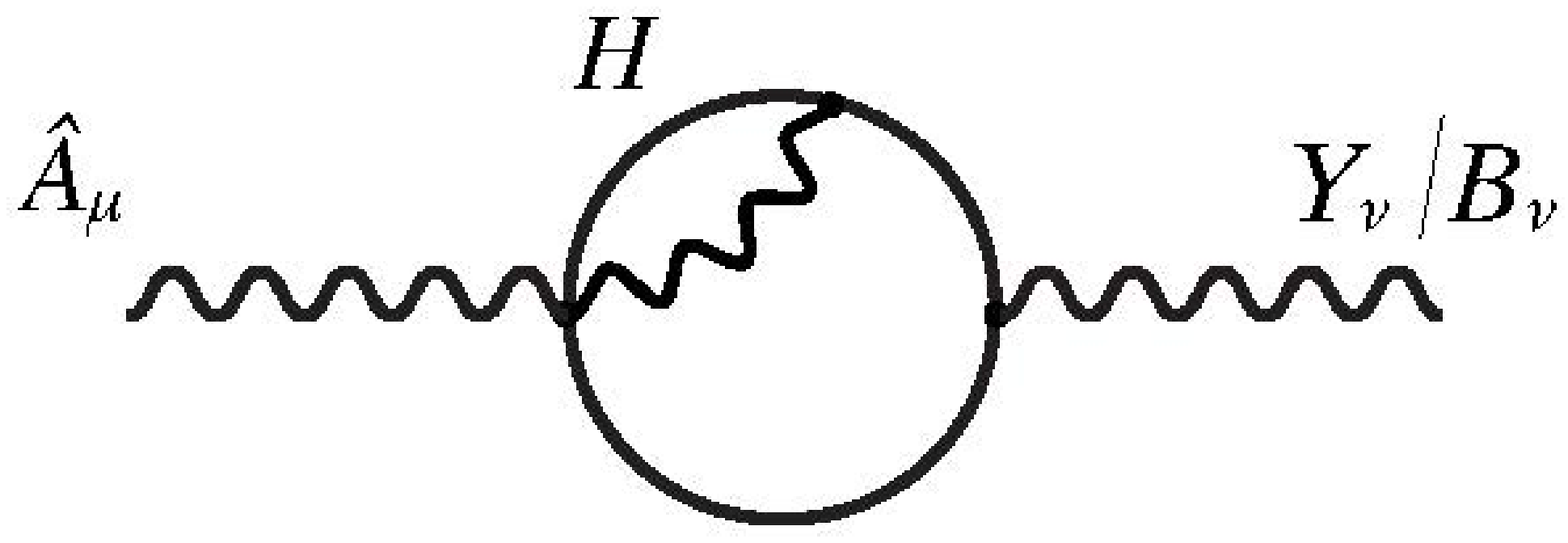} ~~~~~~~~~~
\caption[]{Contributions $V_5$ to the mixing in the unbroken phase.}
\label{AcouplingstoSMV5}
\end{figure}

In this section, we estimate the $V_5$ vev. The leading contribution is coming from two diagrams, which are shown in fig \ref{AcouplingstoSMV5}.
In the internal lines, we have Higgs and hypercharge gauge bosons. In all diagrams, the coupling between ${\widehat{A}}$, $Y$ and $H$ comes from the second term in \eqref{a_1}, while the other couplings are the minimal coupling between Higgs and SM gauge bosons.

The two diagrams yield the following integrals
\begin{align}
M_{5,i} & =
\frac{\Lambda^2}{N M^2} (Q_Y^H)^2 \int d^4 p \int \frac{d^4 \ell_1}{(2 \pi)^4} \frac{d^4 \ell_2}{(2 \pi)^4}
\frac{\eta^{\mu \nu} \ell_2{\cdot}p -\ell_2^\mu p^\nu }{[(\ell_1{+}\ell_2{+}p)^2+m^2](\ell_1^2+m^2) \ell_2^2}
\label{a_40a}\\
M_{5,ii} & = \frac{\Lambda^2}{N M^2} (Q_Y^H)^2 \int d^4 p \int \frac{d^4 \ell_1}{(2 \pi)^4} \frac{d^4 \ell_2}{(2 \pi)^4}
\frac{(\eta^{\mu \rho} \ell_2{\cdot}p -\ell_2^\mu p^\rho)( 2 \ell_1+2p+\ell_2)^\rho( \ell_1+p+\ell_2)^\nu }{[(\ell_1{+}\ell_2{+}p)^2+m^2](\ell_1^2+m^2) \ell_2^2[(\ell_1{-}p)^2+m^2]}
\label{a_41a}
\end{align}
The procedure to compute these diagrams is the same used for $M_2$. Taking $m=0$ and expanding around $p=0$, the leading contribution of both integrals is proportional to the integral \eqref{a_25}, therefore we obtain
\be
M_5\sim \frac{\Lambda^2}{N M^2} (Q_Y^H)^2
\int d^4 p \, F^{\widehat{A}}_{\m\n}(p)F_Y^{\mu \nu}(p)
\sim \frac{(Q_Y^H)^2}{N}
\int d^4 p \, F^{\widehat{A}}_{\m\n}(p)F_Y^{\mu \nu}(p)
\label{a_37}\ee
Therefore this contribution is competing with the one due to $V_1$ in (\ref{a_11}).

\subsubsection{The $V_6$ and the rest of the vevs: higher-loop contribution.}

The $V_6$ vev has at least three-loop contributions that is further suppressed and we will therefore neglect it. This is also the case for every vev, which contains the last coupling in \eqref{a_1}.

\subsection{The broken phase of the SM \label{SMbp}}

\begin{figure}[h]
\centering
\includegraphics[width=0.80\textwidth]{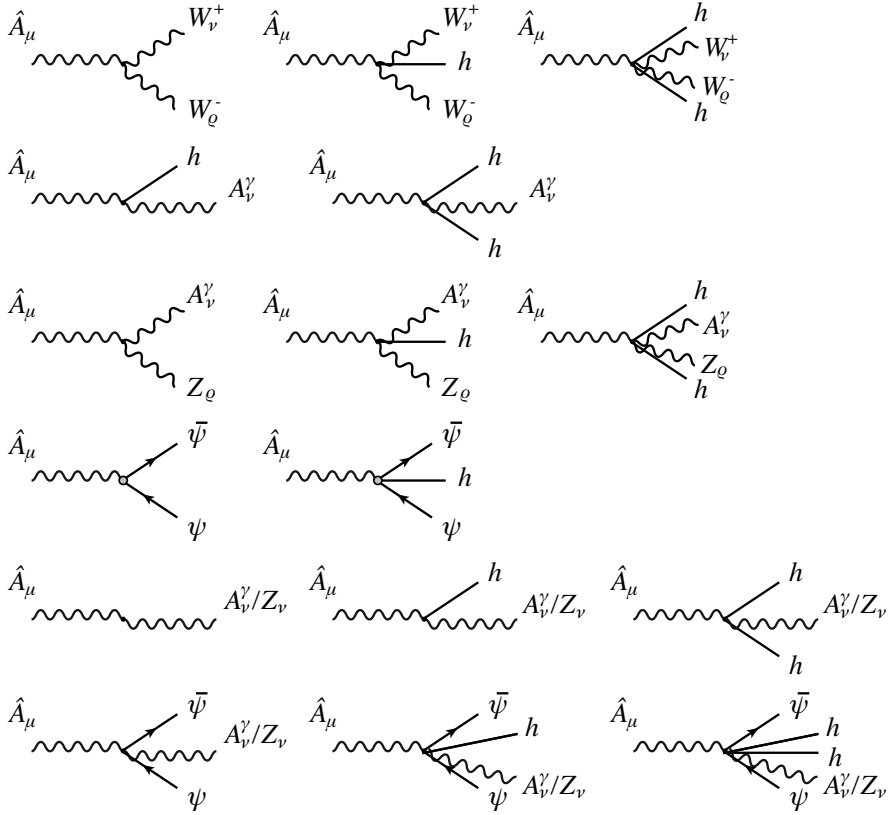}
\\
\caption[]{Couplings at the broken-phase.}
\label{AcouplingstoSMbroken}
\end{figure}

In this subsection, we study the mixing between the photon $\gamma$ and hidden $U(1)$ $\widehat A$ in the broken phase of the electroweak symmetry. In that phase, the \eqref{a_1} becomes
\begin{align}
W_{BROKEN} &\sim
\frac{4 g_w^2}{N M^2}(h{+}v)^2 \,F_{\mu \nu}^{\widehat{A}}W^\mu_+ W^\nu_- + \frac{4 i e}{N M^2} \, (h{+}v) F_{\mu \nu}^{\widehat{A}} A^\mu_\gamma \partial^\nu h\nn \\
&+\frac{4 e}{N M^2} \sqrt{g_w^2{+}g_Y^2} (h{+}v)^2 F_{\mu \nu}^{\widehat{A}} A^\mu_\gamma Z^\nu
+{1\over N^{3\over 2}M^2}F^{\m\n}_{\widehat{A}}\left[(h+v) \bar\psi\gamma_{\m\n}\psi+c.c.\right]\nn\\
&+{1\over N M^2}F_{\m\n}^{\widehat{A}} (\cos \theta_w F^{\gamma, \mu \nu} -\sin \theta_w F^{Z, \mu \nu})  (h+v)^2\nn\\
&
+{1\over N^{ 2}M^4} F_{\m\n}^{\widehat{A}} (\cos \theta_w F^{\gamma, \mu \nu} -\sin \theta_w F^{Z, \mu \nu}) \left[\bar \psi \psi (h+v)+c.c.\right]\cdots
\label{a_38}
\end{align}
There is a long list of couplings between the $\widehat A$ and the SM fields coming from \eqref{a_38} schematically drawn in fig \ref{AcouplingstoSMbroken}.
These couplings should be combined with the SM action to get the mixings.
Since there are too many diagrams, we will focus on the leading contributions, at most at one-loop.

\subsubsection{The direct mixing}

In the broken phase, there is a direct kinetic mixing term coming from the third line in \eqref{a_38} and it is
\bea
M_0^\textrm{broken}=
\frac{v^2}{N M^2}
\int d^4 p \, F^{\widehat{A}}_{\m\n}(p) \Big(\cos \theta_w F^{\gamma, \mu \nu} -\sin \theta_w F^{Z, \mu \nu}\Big) (-p)
\label{a_39}\eea

\subsubsection{The fermionic loop}

\begin{figure}[h]
\centering
\includegraphics[width=0.40\textwidth]{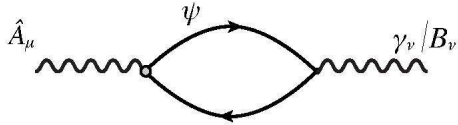}
\\
\caption[]{One-loop contributions from SM fermions and $W^\pm$ to the kinetic mixing of the visible and the hidden gauge bosons in the broken phase.}
\label{AcouplingstoSMV3V4n}
\end{figure}

The one-loop contribution of the SM fermions (fig \ref{AcouplingstoSMV3V4n}.i) to the kinetic mixing is
\be
M_{\psi}^\textup{broken}={Q_Y v\over N^{\frac{3}{2}} M^2}\int d^4q~F^{\widehat{A}}_{\m\n}(q)~A^Y_{\rho}(-q) I^{\m\n\rho}
\label{a_40}\ee
where $v$ is the Higgs vev, and the integral
\be
I^{\m\n\rho}=\int d^4p\, {Tr[\gamma^{\m\n}(\pslash-m_Y)\gamma^{\rho}(\qslash-\pslash-m_Y)]\over (p^2-m_Y^2)[(p-q)^2-m_Y^2]}
\label{a_41}
\ee
where $m_Y$ is the mass of the SM fermion $\psi$. After some massage we have
\be
I^{\m\n\rho}=-4m_Y\int_0^1dx \int d^4p\, {\eta^{\m\rho}(2p^{\n}-q^{\n})-\eta^{\n\rho}(2p^{\m}-q^{\m})\over \left[(p+xq)^2+x(1-x)q^2-m_Y^2\right]^2}
\label{a_42}\ee
Shifting variables and dropping linear terms we eventually obtain
\be
I^{\m\n\rho}=4m_Y(\eta^{\m\rho}q^{\n}-\eta^{\n\rho}q^{\m})\int_0^1dx (2x+1)\int {d^4p \over \left[p^2+x(1-x)q^2-m_Y^2\right]^2}
\label{a_43}\ee
Rotating to Euclidean space we obtain
\begin{align}
I^{\m\n\rho}&=
2im_Y(\eta^{\m\rho}q^{\n}-\eta^{\n\rho}q^{\m})\Omega_3\int_0^1dx (2x+1)\left[\log{\Lambda^2\over m_Y^2-x(1-x)q^2}-1\right]+{\cal O}(\Lambda^{-2})\nn\\
&=
4im_Y(\eta^{\m\rho}q^{\n}-\eta^{\n\rho}q^{\m})\Omega_3\log{\Lambda^2\over m_Y^2}+{\cal O}(q^2)+{\cal O}(\Lambda^{-2})
\label{a_44}
\end{align}
We finally obtain
\be
M_{\psi}^\textup{broken}=4\Omega_{3}Tr_Y\left[{Q_Y m_Y v\over N^{\frac{3}{2}} M^2}\right]\log{\Lambda^2\over m_Y^2}\int d^4q~F_{\widehat{A}}^{\m\n}(q)~F^Y_{\m\nu}(-q)+\cdots
\label{a_45}\ee

\subsubsection{The $W^+W^-$ loop}

\begin{figure}[h]
\centering
\includegraphics[width=0.40\textwidth]{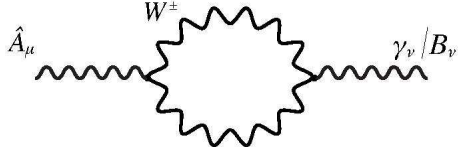}
\\
\caption[]{one-loop contributions from SM fermions and $W^\pm$ to the kinetic mixing of the visible and the hidden gauge bosons in the broken phase.}
\label{AcouplingstoSMV3V4}
\end{figure}

In order to obtain the one-loop correction to $F^{\mu \nu}_{\widehat{A}} F_{\mu \nu}^\gamma$, we consider only the terms with three fields, which appear in the first and the third term taking $h=0$, however, only the first term can be coupled to SM to obtain a one-loop diagram, as shown in figure \ref{AcouplingstoSMV3V4}.ii. The $W$-bosons contribution is then
\be
M_W^\textup{broken}=-\frac{ 16 i e m_W^2}{N M^2} \int d^4 x\, d^4 y \, F_{\mu \nu}^{\widehat{A}}(x) F_{\rho \sigma}^\gamma(y) \langle [W_+^\mu W_-^\nu](x) [W_+^\rho W_-^\sigma](y) \rangle
\label{a_46}\ee
Using the Euclidean propagator
\be
\langle W_+^\mu (x) W_-^\nu(0) \rangle= \int \frac{d^4 q}{(2\pi)^4} \frac{\eta^{\mu \nu}+q^\mu q^\nu/m^2}{q^2+m^2} e^{i qx}
\label{a_47}\ee
and integrating the positions $x$ and $y$ we obtain,
\begin{align}
M_W^\textup{broken}&= -\frac{ 16 i e m_W^2}{N M^2} \int \frac{d^4q}{(2\pi)^4}\, F_{\mu \nu}^{\widehat{A}}(q) F_{\rho \sigma}^\gamma({-}q) I^{\mu \nu\rho\sigma} \nn\\
I^{\mu \nu\rho\sigma} &=\int \frac{d^4 p}{(2\pi)^4} \,
\frac{\eta^{\mu \sigma}{+}(q-p)^\mu (q-p)^\sigma/m_W^2}{(q-p)^2+m_W^2}
\frac{\eta^{\nu \rho}{+}p^\nu p^\rho/m_W^2}{p^2+m_W^2}
\label{a_48}
\end{align}
Introducing the Feynman parametrisation, we obtain
\be
I^{\mu \nu\rho\sigma}=
\frac{\eta^{\mu \sigma} \eta^{\nu \rho} {-} \eta^{\mu \rho} \eta^{\nu \sigma}}{2m_W^2}
\int \frac{d^4 p}{(2\pi)^4} \,
\frac{p^2+2 m^2_W}{(p^2+m^2_W)^2} +\mathcal{O}(q)=
\frac{\eta^{\mu \sigma} \eta^{\nu \rho} {-} \eta^{\mu \rho} \eta^{\nu \sigma}}{4 m_W^2} \frac{\Omega_3}{(2\pi)^4} \frac{\Lambda^4}{m_W^2{+}\Lambda^2}+\mathcal{O}(q)
\label{a_49}\ee
where we have neglected the external momentum $q$ which generates higher derivative corrections to $F_A F_\gamma$. The leading contribution is the following
\be
M_W^\textup{broken}= -e \frac{ \Lambda^2 }{N M^2} \frac{8 i \Omega_3}{(2\pi)^4} \int \frac{d^4 q}{(2\pi)^4}\, F^{\widehat{A}}_{\mu \nu}(q) F_\gamma^{\nu \mu}({-}q)
\label{a_50}\ee
This is parametrically similar to that of the unbroken phase.

\subsection{Mixing computations in the anomalous U(1) case}\label{anomalousU1}

In this subsection, we assume the presence of an extra `anomalous' $U(1)$ $B^\m$, in the extension of the SM that is anomalous, and, in addition, of an axion $\b$ associated to $B^\m$ \cite{Anastasopoulos:2008jt,bianchi} as this realized in string-theory effective actions for anomalous U(1)'s, \cite{rev}.

We also assume that the anomalous $U(1)$ $B^\mu$ couples with SM fields similarly to the hypercharge $Y^\m$, but with different charges denoted by $Q_B^{X}$, depending on the field $X$,
\be
W_B=\int d^4x~ B^{\m}
\left( iQ_B^H Tr[H\pa_{\mu}H^{\dagger}-H^{\dagger} \partial_{\mu} H]
+ Tr[Q_B^{\psi}\bar \psi \gamma_{\m}\psi]\right)
\label{a_51}\ee
The anomalous $U(1)$ couples to the hidden $U(1)$ $\hat A^\m$ through the gauge invariant combination $(\partial \beta + m B)_\mu$, with $\beta$ a St\"uckelberg field, like
\be
\frac{1}{N^{3/2} M^3} F_{\mu \nu}^{\hat A} (\partial \beta + m B)^\mu \bar{\psi} \gamma^{\nu} \psi+\frac{1}{N^{3/2} M^3} \,(\partial \beta + m B)^\mu (H^\dagger D^\nu H-D^\nu H^\dagger H) F_{\mu \nu}^{\hat A}
\label{a_52}\ee

Next, we want to compare the mixing between the hidden gauge field $\hat A$ and the non-anomalous $Y$/anomalous $B$ gauge fields of the SM, via the couplings \eqref{a_4} (fig \ref{AcouplingstoSM}).

The mixings of the anomalous $U(1)$ $B^\m$ with the hidden $\hat A^\m$ are similar to those between $\hat A^\m$ and the hypercharge. The only difference appears due to the different charges $Q_B^{X}$ and $Q_Y^{X}$ each field $X$ has.

\begin{figure}[h]
\centering
\includegraphics[width=0.55\textwidth]{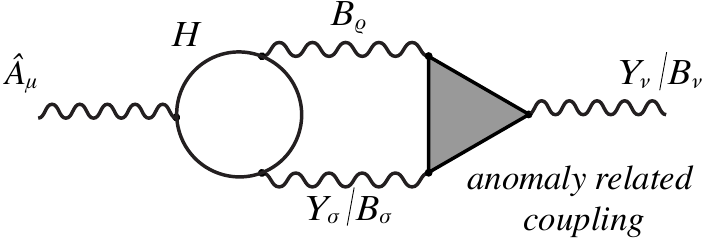}\\
\vspace{.90cm}
\centering
\includegraphics[width=0.75\textwidth]{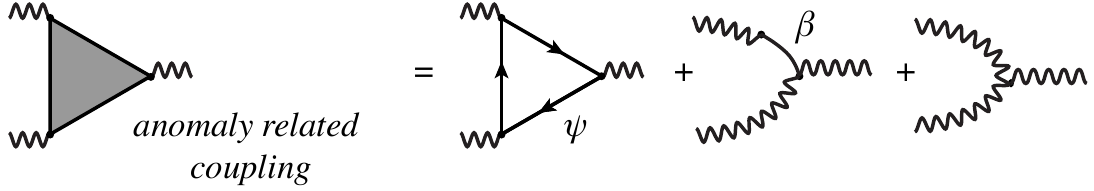}\\
\caption[]{The lowest diagram that includes the anomaly related coupling. That coupling includes the fermionic loop, the axion diagram and the GCS coupling.
}
\label{AtoBHiggsAndAnomaliesALL}
\end{figure}

The additional diagrams that can appear in the anomalous case contain the anomalous fermionic loop (which is absent in the non-anomalous case).
The first contribution related to the anomaly is a three-loop diagram provided in fig \ref{AtoBHiggsAndAnomaliesALL}.
Therefore, we do not expect any significant difference in the mixing of $\widehat{A}^{\mu}$ with non-anomalous $Y^\mu$ or anomalous $B^\mu$ gauge fields of the SM. Consequently, the mixings of the graviphoton/dark photon $\widehat{A}^{\mu}$ to the non-anomalous $Y^\mu$ and anomalous $B^\mu$ gauge fields depend on the charges of the Higgs under these SM gauge bosons. The same arguments apply to the broken phase of the SM.

\subsection{Summary}

To conclude, we summarize the results of this appendix. In the unbroken phase, the couplings in \eqref{2_18} between SM fermions and the hidden gauge boson are absent at all orders, therefore the kinetic mixing between SM and hidden gauge bosons is only due to the Higgs field. The leading term appears at one loop, \eqref{a_18}, and it generates the following effective mixing
\be
{\Lambda^2\over N~M^2} F_{\m\n}^{\widehat{A}}F^{\m\n}_Y\sim {1\over N} F_{\m\n}^{\widehat{A}}F^{\m\n}_Y
\label{a_53} \ee
up to $\mathcal{O}(1)$ coefficients were in the second expression, we took the cutoff to be the messenger scale $M$. We have also computed the leading two-loop corrections to the mixing, \eqref{a_28} and \eqref{a_29}, which show further suppressions in SM couplings.

In the broken phase, additional couplings are generated due to spontaneous symmetry breaking. The hidden gauge boson couples to the massive bosons $W^\pm$ and $Z$ \eqref{a_38} and all the (electromagnetically) charged fermions. The leading contribution to the kinetic mixing with the photon is given by the one-loop diagram with the $W$-bosons circulating in the loop
\be
{\Lambda^2\over N~M^2} F_{\m\n}^{\widehat{A}}F^{\m\n}_\gamma\sim {1\over N} F_{\m\n}^{\widehat{A}}F^{\m\n}_\gamma
\label{a_54}
\ee
up to $\mathcal{O}(1)$ coefficients. As for the unbroken phase, two-loop corrections are further suppressed by SM couplings.

\section{SM renormalization of the graviphoton gauge coupling\label{ggc}}

In this appendix we calculate the renormalization of the emergent vector gauge coupling due to SM quantum effects.
This is associated to the effective interaction terms in (\ref{a_1}).
It is given by
\begin{align}
&{1\over N^2 M^4}\int d^4 x d^4 y~F^{\m\n}_{\widehat{A}}(x)\langle Tr[D_{\m}H D_{\n}H^{\dagger} ](x) Tr[D_{\r}H D_{\s}H^{\dagger} ](y)~F^{\r\s}_{\widehat{A}}(y)+ \label{b_1} \\
+\,&
{1\over N^3M^4}\int d^4 x d^4 y~F^{\m\n}_{\widehat{A}}(x)\langle Tr[F_{\m\n}^Y HH^{\dagger}](x) Tr[F_{\m\n}^Y HH^{\dagger}](y)~F^{\r\s}_{\widehat{A}}(y) + \nn\\
+\,&
{2\over N^{5/2}M^4}\int d^4 x d^4 y~F^{\m\n}_{\widehat{A}}(x)\langle Tr[F_{\m\n}^Y HH^{\dagger}](x) Tr[D_{\r}H D_{\s}H^{\dagger}](y)~F^{\r\s}_{\widehat{A}}(y) \nn
\end{align}

In the unbroken phase, the one-loop contribution is given by the first line.
\be
M_1=\frac{4}{N^2 M^4}
\int \frac{d^4 k}{(2 \pi)^4} F^{\mu \nu}_{\widehat{A}}(k) F^{\rho \sigma}_{\widehat{A}}(-k)
\int \frac{d^D \ell}{(2 \pi)^4} \frac{\ell_\mu (k+\ell)_\nu \ell_\rho (\ell-k)_\sigma
}{\ell^2 (\ell+k)^2}
\label{b_2}\ee
Using Feynman parametrisation and Lorentz invariance we obtain
\be
M_1=\frac{4}{N^2 M^4}
\int \frac{d^4 k}{(2 \pi)^4}\, k^2 F^{\mu \nu}_{\widehat{A}}(k) F_{{\widehat{A}},\mu \nu}(-k)
\int \frac{d^4 \ell}{(2 \pi)^4} \frac{\ell^2 }{[\ell^2+x(1-x)k^2]^2}
\label{b_3}\ee
Integrating the loop momentum $\ell$ and keeping the leading term in the cut-off regularization we obtain a quadratic dependence on the cut-off $\Lambda$
\be
M_1=\frac{1}{N^2 M^4}\left(\Lambda^2-2 m^2 \log \frac{\Lambda^2}{m^2}\right) \frac{2 \Omega_4}{(2\pi)^4}
\int \frac{d^4 k}{(2 \pi)^4}\, k^2 F^{\mu \nu}_{\widehat{A}}(k) F_{{\widehat{A}},\mu \nu}(-k)
\label{b_4}\ee
where $m^2$ is the Higgs 'tachyonic' mass.

In the broken phase there are four one-loop contributions to compute. They are generated by the 3-{point} interaction vertices in \eqref{a_38}. Computing the associated diagrams, as done in section \ref{SMbp}, we obtain the following contributions that are identified by the fields circulating in the internal lines
\begin{itemize}
\item SM fermions contribution
\be
M_{\psi\psi}\sim {m_\psi^2\over N^2 M^2} \log{\Lambda^2\over m_\psi^2}
 \int \frac{d^4 q}{(2\pi)^4}\, F_{\mu \nu}^{\widehat{A}}(q) F_{\widehat{A}}^{\nu \mu}({-}q)
\label{b_5}\ee
\item $W$'s contribution
\be
M_{WW}\sim \frac{ m_W^2}{ M^2} \frac{ \Lambda^2 }{N^2 M^2}
 \int \frac{d^4 q}{(2\pi)^4}\, F_{\mu \nu}^{\widehat{A}}(q) F_{\widehat{A}}^{\nu \mu}({-}q)
\label{b_6}\ee
\item $Z$ and $\gamma$ contribution
\be
M_{Z \gamma}\sim \frac{ e^2 v^2}{ M^2} \frac{ 1}{N^2 M^2} \left(\Lambda^2-3m_Z^2 \log \frac{\Lambda^2}{m_Z^2}\right)
 \int \frac{d^4 q}{(2\pi)^4}\, F_{\mu \nu}^{\widehat{A}}(q) F_{\widehat{A}}^{\nu \mu}({-}q)
\label{b_7}\ee
\item Higgs and $\gamma$ contribution
\be
M_{h \gamma}\sim \frac{ e^2 v^2}{ M^2} \frac{ 1 }{N^2 M^2} \left(\Lambda^2+m_h^2 \log \frac{\Lambda^2}{m_h^2}\right)
 \int \frac{d^4 q}{(2\pi)^4}\, F_{\mu \nu}^{\widehat{A}}(q) F_{\widehat{A}}^{\nu \mu}({-}q)
\label{b_8}\ee
\end{itemize}
In \eqref{b_6} the logarithmic divergent term is neglected because it is proportional to $k^2$.
Notice that the $W$'s contribution cannot cancel the third and the fourth contributions being proportional to $m_W^2 \sim v^2 g_2^2$.

We summarise these calculations as follows. Assuming that the cutoff $\Lambda\sim M$ is at the messenger scale, the corrections have the structure
\be
 {\delta g_{hidden}^2 \over g_{hidden}^2}\sim {1\over N^2}{m_{SM}^2\over M^2}
 \label{b_9}
\ee
where $m_{SM}$ stands for a SM mass scale, typically a mass scale of one of the SM particles.
 Therefore the correction has an extra suppression (beyond the large N suppression) due to the hierarchy, $m_{SM}\ll M$.

\section{Mixing computations in the string theory setup}
\label{Mixing_string}

In this section, we fix the normalizations for the Vertex Operators (VO's) and compute the relevant amplitudes, shown in section \ref{stringsetup}.
We are interested in evaluating the dependence on the scale of the problem and the string coupling, therefore we neglect ${\cal O}(1)$ numerical factors.
{Although we work in flat non-compact space-time with internal tori, orbifolds or CY's that admit a CFT description on the world-sheet, we consider the effect of closed-string fluxes at linear order both in the NSNS and RR sector. This represents a first step towards an analysis in warped spaces such as AdS.}

\subsection{Normalizations}
\label{string_normalizations}

We use a normalisation such that the world-sheet fields $X$ and ${\Psi}$ have dimension of a length $[X], [{\Psi}] \approx \sqrt{\alpha'} =\ell_s $ so much so that $[\partial X] = [k{\cdot}{\Psi}\, {\Psi}]$. It turns out to be convenient to normalise also the spin fields so that $[S] \approx \sqrt{\alpha'} =\ell_s$ as suggested by triality of $SO(8)$ in the light-cone gauge or G-S formalism.

Up to factors of 2 and $\pi$ coming from volumes of spheres $S^n$, the Newton constant and the Plank length in $D=10$ are given by $G_{10} = g_s^2 \ell_s^8$ and $\ell_{10}^8 = g_s^2 \ell_s^8$, where $g_s$ is the closed-string coupling ($g_s{\,=\,}g_{op}^2$) and $\ell_s$ is the string length. In lower dimension
\be
G_{D} = {g_s^2 (\alpha')^4\over V_{10-D}} \quad i.e. \quad \ell_{D}^{D-2} = g_s^2 \ell_s^{D-2}
{\ell_s^{10-D}\over V_{10-D}}
\label{c_1}\ee
where $V_{10-D}$ is the compactification volume and $\ell_{D}$ is the Planck length in $D$ dimensions. Similarly the tension/charge of a Dp-brane reads
\be
T_{Dp} = g_s^{-1} (\alpha')^ {-{p+1\over 2}} = g_s^{-1} \ell_s^{-p-1}
\label{c_2}\ee
which is related to the normalisation of the (empty) disk amplitude
\be
C_{Dp} = {V_{p+1}\over g_s (\alpha')^{p+1\over 2}} = {V_{p+1} \over g_s \ell_s^{p+1}}
\label{c_3}
\ee
where $V_{p+1}$ is the volume occupied by the brane. The gauge coupling of the resulting (low-energy) SYM theory in $p+1$ dimension is
\be
g^{2}_{YM_p} = g_s (\alpha')^{p-3\over 2} = g_s \ell_s^{p-3}
\label{c_4}\ee
When the D$p$-brane is wrapped on a $D{\,-\,}p{\,-\,}1$ cycle (so as to obtain $D$ dimensional SYM theory at low energies) and the VO's are inserted, one has
$V_{p+1}= V_D V_{p+1-D} \rightarrow \delta^{D}(\sum_i k_i) V_{p+1-D} $. As a consequence
\be
g^{2}_{YM^{Dp}_{D}} = g_s (\alpha')^{D-4\over 2} {(\alpha')^{p+1-D\over 2}\over V_{p+1-D}}
= g_s {\ell_s^{p-3}\over V_{p+1-D}}
\label{c_5}\ee

The normalizations of the VO's for vector bosons can be obtained comparing the 3- and 4-{point} amplitudes with the correspondent amplitudes in YM. The VO's read
\be
V_A^{(-1)} = C_A e^{-\varphi} A{\cdot}{\Psi} e^{ik{\cdot}X} \quad , \quad V_A^{(0)} = C_A \left(A{\cdot} i\partial X + \frac{1}{2} F_{\mu \nu}{\Psi}^\mu {\Psi}^\nu\right) e^{ik{\cdot}X}
\label{c_6}\ee
We obtain the following 3- and 4-{point} amplitudes on the disk:
\begin{align}
{\cal A}_3 &=
\langle cV_A^{(-1)}(1) cV_A^{(-1)}(2) cV_A^{(0)}(3) \rangle
= \widehat{C}_{Dp} C_A^3
\ell_s^4 [A_1{\cdot}A_2 A_3{\cdot}{(k_1{-}k_2)} + {...}] \label{c_7}\\
{\cal A}_4 & =\!\! \int \!dz \langle cV_A^{(-1)}(1) cV_A^{(-1)}(2) cV_A^{(0)}(z) cV_A^{(0)}(3) \rangle
=\widehat{C}_{Dp} C_A^4
\ell_s^4 [A_1{\cdot}A_2 A_3{\cdot}A_4 {+} {...} ] \label{c_8}
\end{align}
where the ellipsis are permutations. Imposing the amplitudes to be equal to YM amplitudes, we obtain $\widehat{C}_{Dp} C_A^3 (\alpha')^2{\,=\,} g_{YM} $ and $\widehat{C}_{Dp} C_A^4 (\alpha')^2 {\,=\,} g_{YM}^2$ where $\hat C_{Dp}{\,=\,}C_{Dp}/V_D$. These relations yield
\be
\widehat{C}_{Dp} = {1 \over g_{YM}^2 (\alpha')^2}
\quad , \quad
C_A = g_{YM}
= \sqrt{g_s} {(\alpha')^{p-3\over 4}\over \sqrt{V_{p+1-D}}}
\label{c_9}\ee

With the normalisations of vector bosons and the disk fixed, we can obtain the normalisation of the graviton vertex
\be
W_G^{(-1,-1)} {\,=\,} C_G e^{-\varphi} {\Psi}{\cdot}h{\cdot}\tilde{\Psi} e^{ik{\cdot}X}
~ , ~
W_G^{(0,0)} {\,=\,} C_G (i\partial X {+} {\Psi} k{\cdot}{\Psi}){\cdot}h{\cdot}(i\bar\partial X {+} \tilde{\Psi} k{\cdot}\tilde{\Psi}) e^{ik{\cdot}X}
\label{c_10}\ee
using the mixed 3-{point} amplitude
\be
{\cal A}_{VVG} {=} \langle cV_A^{(-1)}(1) cV_A^{(-1)}(2) (c{+}\tilde{c})W_G^{(0,0)} \rangle {\,=\,} \widehat{C}_{Dp} C_A^2 C_G \ell_s^6 [A_1{\cdot}A_2 \,{k_{12}}{\cdot}h_3{\cdot}{k_{12}}{+}...]
\label{c_11}\ee
where $k_{12}{\,=\,}k_1{\,-\,}k_2$. Imposing the minimal coupling, $\ell_s^6 \,C_{Dp} \,C_A^2\, C_G {\,=\,} \sqrt{G_D}$,
we obtain
\be
C_G = {\sqrt{G_D}\over \alpha'} = {\ell_D^{D-2\over 2} \over \ell^2_s} = {g_s \alpha' \over \sqrt{V_{10-D}}}
\label{c_12}\ee

To obtain the normalization for the RR VO's we need the normalization of the sphere, which can be obtained computing the 3-graviton amplitude:
\be
{\cal M}_{3G} = \langle c\tilde{c}W_G^{(-1,-1)}c\tilde{c}W_G^{(-1,-1)}c\tilde{c}W_G^{(0,0)}(3) \rangle = \widehat{C}_{S^2} C_G^3 \ell_s^8 [h_1{\cdot}h_2 \,{k_{12}}{\cdot}h_3{\cdot}{k_{12}} {\, +\,}...]
\label{c_13}\ee
Imposing again the minimal coupling, $\ell_s^8\,\widehat{C}_{S^2} \,C_G^3 {\,=\,} \sqrt{G_D}$, we obtain
\be
\widehat{C}_{S^2} = {V_{10-D} \over g_s^2 (\alpha')^5}
\label{c_14}\ee
We can also fix the normalisation of RR vertices
\be
W_{RR}^{(-1/2,-1/2)} = C_{RR} \,F_{RR,AB} S^A \tilde{S}^B e^{-\varphi/2} e^{-\tilde\varphi/2} e^{ik{\cdot}X}
\label{c_15}\ee
from 3-{point} amplitude on the sphere
\be
{\cal M}_{CCG} = \langle c\tilde{c}W_{RR}^{(-1/2,-1/2)}c\tilde{c}W_{RR}^{(-1/2,-1/2)}c\tilde{c}W_G^{(-1,-1)}(3) \rangle = \widehat{C}_{S^2} C_G C_{RR}^2 \ell_s^6 [F_{RR,1}h_3 F_{RR,2} + \dots]
\label{c_16}\ee
Comparing this result with the minimal coupling, $\ell_s^6\,\widehat{C}_{S^2} \,C_G C_{RR}^2 {\,=\,} \sqrt{G_D}$, we find
\be
C_{RR} = \sqrt{G_D\over \alpha'}  = {\ell_D^{D-2\over 2} \over \ell_s} = {g_s (\alpha')^{3/2} \over \sqrt{V_{10-D}}}
\label{c_17}\ee

\subsection{Mixings and interactions in absence of RR and NSNS fluxes}
\label{no_fluxes_computations}

\paragraph{Mixings.}

The simplest amplitude related to the kinetic mixing is the two-{point} `mixing' amplitude of a RR graviphoton on the bulk of a disk with a SM gauge boson. Using the VO's in four dimensions
\begin{align}
W_{RR}^{(-\frac{1}{2},-\frac{1}{2})}&= C_{RR} F^{(\alpha\beta)}_{RR, i} \Gamma^i_{(cd)}
e^{-\varphi/2 -\tilde\varphi/2} S_\alpha \Sigma^c \, \tilde{S}_\beta \tilde\Sigma^d \,e^{ip\,X} \label{c_18} \\
V_{A}^{(-1)}&= C_A e^{-\varphi} a_\mu {\Psi}^\mu e^{i k {\cdot} X (x)}
\label{c_19}
\end{align}
it is straightforward to show that the amplitude vanishes
\be
\langle W^{(-1/2,-1/2)}_{RR}(z,\bar z) V^{(-1)}_A (x) \rangle
 \sim F_{RR,(\alpha \beta)} a_\mu \langle S^{(\alpha}(z)\, \tilde{S}^{\beta)}(\bar{z})\, {\Psi}^\mu(x) \rangle
=0
\label{c_20}
\ee
This result is independent on the internal part and the number of preserved supersymmetries.

In order to obtain a non-vanishing disk contribution, we have to insert an open string scalar on the boundary
\be
V_{\phi}^{(-1)}= C_A  e^{-\varphi} \phi_i \, {\Psi}^i e^{i k {\cdot} X (x)}
\label{c_21}\ee
The amplitude reads
\be
{\cal A}_{V\phi \,C} = \langle cV_A^{(0)}(1) cV_\phi^{(-1)}(2) (c{+}\tilde{c})W_{RR}^{(-1/2,-1/2)} \rangle=
 g_s\, \frac{\ell_s}{\sqrt{{\mathcal{V}_6}}} F_{\mu \nu} F_{RR,i}^{\mu \nu} \phi^i
\label{c_22}
\ee
where we have used the normalisations obtained in \ref{string_normalizations}.

\paragraph{Interaction terms.}

We focus now on the interaction terms. To reproduce two couplings with the Higgs shown in \eqref{2_18} we consider two disk amplitudes, both with insertions of a RR graviphoton in the bulk and two scalars (Higgs) on the boundary, however in the second amplitude we also consider a further insertion of a SM gauge boson. Using the VO's \eqref{c_18}, \eqref{c_19}, \eqref{c_21} and
\be
V_{\phi}^{(0)}= C_A \phi_i (i \partial X^i+ k{\cdot} {\Psi} {\Psi}^i )(x) e^{i k {\cdot} X (x)}
\label{c_23}\ee
it is straightforward to prove that these two amplitudes vanish
\begin{align}
\langle &W^{(-1/2,-1/2)}_{RR}(z,\bar z) V^{(-1)}_\phi (x_1) V^{(0)}_\phi (x_2) \rangle \sim \nonumber\\
& ~~~~~ \sim (F_{RR})_{(\alpha \beta)} k_{3,\rho} \langle S^{(\alpha}(z)\, \tilde{S}^{\beta)}(\bar{z}){\Psi}^\rho(x_3) \rangle
=0
\label{c_24}\\
\langle &W^{(-1/2,-1/2)}_{RR}(z,\bar z) V^{(-1)}_A (x_1) V^{(0)}_\phi (x_2) V^{(0)}_\phi (x_2) \rangle \sim \nonumber\\
& ~~~~~ \sim (F_A)_{\mu \nu} (F_{RR})_{(\alpha \beta)} k_{4,\rho} \langle S^{(\alpha}(z)\, \tilde{S}^{\beta)}(\bar{z})\, {\Psi}^{[\mu}(x_1){\Psi}^{\nu]}(x_1){\Psi}^\rho(x_3) \rangle
=0
\label{c_25}
\end{align}
where we have neglected the internal correlators.

The amplitudes associated to couplings in \eqref{2_18} with SM fermions will not vanish. We start from the dipole coupling. Using the VO's \eqref{c_26} and
\be
V_\l^{(-\frac{1}{2})}= \ell_s C_A u^a_{\gamma}\, e^{-\varphi/2} S^\gamma \Sigma^a \,e^{ik X}\label{c_26}
\ee
where $u$ is the (left-handed) gaugino polarization, {and $\Sigma^a$ are internal parts of the spin fields.}
The dipole amplitude reads
\be
{\cal A}_{F\lambda\lambda} =
\int_0^\infty dx
\langle c(i)\tilde{c}(-i) W_{RR}^{(-\frac{1}{2},-\frac{1}{2})}(i,{-}i) \,
c({\pm} x) V_\l^{(-\frac{1}{2})}(x)\,
V_\l^{(-\frac{1}{2})}({-}x) \rangle \label{c_27}
\ee
The symmetrization of indices $\alpha$, $\beta$ and $c$, $d$ in the RR VO implies that the pairs of indices appearing in the fermionic VO's must be symmetrized too. This is the only possibility to obtain a singlet in the correlators. We show the relevant correlators
\begin{align}
&\< S_{(\alpha}(i) \tilde{S}_{\beta)}(-i) S^\gamma(x) \tilde{S}^\delta(-x) \> =-(-i)^{1/2} (\e^{\alpha \gamma} \e^{\beta \delta}+\e^{\alpha \delta} \e^{\beta \gamma}) \frac{\sqrt{x}}{1+x^2} \label{c_28}\\
&\< \S^{(c}(i) \tilde{\S}^{d)}(-i) \S^{(a}(x) \tilde{\S}^{b)}(-x) \>=
(\Gamma^i)^{(ab)} (\Gamma_i)^{(cd)} [-4i x (1+x^2)^2]^{-1/4} \label{c_29}
\end{align}
Multiplying all the correlators, we obtain
\be
(\Gamma^i)^{(ab)} (\Gamma_i)^{(cd)} \left[\e^{\alpha \gamma} \e^{\beta \delta}+\e^{\alpha \delta} \e^{\beta \gamma}\right] x^{\alpha^\prime k_1 k_2/2} |i - x|^{-1+\alpha^\prime p k_1/2} |i + x|^{-1+\alpha^\prime p k_2/2}
\label{c_30}\ee
In the low-energy limit we can take the leading term in the $\alpha'$-expansion, therefore all the scalar products of momenta disappear and the remaining terms are:
\be
(\Gamma^i)^{(ab)} (\Gamma_i)^{(cd)} \left[\e^{\alpha \gamma} \e^{\beta \delta}{+}\e^{\alpha \delta} \e^{\beta \gamma}\right]\int_0^\infty \frac{dx}{1{+}x^2}
\label{c_31}\ee
Integrating over $x$ and multiplying \eqref{c_31} by the normalizations and the kinematic factors we obtain (up to numerical factors)
\be
\calA_{\calF \l\l} = \frac{g_s}{\sqrt{ {{\mathcal{V}_6}} }} u_{\a, a} F^{\a \b, a b}_{RR} u_{\b, b}
\label{c_32}\ee

The second interaction term with SM fermions in \eqref{2_18} can be computed from the amplitude
\be
\calA= \langle c V^{(0)}_A(x_1)\, c V^{(-1/2)}_{\tilde{\lambda}}(x_2)\, c V^{(-1/2)}_\lambda(x_3) \, \int d^2z\, W^{(-1/2,-1/2)}_{RR}(z,\bar{z})
\rangle
\label{c_33}\ee
Following the same steps, as done for the dipole coupling, we obtain, up to numerical factors, the following result
\be
\calA= \,\frac{ g_s^{3/2} }{\sqrt{ {{\mathcal{V}_6}} }}\, \ell_s^3  F_{RR,(ab)}{\cdot} F_A u_\alpha^a u^{b,\alpha}
\label{c_34} \ee

\subsection{Mixing and interactions in presence of RR and NSNS fluxes}
\label{string_setup_mixing_RR_fluxes}

To compute the relevant amplitudes in presence of bulk fluxes we have to find the VO's for the fluxes. The relevant VO's in ten-dimensions read
\begin{align}
W_\textup{RR,flux}^{(-\frac{1}{2},-\frac{1}{2})}&= C_{RR}e^{-\frac{\varphi(z)}{2}} e^{-\frac{\tilde{\varphi}(\bar{z})}{2}} \calF_{AB} S^A(z) \tilde{S}^B(\bar{z})e^{i K_2 {\cdot} X_L (z)} e^{i \tilde{K}_2 {\cdot} X_R (\bar{z})}
\label{c_35}\\
W_\textup{NSNS,flux}^{(-1,0)}&= C_G \mathcal{H}_{PQR} \left[ e^{-\varphi} \Psi^P
(\tilde{X}^Q \bar{\partial} \tilde{X}^R{+}\tilde{\Y}^Q \, \tilde{\Psi}^R){+}
(L{\leftrightarrow}R) \right] e^{i K_2 {\cdot} X_L} e^{i \tilde{K}_2 {\cdot} X_R}
\label{c_36}
\end{align}
{where $\Psi^M$ are world-sheet fermions and $S^A(z)$, $\tilde{S}^B(\bar{z})$ denote their
spin fields}\footnote{For `twisted' fluxes, vertex operators involve `twist fields' and `twisted spin fields" as well. In the amplitudes we consider their correlators are known and reproduce the same results as the ones for `untwisted' fluxes that we describe here in some details.}.
Notice that we have insert the momenta $K_2$ and $\tilde{K}_2$, which has to be zero for background fields. They are kept to regulate the amplitudes later on. In $D=4$ the VO's are determined by choosing the relevant components for the RR and NSNS fluxes, they read
\begin{align}
{\cal F}_{[AB]} &= {\cal F}_{MNP}\Gamma^{MNP}_{[AB]} \rightarrow
{\cal F}_{ijk}\varepsilon_{\a\b} \Gamma^{ijk}_{(rs)} + ...
={\cal F}_{(rs)} \varepsilon_{\a\b} + ...
\label{c_37}\\
\mathcal{H}_{PQR} & \rightarrow
\mathcal{H}_{ijk} +\dots \qquad, \qquad \mathcal{H}_{(ab)}= \Gamma^{ijk}_{(ab)} \mathcal{H}_{ijk}
\label{c_38}
\end{align}
where `$\dots$' denotes all the terms irrelevant for our analysis.

\paragraph{Mixing in the presence of RR fluxes.}

\begin{figure}[t]
\centering
\includegraphics[width=0.4\textwidth]{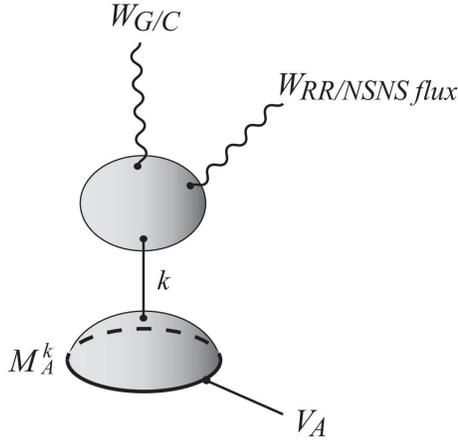}\\
\caption[]{The factorization of the disk amplitude with the insertion of a closed-string $W_{G/C}$ and the flux $W_{RR/NSNS~flux}$ in the bulk and an open-string $V_{A/\phi}$ on the boundary, which exposes a simple pole due to axion exchange.
}
\label{WWVdisk2}
\end{figure}

The relevant amplitude for the kinetic mixing in presence of RR fluxes is
\be
{\cal A}_k= \big\langle c\, V^{(0)}_{A}(1) \, c\tilde{c}\, W^{(-1/2,-1/2)}_{RR, flux}(2,\tilde{2}) \int W^{(-1/2,-1/2)}_{RR}(3,\tilde{3}) \,d^2z_3 \big\rangle
\label{c_39}
\ee
The momenta are taken four-dimensional, for the closed strings we take $K_2=\bar{K}_2=k_2/2$ and $K_3=\bar{K}_3=k_3/2$, where $K_i$ and $k_i$ are the $D=10$ and $D=4$ momenta respectively.

We are interested in the internal components of $\calF_{AB}$ and the four-dimensional field strength of $F_{CD}$, the relevant components in the lower dimension decomposition are
\be
\calF_{AB} \rightarrow \epsilon_{[\alpha \beta]} \calF_{(a b)} \text{ or } \epsilon^{[\dot{\alpha} \dot{\beta}]} \calF^{(a b)} ~~, \qquad
F_{RR,CD} \rightarrow  F_{RR,(\gamma \delta),(cd)} \text{ or } F^{(\dot{\gamma} \dot{\delta}),(cd)}_{RR}
\label{c_40}
\ee
The fact that left and right spin fields in four-dimensions do not correlate forces us to consider two possible combinations:
\be
\calF_{AB} F_{RR,CD} \rightarrow
\epsilon_{[\alpha \beta]} \calF_{(a b)} F^{(\dot{\gamma} \dot{\delta}),(cd)}_{RR}
\text{ or }
\epsilon^{[\dot{\alpha} \dot{\beta}]} \calF^{(a b)} F_{RR,(\gamma \delta),(cd)}
\label{c_41}
\ee
For simplicity we focus on the contributions in which the graviphoton field strength $F_{RR}$ is left handed. Notice also that correlators involving $\partial X$ do not contribute to the amplitude, due to the impossibility of obtaining a Lorentz invariant answer.

The non trivial correlators are the ones with the spin fields in the internal dimensions. To evaluate them we choose specific weights in the representations of spin fields. In this choice we must take into account that the flux polarization $\calF_{(ab)}$ may break supersymmetry. In the complex basis, purely holomorphic and antiholomorphic components ($\calF_{(3,0)}$ and $\calF_{(0,3)}$) break supersymmetry and mixed components ($\calF_{(2,1)}$ and $\calF_{(1,2)}$) do not. We want to select the mixed ones, therefore the weights should have two (one) positive and one (two) negative entries. Finally, using the symmetries of the internal indices $a$, $b$ and $c$, $d$, we choose the following weights:
\begin{align}
&2\< C_{(a}(z_2) \tilde{C}_{b)}(\tilde{z}_2) S^{(c}(z_3) S^{d)}(\tilde{z}_3) \>
= \nn\\
& =
(\delta_a^c \delta_b^d{+}\delta_a^d \delta_b^c)
\< e^{\frac{i}{2} (-\varphi_1+\varphi_2+\varphi_3)}(z_2)
e^{\frac{i}{2} (-\tilde{\varphi}_1+\tilde{\varphi}_2+\tilde{\varphi}_3)}(\tilde{z}_2) e^{\frac{i}{2} (\varphi_1-\varphi_2-\varphi_3)}(z_3)
e^{\frac{i}{2} (\tilde{\varphi}_1-\tilde{\varphi}_2-\tilde{\varphi}_3)}(\tilde{z}_3) \>
\nn\\
&=
(\delta_a^c \delta_b^d{+}\delta_a^d \delta_b^c) \left[\frac{z_{2 \tilde{2}}z_{3 \tilde{3}}}{ z_{23} z_{2 \tilde{3}} z_{ \tilde{2} 3} z_{ \tilde{2} \tilde{3}} } \right]^{\frac{3}{4}}
\label{c_42}
\end{align}
Evaluating the other correlators and setting $x=0$, $z_2=i$ and $\bar{z}_2=-i$ we obtain
\begin{align}
\calA &=
g_s^{3/2}\, \frac{\ell_s^2}{ {{\mathcal{V}_6}} }\, \,
\calF_{a b} F^{\m \n} F_{RR, \mu \nu}^{a b}\label{c_43}\\
&~~~~~\times \int d^2 z_3 \, |i{-}z_3|^{-2+\alpha^\prime k_2{\cdot}k_3/4} |i{+}z_3|^{-2+\alpha^\prime k_2{\cdot}k_3/4} (z_3{-}\bar{z}_3)^{1+\alpha^\prime k_3^2/8} |z_3|^{-2+\alpha^\prime k_1{\cdot}k_3/2} \nn
\end{align}
where $F^{\mu \nu}$ is the field strength of the SM gauge boson. Mapping the upper half plane onto the disk, we obtain
\be
\calA = g_s^{3/2} \frac{\ell_s^2}{ {{\mathcal{V}_6}} }\,
\calF_{a b} F^{\m \n} F_{RR, \mu \nu}^{a b}
\int d^2 w \, |w|^{-2+\frac{\alpha^\prime k_2 {\cdot}k_3}{4}} (1-|w|^2) |1-w|^{-2+\frac{\alpha^\prime k_1 {\cdot}k_{3}}{2}}
~~~~\label{c_44}
\ee
Using the change of variables $w= r e^{i \q}$ and the integrals
\begin{align}
\int_0^{2 \pi} d\theta \, (1+x^2-2 x \cos \q)^{-a}=& 2 \pi
{}_2 F_1 \left(\begin{matrix}
a, &a&\\
~~~~1&&\end{matrix}~;~x^2\right)
\label{c_45}\\
\int_0^1 dx\, x^p (1-x)^q {}_2 F_1 \left(\begin{matrix}
a_1,& a_2&\\
 ~~~~b& &\end{matrix};\,x\right){=}& B(1{\,+\,}p,1{\,+\,}q)
{}_3 F_2 \left(\begin{matrix}
1{+}p,~~ a_1,~~ a_2&&\\
 2{+}p{+}q, ~~ b_1 && \end{matrix}\!\!\!;1\right)
\label{c_46}
\end{align}
we compute the worldsheet integrals obtaining
\begin{align}
\calA &= g_s^{3/2} \frac{\ell_s^2}{ {{\mathcal{V}_6}} }\,
\calF_{a b}  F^{\m \n} F_{RR, \mu \nu}^{a b} \frac{\Gamma(\alpha^\prime k_2{\cdot}k_3/8)}{\Gamma(2+\alpha^\prime k_2 {\cdot}k_3/8)}
\setlength\arraycolsep{1pt}\nn\\
&~~~~~~~~~~~~~~~~~~~~\times {}_3 F_2\left(\begin{matrix}1{-}\alpha^\prime k_1{\cdot}k_3/8,& &1{-}\alpha^\prime k_1{\cdot}k_3/8,& &\quad \alpha^\prime k_2{\cdot}k_3/8\\&1, &2{+}\alpha^\prime k_2{\cdot}k_3/8& &\end{matrix};1\right)~~
\label{c_47}
\end{align}
Expanding the amplitude in $k_2$ we expose a massless pole and a finite contribution:
\be
\calA =
g_s^{3/2} \frac{\ell_s^2}{ {{\mathcal{V}_6}} }\, \calF_{a b}  F^{\m \n} F_{RR, \mu \nu}^{a b} \left[ \frac{16}{\alpha^\prime k_1^2}+1\right]
\label{c_48}
\ee
The $k^2_1$ pole is due to the factorization of the amplitude shown in fig \ref{WWVdisk2} where an axion is propagating with momentum $k_1$. In deriving the effective action, this on-shell pole has to be subtracted since it is already included in the closed-string open-string mixing at disk level. The finite term correspond to the result shown in \eqref{3_23}.

\subsubsection*{Mixing in presence of NSNS fluxes}
\label{string_setup_mixing_NSNS_fluxes}

The mixing amplitude in presence of a NSNS flux is similar to the previous case. The amplitude reads
\be
{\cal A}_k= \big\langle c\, V^{(0)}_{A}(1) \, c\tilde{c}\, W^{(-1/2,-1/2)}_{NSNS, flux}(2,\tilde{2})  \int W^{(-1/2,-1/2)}_{RR}(3,\tilde{3}) \,d^2z_3 \big\rangle
\label{c_49}
\ee
Again the only non trivial correlator is the one between the three worldsheet fermions and two spin fields coming from $W_\textup{NSNS,flux}$ and $W_\textup{RR}$ respectively. We choose again specific weights according to the symmetries of the indices obtaining
\begin{align}
&\< {\Psi}^{[i}(z_2) (\tilde{{\Psi}}^j \tilde{{\Psi}}^{k]})(\bar{z}_2) \S^{(c}(z_3) \tilde{\S}^{d)}(\bar{z}_3)\>= \nn \\
&~~~~~=(\Gamma^{ijk})^{cd}
\< e^{-i\varphi_1}(z_2)
e^{-i\tilde{\varphi}_2}(\bar{z}_2)
e^{\frac{i}{2} (\varphi_1-\varphi_2-\varphi_3)}(z_3)
e^{\frac{i}{2} (\tilde{\varphi}_1-\tilde{\varphi}_2-\tilde{\varphi}_3)}(\tilde{z}_3) \> ~~~~~~~~~~~~~~~~~~~~~\nonumber
\\
&~~~~~= (\Gamma^{ijk})^{cd} \frac{z_{3 \bar{3}}^{3/4}}{z_{23}^{1/2} z_{2\bar{3}}^{1/2} z_{\bar{2}3} z_{\bar{2}\bar{3}}} \label{c_50}
\end{align}
Taking in account the other correlators and mapping the upper half-plane onto the disk, we obtain
\be
\calA=g_s^{3/2}\frac{\ell_s^2}{ {{\mathcal{V}_6}} }
F^{\m \n} F_{RR,\m \n}^{cd} \mathcal{H}_{ijk} \Gamma^{ijk}_{cd}
\int d^2 w \,|w|^{\frac{k_2 {\cdot}k_3}{2}-1} (1-|w|^2)^{-\frac{1}{2}}|1-w|^4 |1+w|^{-2}
\label{c_51} \ee
where we have neglected terms not involving $k_2$ and numerical factors. Integrating $w$, using \eqref{c_45} and \eqref{c_46}, and expanding in $k_2$ we expose a massless pole as done in the RR case. Subtracting the pole we obtain, up to numerical constants, the effective coupling
\be
\calA=g_s^{3/2} \frac{\ell_s^2}{ {{\mathcal{V}_6}} }
F^{\m \n} F_{RR,\m \n}^{cd} \mathcal{H}_{ijk} \Gamma^{ijk}_{cd}
\label{c_52} \ee

\paragraph{Interaction terms.}

In this case we start from the coupling with SM fermions. The relevant amplitudes to consider in order to reproduce the couplings shown in \eqref{2_18} are the following:
\begin{align}
{\cal A}_{\bar{\psi} \hat{F} {\psi}}&{=} \big\langle c\tilde{c} W^{(-\frac{1}{2},-\frac{1}{2})}_{RR, flux}(i,{-}i) \!\int\! W^{(-\frac{1}{2},-\frac{1}{2})}_{RR}(z, {\bar z}) d^2 z\, V^{(-\frac{1}{2})}_{\tilde{\lambda}}(x_1)\, c V^{(+\frac{1}{2})}_\lambda(x_2) \, \big\rangle \label{c_53}\\
{\cal A}_{\hat{F} F \bar{\psi} \psi}&{=} \big\langle c\tilde{c} W^{(-\frac{1}{2},-\frac{1}{2})}_{RR, flux}(i,{-}i) \!\int \! W^{(-\frac{1}{2},-\frac{1}{2})}_{RR}(z, {\bar z}) d^2 z\, V^{(-\frac{1}{2})}_{\tilde{\lambda}}(x_1)\, V^{(+\frac{1}{2})}_\lambda(x_2) \, c V^{(0)}_A(x_3) \big\rangle \label{c_54}
\end{align}
The VO's $W^{(-1/2,-1/2)}_{RR}$ and $V^{(0)}_A$ already contain field strengths, which we expect to appear in the amplitude. In the VO $V_\lambda^{(+1/2)}$ there is an extra momentum coming from the term $k{\cdot} \psi$ (as shown for the kinetic mixings $\partial X$ does not contribute)
\be
[V_\lambda^{(+1/2)}]_{4D} \,= e^{\varphi/2} \lambda_\alpha
(i \partial X^\mu {\,+\,}k{\cdot} {\Psi} {\Psi}^\mu )
\gamma_\mu^{\alpha \dot{\alpha}} C_{\dot{\alpha}} \,e^{ik X}
\label{c_55} \ee
This extra momentum cannot be used to form field strengths, which are already present, therefore in the low-energy limit the amplitudes above are subleading in comparison to the couplings \eqref{2_18}.


The interaction terms involving the Higgses can be computed from
\begin{align}
{\cal A}&= \big\langle c V_{\phi}^{(0)}({-}\infty) c V_{\phi}^{(0)}(1)\! \int\! c W^{(-\frac{1}{2},-\frac{1}{2})}_{RR, flux}(ix,{-}ix) \,dx \!\!\int W^{(-\frac{1}{2},-\frac{1}{2})}_{RR}(z,\bar{z}) d^2z \big\rangle \label{c_56}\\
{\cal A}&= \big\langle c V_{\phi}^{(0)}(-\infty) c V_{\phi}^{(0)}(1) V_{A}^{(0)}(x_1) \!\int\! c W^{(-\frac{1}{2},-\frac{1}{2})}_{R{-}R, flux}(ix_2,{-}ix_2) \,dx\!\! \int \! W^{(-\frac{1}{2},-\frac{1}{2})}_{RR}(z,\bar{z}) d^2z \big\rangle \label{c_57}
\end{align}
Avoiding involved computations, we can estimate the normalizations and the kinematic structures obtaining
\begin{align}
\D S &= \int d^4x~ g_s^2 \frac{\ell_s^4}{ {{\mathcal{V}_6}} } \calF_3 F_{RR}^{\mu \nu} \, \partial_\mu \tilde{\phi}\, \partial_\nu \phi
\label{c_58}\\
\D S &= \int d^4x~ g_s^{5/2} \frac{\ell_s^6}{ {{\mathcal{V}_6}} }\,
\partial_\rho\tilde{\phi}{\cdot}\partial^\rho \phi \,
\calF_3 F_{RR,\mu \nu} F^{\mu \nu}
\label{c_59}
\end{align}

\subsection{Closed-string NSNS graviphotons and their couplings}
\label{NSNSgraviphotons}

NSNS gravi-photons are present in the massless spectrum iff the internal manifold has some non-trivial one-cycle, ie. there is some `untwisted' direction (possibly `shifted' a' la Scherk-Schwarz).

The relevant vertex operators in the canonical $(-1,-1)$, $(0,0)$ and mixed $(0,-1)$ pictures read,
\begin{align}
W^{(-1,-1)}_{NSNS} &= e^{-(\varphi+\tilde\varphi)} \hat{A}^\mu_u \psi_\mu \tilde\chi^u\label{c_60}\\
W^{(0,-1)}_{NSNS} &= (\hat{A}i\partial X + \hat{F}\psi\psi) e^{-\tilde\varphi}\tilde\chi^u \label{c_61}\\
W^{(0,0)}_{NSNS} &= (\hat{A}i\partial X + \hat{F}\psi\psi) (i\bar\partial X^u + k{\cdot}\tilde\psi\tilde\chi^u) \label{c_62}
\end{align}
where $\chi^u$ is an internal world-sheet fermion along some `untwisted' direction $u$ (present in the case of $T^6$, $K3\times T^2$ or shift orbifolds with $N=1,0$ susy).

\paragraph{Zero bulk flux}

The couplings between closed-string NSNS graviphotons (with `untwisted' vertex operator as above) and open-string SM fields are the same as for RR graviphotons (with both `untwisted' and `twisted' vertex operators).

Indicating with $\hat{F}$ the NSNS field strength and barring the $z$ dependence, the relevant correlators are
\begin{gather}
\langle \hat{F}\psi\psi(z) S^{\alpha}(z_1) S^\beta(z_2)\rangle = \hat{F}^{\alpha\beta} \quad , \quad
\langle \hat{F}\psi\psi(z) S^{\alpha}(z_1) S^\beta(z_2) {F}\psi\psi(z_3)\rangle = \varepsilon^{\alpha\beta} \hat{F}^{\mu\nu}F_{\mu\nu} +... \nn \\
\langle \tilde\chi^u(\bar{z}) \Sigma_c(z_1) \Sigma_d(z_2)\rangle = \Gamma^u_{cd} \quad , \quad \langle \tilde\chi^u(\bar{z}) \chi_v (z_2)\rangle = \delta^u{}_v
\label{c_63}
\end{gather}
where the ellipsis represents terms irrelevant in our calculations.

Using the same order as in table \ref{tab_closed_noflux}, one finds the following amplitudes
\begin{itemize}
\item[1)] $\hat{F}^{\mu\nu} D_\mu H D_\nu H^\dagger \leftrightarrow \langle W_{NSNS}^{(0,0)} V^{(-1)}_H V^{(-1)}_{H^\dagger}\rangle = 0$
\item[2)] $\hat{F}^{\mu\nu} \lambda \sigma_{\mu\nu} \lambda \leftrightarrow \langle W^{(0,-1)}_{NSNS}V^{(-1/2)}_\Lambda V^{(-1/2)}_{\Lambda}\rangle = {g_s\ell_s\over \sqrt{\mathcal{V}_6}} \lambda_\alpha \hat{F}^{\alpha\beta} \lambda'_\beta$ \\
where $\lambda_\alpha$ denote an open-string gaugino or a non-chiral fermion with vertex operator containing the internal spin field $\Sigma^{\pm}_u=e^{\pm i \varphi_u/2}$ in the notation where $\chi^u= e^{\pm i \varphi_u}$. Hyperini or (non)chiral fermions, whose vertex operators do not contain the relevant $\Sigma^{\pm}_u$, give zero.

\item[3)] $\hat{F}^{\mu\nu} F_{\mu\nu}HH^\dagger \leftrightarrow \langle W_{NSNS}^{(0,0)} V_A^{(0)}V^{(-1)}_H V^{(-1)}_{H^\dagger}\rangle = 0$
\item[4)] $\hat{F}^{\mu\nu} F_{\mu\nu} \lambda\lambda\leftrightarrow \langle W^{(0,-1)}_{NSNS}V_A^{(0)}V^{(-1/2)}_\Lambda V^{(-1/2)}_{\Lambda}\rangle = {g^{3/2}_s\ell^2_s\over \sqrt{\mathcal{V}_6}} \lambda^\alpha\lambda'_\alpha \hat{F}^{\mu\nu} F_{\mu\nu} $ \\
where, as above, $\lambda_\alpha$ denote an open-string gaugino or a non-chiral fermion.

\item[5)] $\hat{F}^{\mu\nu} F_{\mu\nu} \rightarrow 0$
\item[6)] $\hat{F}^{\mu\nu}_u F_{\mu\nu} \phi^u \leftrightarrow \langle W_{NSNS}^{(0,0)} V^{(-1)}_{\phi^u} V^{(-1)}_{A}\rangle ={g_s\ell^2_s\over \sqrt{\mathcal{V}_6}} \langle\phi^u\rangle \hat{F}^{\mu\nu}_u F_{\mu\nu}$ \\
where $\phi^u$ is an open-string scalar in an $N=2$ vector multiplet or alike along the 'untwisted' direction $u$.
\end{itemize}

\paragraph{Non-Zero (RR) bulk flux}

Even in the presence of bulk fluxes, the interactions of a NSNS graviphoton are essentially identical to those of RR photons.

Barring the $z$ dependence, the crucial correlators are
\be
<\tilde\chi^u(\bar{z}) \Sigma^\dagger(w)\tilde\Sigma^\dagger(\bar{w})> = \Gamma^u_{c,c'}
\label{c_64}
\ee
where $\Sigma^\dagger(w), \tilde\Sigma^\dagger(\bar{w})$ are internal spin fields appearing in the bulk RR flux or
\be
<\tilde\chi^u(\bar{z}) \Sigma^\dagger(w)\tilde\Sigma^\dagger(\bar{w}) \Phi(z_3)\Phi'(z_4)> =
\Gamma^u_{c,c'} \delta_{34}
\label{c_65}
\ee
where $\Phi(z_3), \Phi'(z_4)$ are dimension $h=1/2$ world-sheet fields appearing in the definition of the vertex operators of open-string scalars (Higgs or alike).

Indicating with ${\cal F}_3$ the 3-form flux and $\hat{F}$ the NSNS field strength, we obtain (using the same order as in table \ref{tab_closed_withflux})
\begin{itemize}

\vskip 1cm

\item[1)] ${\cal F}_3 \hat{F}^{\mu\nu} D_\mu H D_\nu H^\dagger ~~~\leftrightarrow~~~ \langle W_{NSNS}^{(0,0)} W_{flux}^{(-1/2,-1/2)}V^{(-1)}_H V^{(0)}_{H^\dagger}\rangle ~~=~~ {g^2_s \ell^4_s {\cal F}_3 \over \mathcal{V}_6}
\hat{F}^{\mu\nu} k_\mu k'_\nu $

\vskip 1cm

\item[2)] ${\cal F}_3\hat{F}^{\mu\nu} \lambda \sigma_{\mu\nu} \lambda ~~~\leftrightarrow~~~ \langle W^{(0,0)}_{NSNS}W_{flux}^{(-1/2,-1/2)}V^{(-1/2)}_\Lambda V^{(-1/2)}_{\Lambda}\rangle ~~=~~ {\rm subleading}$ \\
where `sub-leading' means vanishing at this order in $g_s$ (disk level) and in $\ell_s$ (derivative expansion). One cannot exclude higher order corrections at one loop or beyond or higher derivative corrections.

\vskip 1cm

\item[3)] ${\cal F}_3\hat{F}^{\mu\nu} F_{\mu\nu}HH^\dagger ~~~\leftrightarrow~~~ \langle W_{NSNS}^{(0,0)} W_{flux}^{(-1/2,-1/2)} V_A^{(0)}V^{(-1)}_H V^{(0)}_{H^\dagger}\rangle ~~=~~ {g^{5/2}_s \ell^6_s {\cal F}_3 \over \mathcal{V}_6}
\hat{F}^{\mu\nu} F_{\mu\nu} kk'$

\vskip 1cm

\item[4)] ${\cal F}_3\hat{F}^{\mu\nu} F_{\mu\nu} \lambda\lambda~~~\leftrightarrow~~~ \langle W^{(0,0)}_{NSNS}W_{flux}^{(-1/2,-1/2)}V_A^{(0)}V^{(-1/2)}_\Lambda V^{(-1/2)}_{\Lambda}\rangle ~~=~~ {\rm subleading} $ \\
where `subleading' has the same meaning as above.

\vskip 1cm

\item[5)] ${\cal F}_3\hat{F}^{\mu\nu} F_{\mu\nu} ~~~\leftrightarrow~~~ \langle W_{NSNS}^{(0,0)} W_{flux}^{(-1/2,-1/2)}V^{(-1)}_A\rangle ~~=~~ {g^{3/2}_s \ell^2_s {\cal F}_3 \over \mathcal{V}_6}
\hat{F}^{\mu\nu} F_{\mu\nu} $

\vskip 1cm

\end{itemize}

The case of non-zero NSNS bulk fluxes is very similar and gives essentially the same results. Therefore, we shall not describe here it in any detail.

\subsection{Mixing between visible and dark open-string vectors}
\label{string_setup_mixing_scalars}

\paragraph{Mixing contribution from scalar messengers.}

The amplitude can be obtained from the disk 4-{point} gauge boson amplitude in ten dimensions with a dimensional reduction of the four gauge boson amplitude. The color-ordered amplitude is the following
\be
\calA[A_{ab},\tilde{\phi}_{bi},A_{ij},{\tilde{\phi}}_{ja}]{=}{\,-}\ell_s^4 g_{SM} g_{H} \phi{\cdot} \tilde{\phi} \left[2 a_1 {\cdot} k_2 a_3 {\cdot} k_4 t {\,+\,}2 a_1 {\cdot} k_4 a_3 {\cdot} k_2 s{\,+\,}a_1 {\cdot} a_3 s t \right]\frac{\Gamma(-\alpha^\prime s) \Gamma(-\alpha^\prime t)}{\Gamma(1+\alpha^\prime u)}
\label{c_60a}
\ee
In the limit $k_2, k_4\rightarrow 0$ only one term survives:
\be
\calA[A_{ab},\tilde{\phi}_{bi},A_{ij},{\tilde{\phi}}_{ja}]=- g_{SM} g_{H} \phi{\cdot} \tilde{\phi}\, a_1 {\cdot} a_3=- g_s \phi{\cdot} \tilde{\phi}\, a_1 {\cdot} a_3
\label{c_61a}
\ee

\paragraph{Mixing at one loop.}

The one-loop contribution for the kinetic mixing of the vis/dark photons can be extracted from a 2-{point} `amplitude' with the insertion on gauge vector bosons on the two boundaries of the annulus. To have a non zero result we have to relax momentum conservation. The result of this computation can be found in \cite{MBConsoli,Bianchi:2006nf} (in the helicity basis) and it reads
\begin{align}
{\cal A}^{\rm one-loop}_{a,i} &= \langle V_{A_a}^{(0)} \, V_{A_i}^{(0)} \rangle= \label{c_62a}\\
&= g_s \ell_s^4 {\rm tr} (Q_a) {\rm tr} (Q_i) F_a^{\pm}(k_1) F_i^{\pm}(k_2) \int_0^\infty T dT [ {\cal E}_{\cal N} \pm i {\cal C}_{{\cal N}}] \int_0^1 d{{z}} \, e^{ \ap k_1{\cdot}k_2 \mathcal{G}({{z}})}\nonumber
\end{align}
where $F^{\pm}(k)$ denote complex (anti)self-dual field strengths, $\mathcal{G}({{z}})$ is the bosonic propagator on the annulus, ${\cal N}$ denotes the number of SUSY preserved in the $(a,i)$ `messenger' sector. The function ${\cal E}_{\cal N}$ is defined as
\be
{\cal E}_{\cal N} = -\sum_{\a=2,3,4} c_\a e_{\a-1} {\cal Z}^{\cal N}_\alpha
\label{c_63a}
\ee
where the sum over $\alpha$ is the sum over the even spin structures, $c_\a$ the GSO projection coefficients, $e_{\alpha-1}(\tau)=4\pi i \partial_\tau \ln [\theta_{\alpha+1}(0|\tau)/\eta (\tau)]$ and ${\cal Z}^{\cal N}_\alpha$ is the one-loop partition function. For ${\cal N} = 4,2,1$ the function ${\cal E}_{\cal N}$ reduces to
\be
{\cal E}_{{\cal N} = 4} = 0 \quad , \quad
{\cal E}_{{\cal N} = 2} = { \pi^2 \Lambda_{a,i}^\parallel I^\perp_{a,i} \over 4 n_\textup{orb} (\alpha' T)^2} \quad , \quad
{\cal E}_{{\cal N} = 1} = { \pi^2 I_{a,i} \over 4 n_\textup{orb} (\alpha' T)^2 }{{\cal H}' \over {\cal H}}(0)
\label{c_64a}\ee
where ${\cal H}(z) = \prod_{I=1}^3 \theta_1(z+ u^I_{a,i})$ with $u^I_{a,i}$ denoting the `twist' in the $(a,i)$ sector while $I_{a,i}$ and $\Lambda_{a,i}$ are the number of intersections and the lattice sum respectively.

The function ${\cal C}_{\cal N}$ is the contribution of the odd spin structure, which is relevant only in ${\cal N}=1$ sectors due to uncancelled fermionic zero-modes in ${\cal N}=4,2$ sectors.
\be
{\cal C}_{{\cal N} =1} = {\pi^2 I_{a,i} \over 4 n_\textup{orb} (\alpha' T)^2}
\label{c_65a}
\ee

In the IR limit $T \to \infty$ the contributions from ${\cal E}_{\cal N}$ and ${\cal C}_{{\cal N}}$ are dominated by the lightest particles circulating in the loop, which are the messengers in our case. The lightest messengers have masses depending on the branes separation, $M = \Delta x/\alpha'$, therefore ${\cal E}_{\cal N}$ and ${\cal C}_{{\cal N}}$ reduce to
\be
{\cal E}_{\cal N},{\cal C}_{{\cal N}} \sim \frac{e^{-\ap \pi M^2 T}}{(\alpha' T)^2}
\label{c_66a}\ee
In the IR limit the amplitude \eqref{3_33} becomes
\be
{\cal A}^{\rm one-loop}_{a,i} \sim g_s{\rm tr} (Q_a) {\rm tr} (Q_i) F_a^{\pm}(k_1) F_i^{\pm}(k_2) \int_0^\infty \frac{dT}{T}\, \int_0^1 d{{z}}\, e^{-\pi \ap [M^2 -k_1{\cdot}k_2 {{z}}(1-{{z}})]}+ \dots
\label{c_67a}\ee
Regarding $T$ as a Schwinger parameter, the integral over $T$ can be written as an integral over a four-dimensional loop momentum and ${{z}}$ assumes the role of a Feynman parameter:
\be
{\cal A}^{\rm one-loop}_{a,i} \sim g_s{\rm tr} (Q_a) {\rm tr} (Q_i) F_a^{\pm}(k_1) f_F^{\pm}(k_2) \int d^4 \ell \, \int_0^1 d{{z}}\, \frac{1}{[\ell^2 +k^2 {{z}}(1-{{z}})+M^2]^2}+ \dots
\label{c_68a}\ee
Integrating the loop momentum using the cut-off regularization we read the leading contribution of the messengers at low energy
\be
{\cal A}^{\rm one-loop}_{a,i} \sim g_s{\rm tr} (Q_a) {\rm tr} (Q_i) F_a^{\pm}(k_1) F_i^{\pm}(k_2) \log \frac{\Lambda^2}{M^2}
\label{c_69a}\ee

\paragraph{Interaction terms.}

As for the case of closed-string graviphotons in presence of fluxes, fermionic couplings with the SM are subleading at low energy, in comparison to the couplings in \eqref{2_18}. The reason is basically the same, the amplitudes to consider are the followings
\begin{align}
{\cal A}_{A \l \l}&=\langle V_{A_{ij}}^{(0)} \, V_{\lambda_{ia}}^{(-1/2)} V_{\lambda_{aj}}^{(+1/2)} \rangle \label{c_70a}\\
{\cal A}_{A \widehat{A} \l \l}&=\langle V_{A_{ij}}^{(0)} \, V_{\lambda_{ia}}^{(-1/2)} V_{\lambda_{bj}}^{(+1/2)} V_{A_{ab}}^{(0)} \rangle \label{c_71a}
\end{align}
The field strengths of the gauge fields are already present in the VO's $V^{(0)}_A$. The fermionic VO in the picture $+1/2$ carries the term $k{\cdot}{\Psi}$, which yields an extra momentum in the amplitude in comparison to the couplings \eqref{2_18}. For instance, if we consider the amplitude \eqref{c_68a}, in the IR expansion, we obtain
\be
{\cal A}^{\rm one-loop}_{a,i} \sim \, \frac{g_s^{3/2} \ell_s}{M^2}
 {\rm tr} (Q_a) {\rm tr} (T_i T_j)
F^{[\mu \nu]} \lambda_\alpha \gamma^{\alpha \dot{\alpha}}_{[\mu} k_{\nu]} \bar{\lambda}_{\dot{ \alpha}}
\label{c_72a}
\ee
which is an higher derivative term compared to $\bar{\psi} F^{\mu \nu} \gamma_{\mu \nu} \bar{\psi}$.
This is agreement with the field theory calculation presented in section \ref{holosetup}.
The absence of the leading (in derivatives) dipole coupling is related to the masslessness of the fermions.

We now focus on the amplitudes involving the Higgses. To obtain the effective coupling $D_\mu H^\dagger D_\nu H \hat{F}^{\mu \nu}$ we consider the 3-{point} amplitude on the annulus with two scalars in one boundary and a `dark photon' in the other boundary. We found the following amplitude (\cite{MBConsoli,Bianchi:2006nf} for more details)
\begin{align}
{\cal A}^{\rm one-loop}_{a,ij} \sim & \,g_s^{3/2} \ell_s^6 {\rm tr} (Q_a) {\rm tr} (T_i T_j) k_1{\cdot}k_2 \varphi{\cdot}\widetilde{\varphi}
\int_0^\infty \, dT [ {\cal E}_{\cal N} \pm i {\cal C}_{{\cal N}}]
\int dz_1\,dz_2
\label{c_73a}\\
&(a{\cdot} k_1 \partial_{z_1} \mathcal{G}(z_1)-a{\cdot} k_2 \partial_{z_2} \mathcal{G}(z_2))
e^{\ap k_1{\cdot} k_2 \mathcal{G}(z_{12})+\ap k_1{\cdot} k_3 \mathcal{G}(z_1)+\ap k_2{\cdot} k_3 \mathcal{G}(z_2)} \nonumber
\end{align}
where $z_{1,2}=\frac{1}{2}+i \tau_2 \nu_{1,2}$ are the positions of the scalars while the position of the gauge field in the other boundary can be set to zero, $z_3=0$. The positions $\nu_{1}$ and $\nu_{2}$ can be replaced by three parameters $\b_i$ which can be interpreted as Feynman parameters:
\be
\nu_1= 1-\b_1 \qquad \nu_2=1-\b_2 \qquad \nu_1-\nu_2=\b_3
\label{c_74a}\ee
Using this parametrization and taking the limit $T \to \infty$, the amplitude becomes
\begin{align}
{\cal A}^{\rm one-loop}_{a,ij} \sim & \,g_s^{3/2} \alpha' {\rm tr} (Q_a) {\rm tr} (T_i T_j) k_1{\cdot}F{\cdot}k_2 \varphi{\cdot}\widetilde{\varphi}
\int_0^\infty \, dT
\int_0^1 d\b_1\,d\b_2 \,d\b_3 \, \delta(1-\sum \b_i) \nonumber\\
& \b_3 \, e^{-\pi T\ap [M^2+k_2{\cdot} k_3 \b_1(1-\b_1)+ k_1{\cdot} k_3 \b_2(1-\b_2)+k_1{\cdot} k_2 \b_3(1-\b_3)]} \label{c_75a}
\end{align}
Integrating over $T$ and neglecting $k_i{\cdot}k_j$ terms, which are subleading compared to $M^2$, we obtain
\be
{\cal A}^{\rm one-loop}_{a,ij} \sim \, \frac{g_s^{3/2}}{M^2} {\rm tr} (Q_a) {\rm tr} (T_i T_j) k_1{\cdot}F{\cdot}k_2 \varphi{\cdot}\widetilde{\varphi}
\label{c_76a} \ee
This term can be easily translated in an interaction term in the effective action
\be
\Delta S = \frac{g_s^{3/2}}{M^2}  {\rm tr} (Q_a) {\rm tr} (T_i T_j) \int d^4 x \, F^{[\mu \nu]} \der_{[\mu} \varphi^i \der_{\nu]} \widetilde{\varphi}_i
\label{c_77a}\ee

The term corresponding to the coupling $H^\dagger H \hat{F}_{\mu \nu} F^{\mu \nu}$ can be obtained from four-point amplitudes with the dark photon on one boundary and two SM scalars and one gauge boson in the other boundary. Following steps similar to the ones taken for the previous amplitudes we can estimate the result without fully computing the amplitude
\be
{\cal A}_{a,i}\sim \frac{g_s^2}{M^4} \, F^{a,\mu \nu} F_{i,\mu \nu} \, \partial_\rho\phi{\cdot}\partial^\rho\tilde{\phi}
\label{c_78a}\ee

\section{The effective dark photon interactions in the holographic setup\label{gau12}}

We consider the bulk plus brane action
 \be
S_{total}=S_{bulk}+S_{brane}
\label{n1}\ee
\be
S_{bulk}=M_P^3\int d^5x\sqrt{g}\left[{Z\over 4} F_{\hat A,mn}F_{\hat A}^{mn}+{\cal J}^{m}\hat A_m\dots\right]
\label{n2}\ee
\be
S_{brane}=\int dz\delta(z-z_0)\int d^4x\sqrt{\gamma}\left[{m_{\hat A}^2\over 4}\widehat{F}_{\m\n}\widehat{F}^{\m\n}+{m_{\hat A}\chi\over 2}\widehat{F}_{\m\n}F_{Y}^{\m\n}+{1\over 4g_Y^2}F_{Y}^2+Y^{\m}J_{\m}+\cdots\right]
\label{n3}\ee
where $m,n$ are five-dimensional indices and $\m,\n$ are four-dimensional indices. $\hat A_{m}$ is the bulk gauge field, dual to a global conserved current of the hidden holographic theory. $\hat J_{m}$ is the bulk current coupled to $\hat A_{m}$ and obtaining contributions from fields dual to operators that are charged under the global symmetry.
$\widehat{A}_{\m}(x)\equiv A_{\m}(z_0,x)$ is the induced gauge-field on the brane and $\widehat{F}_{\m\n}$ its field-strength.
$Y^{\m}$ is the SM hypercharge gauge field, $F_Y$ its four-dimensional field strength and $J_{\m}$ is the gauge invariant SM hypercharge current.
$\widehat A_{m}$ is dimensionless, while the hypercharge $Y_{\m}$ has the canonical mass dimension one. Therefore $m_{\hat A},M_P$ are mass scales while $\chi$ and $g_Y$ are dimensionless. $Z$ is a function of various scalar fields, some of them having a non-trivial radial profile in the fifth (holographic) dimension. For all practical purposes we can therefore consider $Z$ as a function of $z$.

The ellipsis in the bulk action stands for other terms involving different fields.
The ellipsis in the brane action stands for the SM model fields. We have added explicitly a possible mixing terms of $\widehat A_{\m}$ to the hypercharge gauge boson and assumed that no SM particle is charged under $\widehat A_{\m}$.

We would now like to calculate the induced interaction on the hypercharge current $J_{\m}$ and the bulk (hidden) current ${\cal J}_{m}$ generalizing the calculation in \cite{h}.

The classical equations for the two gauge fields at quadratic order read
\be
\pa_{m}\left(Z\sqrt{g}g^{mr}g^{ns}\hat F_{rs}\right)-{\delta(z-z_0)\over M_P^3}\pa_{\m}\left[\sqrt{\gamma}\gamma^{\m \r}\gamma^{n\s}(m_{\hat A}^2\widehat{F}_{\r\s}+m_{\hat A}\chi F^Y_{\r\s})\right]= \sqrt{g}~g^{ns}{\cal J}_{s}
\label{n4}\ee
\be
\pa_{\m}\left[\sqrt{\gamma}\gamma^{\m \r}\gamma^{\n\s}(m_{\hat A}\chi\widehat{F}_{\r\s}+{1\over g_Y^2} F^Y_{\r\s})\right]=\sqrt{\gamma}~\gamma^{\n\r}J_{\r}
\label{n5}\ee
while the standard conservations of currents read
\be
\pa_{m}(\sqrt{g}g^{ms}{\cal J}_s)=\pa^{\m}{\cal J}_{\m}+\pa_{z}{\cal J}_z+3A'{\cal J}_z=0\sp \pa_{\m}J^{\m}=0
\label{n7}\ee

We shall pick a gauge that fixes all gauge degrees of freedom for $\hat A_{m}$, namely the five-dimensional Coulomb gauge, while we pick the Lorentz gauge for $Y_{\m}$. This  amounts to imposing the conditions
\be
 \pa^{\m}\hat A_{\m}=0\sp \pa^{\m}Y_{\m}=0\eqp
\label{n6}\ee
where the indices above are raised with the Minkowski (boundary) metric.
We consider a general bulk metric in conformal coordinates with Poincar\'e invariance
\be
ds^2=e^{2A(z)}(dz^2+dx_{\m}dx^{\m})
\label{n8}
\ee
and the equations in (\ref{n4}) and (\ref{n5}) become
\be
\square_4 A_z={e^{2A}\over Z}{\cal J}_{z}
\label{n10}\ee
\be
\square_{5}\hat A_{\m}-{1\over M_P^3~Z_0 e^{A_0}}(m_{\hat A}^2\square_4 \widehat A_{\m}+m_{\hat A}\chi \square_4 Y_{\m})\delta(z-z_0)={e^{2A}\over Z}\tilde J_{\m}
\label{n9}\ee
\be
(m_{\hat A}\chi\square_4 \widehat A_{\m}+{1\over g_Y^2}\square_4 Y_{\m})=e^{2A_0}J_{\m}
\label{n13}\ee
where
\be
\square_{5}\equiv \pa_z^2+\left(A'+{Z'\over Z}\right)\pa_z+\square_4\sp \square_4\equiv \pa_{\m}\pa^{\m}
\label{n14}\ee
$\square_4$ is the flat four-dimensional Laplacian.
\be
Z_0\equiv Z(z_0)\sp A_0\equiv A(z_0)
\label{n20}\ee
The current
\be
\tilde J_{\m}\equiv {\cal J}_{\m}+\square_{4}^{-1}\pa_{\m}\pa_z {\cal J}_z+3A'\square_4^{-1}\pa_{\m}{\cal J}_z
\label{n11}\ee
is the ``transverse" current satisfying
\be
\pa^{\m}\tilde J_{\m}=\pa^{\m}{\cal J}_{\m}+\pa_{z}{\cal J}_z+3A'{\cal J}_z=0
\label{n12}\ee
The Green's functions are defined as usual
\be
\square_4^{-1}F(z,x)\equiv \int d^4x' G_4(x,x')F(z,x')\sp \square_4 G(x,x')=\delta^{(4)}(x-x')\sp \int d^4x \delta^{(4)}(x-x')=1
\label{n15}\ee
\be
\square_5^{-1}F(z,x)\equiv \int dz'd^4x' G_5(x,z;x',z')F(z',x')\sp \square_5 G(x,z;x',z')=\delta(z-z')\delta^{(4)}(x-x')
\label{n16}\ee
For future convenience we shall rescale the coordinates, $z\to e^{-A_0}z$, $x^{\mu}\to e^{-A_0}x^{\mu}$ in order to reduce the induced metric at $z=z_0$ to the Minkowski metric $\eta_{\m\n}$. The equations (\ref{n10}), (\ref{n9}) and (\ref{n13}) become
\be
\square_4 \hat A_z={e^{2\tilde A}\over Z}{\cal J}_z\sp \tilde A(z)=A(z)-A_0
\label{nn10}\ee
\be
\square_{5}\hat A_{\m}-{1\over M_P^3~Z_0}(m_{\hat A}^2\square_4 \widehat A_{\m}+m_{\hat A}\chi \square_4 Y_{\m})\delta(z-z_0)={e^{2\tilde A}\over Z}\tilde J_{\m}
\label{nn9}\ee
\be
(m_{\hat A}\chi\square_4 \widehat A_{\m}+{1\over g_Y^2}\square_4 Y_{\m})=J_{\m}
\label{nn13}\ee
while (\ref{n14}), (\ref{n11}) and (\ref{n12}) remain as they are.

Fourier transforming (\ref{n9}) in the $x^{\mu}$ coordinates we obtain
\be
\square_{5}\hat A_{\m}(z,p)+{1\over M_P^3~Z_0}(m_{\hat A}^2 \widehat A_{\m}(z_0,p)+m_{\hat A}\chi Y_{\m}(p))p^2\delta(z-z_0)={e^{2\tilde A}\over Z}\tilde J_{\m}(z,p)
\label{n17}\ee
\be
m_{\hat A}\chi \widehat A_{\m}(z_0,p)+{1\over g_Y^2}Y_{\m}(p)=-{J_{\m}(p)\over p^2}
\label{n18}\ee
Substituting $Y_{\m}$ from (\ref{n18}) in (\ref{n17}) we finally obtain
\be
\square_{5}\hat A_{\m}(z,p)+{m_{\hat A}^2(1-\chi^2 g_Y^2) \over M_P^3~Z_0 } \widehat A_{\m}(z_0,p)p^2\delta(z-z_0)={e^{2\tilde A}\over Z}\tilde J_{\m}(z,p)+{m_{\hat A}\chi g_{Y}^2\over M_P^3 Z_0 }J^{\m}(p)\delta(z-z_0)\eqp
\label{n19}\ee
The solution for the bulk and brane gauge fields is given by
\be
\hat A_z(z,p)=-{e^{2\tilde A}\over Z}{{\cal J}_z(z,p)\over p^2}
\label{n21}\ee
\be
\hat A_{\m}(z,p)=\int dz'{e^{2\tilde A(z')}\over Z(z')}{G_5(z,z',p)\tilde J_{\m}(z',p)\over 1+{m_{\hat A}^2(1-\chi^2 g_Y^2) \over M_P^3 Z_0}p^2 G_5(z,z_0,p)}+{{m_{\hat A}\chi g_Y^2 \over M_P^3Z_0} G_5(z,z_0,p)J_{\m}(p)\over 1+{m_{\hat A}^2(1-\chi^2 g_Y^2) \over M_P^3Z_0 }p^2 G_5(z,z_0,p)}
\label{n22}\ee
where $G_5(z,z',p)$ is the solution of (\ref{n16}) in momentum space
\be
\square_5 G_5(z,z',p)=\delta(z-z')
\label{n23}\ee
and Neumann boundary conditions at the AdS boundary.
We also obtain
\be
Y_{\m}(p)=-g_Y^2{J^{\m}(p)\over p^2}{{1 \over m_{\hat A}^2p^2}+{G_5(z_0,z_0,p)\over M_P^3 Z_0}\over {1\over m_{\hat A}^2 p^2}+{(1-\chi^2 g_Y^2)G_5(z_0,z_0,p)\over M_P^3Z_0 } }-
\label{n24}\ee
$$
-m_{\hat A}\chi g_Y^2\int dz'{e^{2\tilde A(z')}\over Z(z')}{G_5(z_0,z',p)\over 1+{m_{\hat A}^2(1-\chi^2 g_Y^2)G_5(z_0,z_0,p)p^2\over M_P^3 Z_0}}\tilde J_{\m}(z',p)
$$

where
\be
\tilde J^{\m}(z,p)={\cal J}^{\m}(z,p)-i{p^{\m}\over p^2}\left[\pa_z {\cal J}_z+3A'{\cal J}_z\right]
\label{n25}\ee

The interactions induced between the currents ${\cal J}^{m}$ and $J_{\m}$ are given by
\be
S_{int}={M_P^3\over 2}\int d^5x\sqrt{g} {\cal J}^m A_m+{1\over 2}\int d^4x\sqrt{\gamma}J^{\m}Y_{\m}=
\label{n26}\ee
$$
={1\over 2}\int {d^4p\over (2\pi)^4}J_{\m}(-p)G_{44}(p)J^{\m}(p)+{1\over 2}\int dz\int {d^4p\over (2\pi)^4}J_{\m}(-p)G_{45}(z,p){\cal J}^{\m}(z,p)+
$$
$$
+{1\over 2}\int dz\int dz'{d^4p\over (2\pi)^4}{\cal J}_m(z,-p)G^{mn}_{55}(z,z',p){\cal J}_{n}(z',p)
$$
where the interaction kernels can be deduced from the classical solutions in
(\ref{n21}), (\ref{n22}), (\ref{n24}),

\be
G_{44}(p)=-{g_Y^2\over p^2}{{1\over m_{\hat A}^2p^2}+{G_5(z_0,z_0,p)\over M_P^3 Z_0}\over {1\over m_{\hat A}^2p^2}+{(1-\chi^2 g_Y^2)G_5(z_0,z_0,p)\over M_P^3Z_0}}
\label{n27}\ee
\be
G_{45}(z,p)=\left({e^{3\tilde A(z)}~G_5(z,z_0,p)\over Z_0}-{e^{2\tilde A(z)}\over Z(z)}G_5(z_0,z,p)\right) {m_{\hat A}\chi g_Y^2\over 1+{m_{\hat A}^2(1-\chi^2 g_Y^2)G_5(z_0,z_0,p)p^2\over M_P^3 Z_0}}
\label{n28}\ee

\be
G_{55}^{zz}(z,z',p)={M_P^3 e^{5\tilde A(z')}\over Z(z')}{1\over p^2}\sp G_{55}^{z\m}(z,z',p)=0
\label{n29}\ee
\be
G_{55}^{\m\n}(z,z',p)={M_P^3 e^{5\tilde A(z')}\over Z(z')}{G_{5}(z,z',p)\over 1+{m_{\hat A}^2(1-\chi^2 g_Y^2) \over M_P^3Z_0 }p^2 G_5(z,z_0,p)}\left(\eta^{\m\n}-{p^{\m}p^{\n}\over p^2}\right)
\label{n30}\ee
where we assumed that the hypercharge current is conserved, $p^{\m}J_{\m}(p)=0$.

The induced interactions of the two sets of charges (hypercharge on the brane, global charge in the bulk), contain two novel effects. One is due to the induced kinetic term for the bulk gauge field on the brane. This phenomenon is akin to the DGP phenomenon, \cite{dgp}, and has been analyzed in \cite{u1}.
The second is the brane mixing of the bulk photon and the hypercharge, proportional to $\chi$. Its main effect is the interaction between hypercharge and bulk charge.

To track the different effects we first set $m=0$, turning off both effects.
then (\ref{n27})-(\ref{n30}) become

\be
G_{44}(p)=-{g_Y^2\over p^2}\sp G_{45}(z,p)=0
\label{n27a}\ee
\be
G_{55}^{zz}(z,z',p)={M_P^3 e^{5\tilde A(z')}\over Z(z')}{1\over p^2}\sp G_{55}^{z\m}(z,z',p)=0
\label{n28a}\ee
\be
G_{55}^{\m\n}(z,z',p)={M_P^3 e^{5\tilde A(z')}\over Z(z')}{G_{5}(z,z',p)}\left(\eta^{\m\n}-{p^{\m}p^{\n}\over p^2}\right)
\label{n29a}\ee
ans show a standard four-dimensional vector (hypercharge) interaction in Lorentz gauge, and a bulk-bulk vector interaction in Coulomb gauge.
The detailed properties of the bulk interaction depend significantly on the background bulk metric. However, at short enough distances, $z\to z'$ and $p\to\infty$, the five-dimensional scalar propagator $G_5$ is that of flat-five dimensional space.
Assuming that the bulk theory has a single mass scale $E_0$, then
\be
G_5(z,z,p)\simeq \left\{ \begin{array}{lll}
\displaystyle ~~~~{z\over L}~{1\over 2p},&\phantom{aa} &p\gg E_0\\ \\
\displaystyle d_0(z)+{p^2\over E_0^2} d_2(z)+{\cal O}\left({p^4\over E_0^4}\right) ,&\phantom{aa}&p\ll E_0.
\end{array}\right\}
 \ee
where $L$ is the UV AdS scale.

Turning-on the induced kinetic term on the brane, without mixing ($m_{\hat A}\not= 0$, $\chi=0$) we obtain instead
\be
G_{44}(p)=-{g_Y^2\over p^2}\sp G_{45}(z,p)=0
\label{n27b}\ee
\be
G_{55}^{zz}(z,z',p)={M_P^3 e^{5\tilde A(z')}\over Z(z')}{1\over p^2}\sp G_{55}^{z\m}(z,z',p)=0
\label{n29b}\ee
\be
G_{55}^{\m\n}(z,z',p)={M_P^3 e^{5\tilde A(z')}\over Z(z')}{G_{5}(z,z',p)\over 1+{m_{\hat A}^2\over M_P^3Z_0 }p^2 G_5(z,z_0,p)}\left(\eta^{\m\n}-{p^{\m}p^{\n}\over p^2}\right)
\label{n30b}\ee
As expected, $m_{\hat A}\not=0$ changes the bulk interaction. For charges at the position of the brane the interaction has the features found for similar interactions mediated by the graviton in \cite{self} and the emergent axion in \cite{axion}. In particular, with sources at the position of the brane, the bulk photon interaction is five dimensional both a large enough, and small enough distances. Depending on scales, there may be an intermediate five-dimensional regime or not, \cite{self}.

Turning on mixing, without an induced kinetic term, $m_{\hat A}\to 0$ with $m_{\hat A}\chi=\tilde\chi\not=0$ and fixed we obtain instead,

\be
G_{44}(p)=-{g_Y^2\over p^2}{1\over 1-{\tilde \chi^2 g_Y^2~G_5(z_0,z_0,p)p^2\over M_P^3Z_0}}
\label{n27c}\ee
\be
G_{45}(z,p)=\left({e^{3\tilde A(z)}~G_5(z,z_0,p)\over Z_0}-{e^{2\tilde A(z)}\over Z(z)}G_5(z_0,z,p)\right) {\tilde\chi g_Y^2\over 1-{\tilde\chi^2 g_Y^2~G_5(z_0,z_0,p)p^2\over M_P^3 Z_0}}
\label{n28c}\ee

\be
G_{55}^{zz}(z,z',p)={M_P^3 e^{5\tilde A(z')}\over Z(z')}{1\over p^2}\sp G_{55}^{z\m}(z,z',p)=0
\label{n29c}\ee
\be
G_{55}^{\m\n}(z,z',p)={M_P^3 e^{5\tilde A(z')}\over Z(z')}{G_{5}(z,z',p)\over 1-{\tilde \chi^2 g_Y^2\over M_P^3Z_0 }p^2 G_5(z,z_0,p)}\left(\eta^{\m\n}-{p^{\m}p^{\n}\over p^2}\right)
\label{n30c}\ee
The expected effect is the interaction between hypercharge and bulk charge in (\ref{n28c}). However, both at low and high momenta, the two vector interactions become ghost-like because ${\tilde \chi^2 g_Y^2\over M_P^3Z_0 }p^2 G_5(z,z_0,p)$ becomes larger than one. Unitarity in fact implies that when $\chi\not=0$ then $m\not=0$ and we are in the original case, (\ref{n27})-(\ref{n30}). In that case, we obtain normal positivity
if
\be
\chi^2 g_Y^2<1
\ee
as expected.

The effect of the induced kinetic term due to the SM corrections and the mixing term for the current-current interactions in (\ref{n26}), have as consequences
\begin{itemize}

\item The hypercharge interaction in (\ref{n27}) is modified. However, in the far UV or far IR it is unchanged to leading order.
  At intermediate distances, there is a new DGP-like resonance mediating the hypercharge interaction with strength proportional to $\chi^2$.
This, however, is heavy. Taking into account the estimates
\be
M_P^3\sim {\cal O}(N^2)\sp Z_0,m_{\hat A}\sim {\cal O}(1)\sp \chi\sim {\cal O}(N^{-1})
\ee
we obtain that the mass scales as ${\cal O}(N^2)$.

\item The interaction of hypercharge via (\ref{n28}) with the bulk global charge is present due to the non-trivial mixing. It is doubly weak: once because of the smallness of $\chi$ and second because of the large-N suppression of bulk interactions.

\item The bulk gauge interaction in (\ref{n30}) is modified both at short and long distances and becomes, in particular, four-dimensional at the position of the brane.

\item Overall, the fact that the dark photon arises in a holographic hidden sector provides additional suppression to the effects of its mixing to hypercharge.
\end{itemize}

\end{appendix}

\newpage



\begin{thebibliography}{99}


\bibitem{Leike}
A.~Leike,
{\em ``The Phenomenology of extra neutral gauge bosons,''}
\href{www.doi.org/10.1016/S0370-1573(98)00133-1}{Phys. Rept. \textbf{317} (1999), 143-250};
\hre{hep-ph}{9805494}.

\bibitem{Rizzo}
T.~G.~Rizzo,
{\em ``$Z^\prime$ phenomenology and the LHC,''}
\hre{hep-ph}{0610104}.

\bibitem{Contino}
R.~Contino,
{\em ``$Z^{\prime}, Z_{KK}, Z^{*}$ and all that: Current bounds and theoretical prejudices on heavy neutral vector bosons,''}
Nuovo Cim. B \textbf{123} (2008), 511-515
\hri{0804.3195}{[hep-ph]}.

\bibitem{Lang}
P.~Langacker,
{\em``The Physics of Heavy $Z^\prime$ Gauge Bosons,''}
Rev. Mod. Phys. \textbf{81} (2009), 1199-1228
doi:10.1103/RevModPhys.81.1199
\hri{0801.1345}{[hep-ph]}.



\bibitem{Bianchi:1990yu}
M.~Bianchi and A.~Sagnotti,
{\em``On the systematics of open string theories,''}
\href{https://doi.org/10.1016/0370-2693(90)91894-H}{Phys. Lett. B \textbf{247} (1990), 517-524}.

\bibitem{Bianchi:1990tb}
M.~Bianchi and A.~Sagnotti,
{\em``Twist symmetry and open string Wilson lines,''}
\href{https://doi.org/10.1016/0550-3213(91)90271-X}{Nucl. Phys. B \textbf{361} (1991), 519-538}.

\bibitem{Bianchi:1991eu}
M.~Bianchi, G.~Pradisi and A.~Sagnotti,
{\em``Toroidal compactification and symmetry breaking in open string theories,''}
\href{https://doi.org/10.1016/0550-3213(92)90129-Y}{Nucl. Phys. B \textbf{376} (1992), 365-386}.

\bibitem{Angelantonj:1996uy}
C.~Angelantonj, M.~Bianchi, G.~Pradisi, A.~Sagnotti and Y.~Stanev,
{\em ``Chiral asymmetry in four-dimensional open string vacua,''}
Phys. Lett. B \textbf{385} (1996), 96-102
\hre{hep-th}{9606169}.



\bibitem{AKT}
  I.~Antoniadis, E.~Kiritsis and T.~N.~Tomaras,
  {\em ``A D-brane alternative to unification,''}
  Phys.\ Lett.\ B {\bf 486} (2000) 186;
  \hre{hep-ph}{0004214};\\
  {\em ``D-brane standard model,''}
  Fortsch.\ Phys.\  {\bf 49} (2001) 573;
  \hre{hep-th}{0111269};\\
  I.~Antoniadis, E.~Kiritsis, J.~Rizos and T.~N.~Tomaras,
  {\em ``D-branes and the standard model,''}
  Nucl.\ Phys.\ B {\bf 660} (2003) 81;
  \hre{hep-th}{0210263}.

\bibitem{Ib}
  G.~Aldazabal, L.~E.~Ibanez, F.~Quevedo and A.~M.~Uranga,
  {\em ``D-branes at singularities: A Bottom up approach to the string embedding of the standard model,''}
  JHEP {\bf 0008} (2000) 002;
  \hre{hep-th}{0005067}.



\bibitem{rev}
  E.~Kiritsis,
{\em ``D-branes in standard model building, gravity and cosmology,''}
Phys. Rept. \textbf{421} (2005), 105-190
doi:10.1016/j.physrep.2005.09.001
\hri{hep-th}{0310001};\\
 {\em ``D-branes in standard model building, gravity and cosmology,''}, Section 8.6.
\href{www.doi.org/10.1002/prop.200310120}{Fortsch. Phys. \textbf{52} (2004) no.2-3, 200-263};
  \hre{hep-th}{0310001v1}.



\bibitem{AKR}
  I.~Antoniadis, E.~Kiritsis and J.~Rizos,
  {\em ``Anomalous U(1)s in type-I superstring vacua,''}
  Nucl.\ Phys.\ B {\bf 637} (2002) 92
  \hre{hep-th}{0204153}.



\bibitem{muon}
E.~Kiritsis and P.~Anastasopoulos,
{\em ``The Anomalous magnetic moment of the muon in the D-brane realization of the standard model,''}
JHEP \textbf{05} (2002), 054
\hre{hep-ph}{0201295}.


\bibitem{AL}
L.~A.~Anchordoqui, I.~Antoniadis, K.~Benakli and D.~Lust,
{\em ``Anomalous $U(1)$ Gauge Bosons as Light Dark Matter in String Theory,''}
\hri{2007.11697}{[hep-th]}.


\bibitem{ADKS}
  P.~Anastasopoulos, T.~P.~T.~Dijkstra, E.~Kiritsis and A.~N.~Schellekens,
  {\em ``Orientifolds, hypercharge embeddings and the Standard Model,''}
  Nucl.\ Phys.\ B {\bf 759} (2006) 83;
  \hre{hep-th}{0605226}.



\bibitem{Bachas} E. Kiritsis, {\em Orientifolds, And The Search For The Standard Model
  In String Theory}, \\ Published in
  C.~Bachas, L.~Baulieu, M.~Douglas, E.~Kiritsis, E.~Rabinovici, P.~Vanhove, P.~Windey
  and L.~F.~Cugliandolo,
  {\em ``String theory and the real world: From particle physics to astrophysics. Proceedings, Summer School in Theoretical Physics, 87th Session, Les Houches, France, July 2-27, 2007,''}\\
\href{http://hep.physics.uoc.gr/~kiritsis/papers4/leshouches-2007.pdf}{Les Houches \textbf{87} (2008), page 46-125}.



\bibitem{anasta}
  P.~Anastasopoulos,
  {\em ``4-D anomalous U(1)'s, their masses and their relation to 6-D anomalies,''}
  JHEP {\bf 0308} (2003) 005
  \hre{hep-th}{0306042}.



\bibitem{IQ}
L.~E.~Ibanez and F.~Quevedo,
{\em ``Anomalous U(1)'s and proton stability in brane models,''}
JHEP \textbf{10} (1999), 001
\hre{hep-ph}{9908305}.



\bibitem{irges}
  C.~Coriano, N.~Irges and E.~Kiritsis,
  {\em ``On the effective theory of low scale orientifold string vacua,''}
  Nucl.\ Phys.\ B {\bf 746} (2006) 77
  \hre{hep-ph}{0510332}.

\bibitem{Anastasopoulos:2008jt}
  P.~Anastasopoulos, F.~Fucito, A.~Lionetto, G.~Pradisi, A.~Racioppi and Y.~S.~Stanev,
  {\em ``Minimal Anomalous U(1)-prime Extension of the MSSM,''}
  Phys.\ Rev.\ D {\bf 78}, 085014 (2008)
  \hri{0804.1156}{[hep-th]}.

\bibitem{AbelGoodsell}
  S.~A.~Abel, M.~D.~Goodsell, J.~Jaeckel, V.~V.~Khoze and A.~Ringwald,
  {\em ``Kinetic Mixing of the Photon with Hidden U(1)s in String Phenomenology,''}
  JHEP {\bf 0807}, 124 (2008)
  \hri{0803.1449}{[hep-ph]}.



\bibitem{bianchi}
  P.~Anastasopoulos, M.~Bianchi, E.~Dudas and E.~Kiritsis,
  {\em ``Anomalies, anomalous U(1)'s and generalized Chern-Simons terms,''}
  JHEP {\bf 0611} (2006) 057
  \hre{hep-th}{0605225}.



\bibitem{Abel}
S.~Abel and B.~Schofield,
{\em ``Brane anti-brane kinetic mixing, millicharged particles and SUSY breaking,''}
Nucl. Phys. B \textbf{685} (2004), 150-170
\hre{hep-th/0311051}{[hep-th]}.

\bibitem{Wit}
M.~Bullimore, J.~P.~Conlon and L.~T.~Witkowski,
{\em ``Kinetic mixing of U(1)s for local string models,''}
\href{www.doi.org/10.1007/JHEP11(2010)142}{JHEP \textbf{11} (2010), 142};
\hri{1009.2380}{[hep-th]}.



\bibitem{Camara}
P.~G.~Camara, L.~E.~Ibanez and F.~Marchesano,
{\em ``RR photons,''}
JHEP \textbf{09} (2011), 110
\hri{1106.0060}{[hep-th]}.



\bibitem{Goodsell}
M.~Goodsell, S.~Ramos-Sanchez and A.~Ringwald,
{\em ``Kinetic Mixing of U(1)s in Heterotic Orbifolds,''}
JHEP \textbf{01} (2012), 021
\hri{1110.6901}{[hep-th]}.

\bibitem{Marchesano}
F.~Marchesano, D.~Regalado and G.~Zoccarato,
{\em ``U(1) mixing and D-brane linear equivalence,''}
JHEP \textbf{08} (2014), 157
\hri{1406.2729}{[hep-th]}.



\bibitem{Fayet}
P.~Fayet,
{\em ``On the Search for a New Spin 1 Boson,''}
\href{www.doi.org/10.1016/0550-3213(81)90122-X}{Nucl. Phys. B \textbf{187} (1981), 184-204}.

\bibitem{Fayet2}
P.~Fayet,
{\em ``U-boson production in e+ e- annihilations, psi and Upsilon decays, and Light Dark Matter,''}
Phys. Rev. D \textbf{75} (2007), 115017
\hre{hep-ph}{0702176}.

\bibitem{Hambye}
T.~Hambye,
{\em ``Hidden vector dark matter,''}
JHEP \textbf{01} (2009), 028
\hri{0811.0172}{[hep-ph]}.

\bibitem{Pospelov}
M.~Pospelov,
{\em ``Secluded U(1) below the weak scale,''}
Phys. Rev. D \textbf{80} (2009), 095002
\hri{0811.1030}{ [hep-ph]}.

\bibitem{Yann}
O.~Lebedev, H.~M.~Lee and Y.~Mambrini,
{\em ``Vector Higgs-portal dark matter and the invisible Higgs,''}
\href{www.doi.org/10.1016/j.physletb.2012.01.029}{Phys. Lett. B \textbf{707} (2012), 570-576};
\hri{1111.4482}{[hep-ph]}.

\bibitem{Graham}
P.~W.~Graham, J.~Mardon and S.~Rajendran,
{\em ``Vector Dark Matter from Inflationary Fluctuations,''}
\href{www.doi.org/10.1103/PhysRevD.93.103520}{Phys. Rev. D \textbf{93} (2016) no.10, 103520}
\hri{1504.02102}{[hep-ph]}.

\bibitem{dark1}
M.~Raggi and V.~Kozhuharov,
{\em ``Results and perspectives in dark photon physics,''}
\href{https://s3.cern.ch/inspire-prod-files-2/235d83cdb212f5e1fd393b72579d89e3}{Riv. Nuovo Cim. \textbf{38} (2015) no.10, 449-505}.

\bibitem{ship}
S.~Alekhin et al.
{\em ``A facility to Search for Hidden Particles at the CERN SPS: the SHiP physics case,''}
\href{www.doi.org/10.1088/0034-4885/79/12/124201}{Rept. Prog. Phys. \textbf{79} (2016) no.12, 124201};
\hri{1504.04855}{[hep-ph]}.


\bibitem{Reece}
M.~Reece,
{\em ``Photon Masses in the Landscape and the Swampland,''}
\hrj{10.1007/JHEP07(2019)181}{JHEP \textbf{07} (2019), 181};
\hri{1808.09966}{ [hep-th]}.


\bibitem{Deli}
M.~Deliyergiyev,
{\em ``Recent Progress in Search for Dark Sector Signatures,''}
\href{www.doi.org/10.1515/phys-2016-0034}{Open Phys. \textbf{14} (2016) no.1, 281-303};
\hri{1510.06927}{[hep-ph]}.

\bibitem{Alex}
J.~Alexander et al.
{\em ``Dark Sectors 2016 Workshop: Community Report,''}
\hri{1608.08632}{[hep-ph]}.

\bibitem{Beach}
J.~Beacham et al.
{\em ``Physics Beyond Colliders at CERN: Beyond the Standard Model Working Group Report,''}
\href{www.doi.org/10.1088/1361-6471/ab4cd2}{J. Phys. G \textbf{47} (2020) no.1, 010501};
\hri{1901.09966}{[hep-ex]}.

\bibitem{Sirunyan:2019xst}
A.~M.~Sirunyan \textit{et al.} [CMS],
{\em ``Search for dark photons in decays of Higgs bosons produced in association with Z bosons in proton-proton collisions at $ \sqrt{s} $ = 13 TeV,''}
JHEP \textbf{10} (2019), 139
\hri{1908.02699}{[hep-ex]}.

\bibitem{Fabbri}
M.~Fabbrichesi, E.~Gabrielli and G.~Lanfranchi,
{\em ``The Dark Photon,''}
\hri{2005.01515}{[hep-ph]}.



\bibitem{h}
B.~Holdom,
{\em ``Two U(1)'s and Epsilon Charge Shifts,''}
\href{www.doi.org/10.1016/0370-2693(86)91377-8}{Phys. Lett. B \textbf{166} (1986), 196-198}.



\bibitem{Heck}
M.~Del Zotto, J.~J.~Heckman, P.~Kumar, A.~Malekian and B.~Wecht,
{\em ``Kinetic Mixing at Strong Coupling,''}
Phys. Rev. D \textbf{95} (2017) no.1, 016007
\hri{1608.06635}{[hep-ph]}.



\bibitem{u1}
P.~Betzios, E.~Kiritsis, V.~Niarchos and O.~Papadoulaki,
{\em ``Global symmetries, hidden sectors and emergent (dark) vector interactions,''}
\hri{2006.01840}{ [hep-ph]}.

\bibitem{Contino}
R.~Contino,
{\em ``The Higgs as a Composite Nambu-Goldstone Boson,''}
\hrj{10.1142/9789814327183\_0005}{TASI 2009, 235-306};
\hrj{1005.4269}{ [hep-ph]}.

\bibitem{Barbieri}
R.~Barbieri, S.~Rychkov and R.~Torre,
{\em ``Signals of composite electroweak-neutral Dark Matter: LHC/Direct Detection interplay,''}
\hrj{10.1016/j.physletb.2010.04.010}{Phys. Lett. B \textbf{688} (2010), 212-215};
\hri{1001.3149}{[hep-ph]}.

\bibitem{Gh}
T.~Gherghetta and B.~von Harling,
{\em ``A Warped Model of Dark Matter,''}
\hrj{10.1007/JHEP04(2010)039}{JHEP \textbf{04} (2010), 039};
\hri{1002.2967}{ [hep-ph]}.

\bibitem{Morriss}
K.~L.~McDonald and D.~E.~Morrissey,
{\em ``Low-Energy Signals from Kinetic Mixing with a Warped Abelian Hidden Sector,''}
\hrj{10.1007/JHEP02(2011)087}{JHEP \textbf{02} (2011), 087};
\hri{1010.5999}{ [hep-ph]}.

\bibitem{Dienes}
K.~R.~Dienes, C.~F.~Kolda and J.~March-Russell,
{\em ``Kinetic mixing and the supersymmetric gauge hierarchy,''}
Nucl. Phys. B \textbf{492} (1997), 104-118
\hre{hep-ph/9610479}{[hep-ph]}.


\bibitem{1}
  E.~Kiritsis,
  {\em ``Gravity and axions from a random UV QFT,''}
  \href{http://www.epj-conferences.org/articles/epjconf/abs/2014/08/epjconf_icnfp2013_00068/epjconf_icnfp2013_00068.html}{EPJ Web Conf.\  {\bf 71} (2014) 00068}
  \hri{1408.3541}{[hep-ph]}.

\bibitem{grav}
P.~Betzios, E.~Kiritsis and V.~Niarchos,
{\em ``Emergent gravity from hidden sectors and TT deformations,''}
\hri{2010.04729}{[hep-th]}.



\bibitem{self}
  C.~Charmousis, E.~Kiritsis and F.~Nitti,
  {\em ``Holographic self-tuning of the cosmological constant,''}
  JHEP {\bf 1709} (2017) 031
  \hri{1704.05075}{[hep-th]};\\
Y.~Hamada, E.~Kiritsis, F.~Nitti and L.~T.~Witkowski,
{\em ``The self-tuning of the cosmological constant and the holographic relaxion,''}
\hri{2001.05510}{[hep-th]}.



\bibitem{axion}
P.~Anastasopoulos, P.~Betzios, M.~Bianchi, D.~Consoli and E.~Kiritsis,
{\em ``Emergent/Composite axions,''}
\href{www.doi.org/10.1007/JHEP10(2019)113}{JHEP \textbf{19} (2020), 113};
\hri{1811.05940}{[hep-ph]}.


\bibitem{Anastasopoulos:2020gbu}
P.~Anastasopoulos, K.~Kaneta, Y.~Mambrini and M.~Pierre,
{\em ``Energy-Momentum portal to dark matter and emergent gravity,''}
Phys. Rev. D \textbf{102}, no.5, 055019 (2020)
\hri{2007.06534}{ [hep-ph]}.







\bibitem{Bianchi:2001kw}
  M.~Bianchi, D.~Z.~Freedman and K.~Skenderis,
  {\em ``Holographic renormalization,''}
  Nucl.\ Phys.\ B {\bf 631} (2002) 159
  \hre{hep-th}{0112119}.

\bibitem{Bianchi:2001de}
  M.~Bianchi, D.~Z.~Freedman and K.~Skenderis,
  {\em ``How to go with an RG flow,''}
  JHEP {\bf 0108} (2001) 041
  \hre{hep-th}{0105276}.



\bibitem{flavour}
J.~Erdmenger, N.~Evans, I.~Kirsch and Threlfall,
{\em ``Mesons in Gauge/Gravity Duals - A Review,''}
Eur. Phys. J. A \textbf{35} (2008), 81-133
\hri{0711.4467}{[hep-th]}.



\bibitem{del}
  A.~Delgado, J.~R.~Espinosa and M.~Quiros,
  {\em ``Unparticles Higgs Interplay,''}
  JHEP {\bf 0710} (2007) 094
  \hri{0707.4309}{[hep-ph]}.


\bibitem{AD}
  I.~Antoniadis and S.~Dimopoulos,
  {\em ``Splitting supersymmetry in string theory,''}
  Nucl.\ Phys.\ B {\bf 715} (2005) 120;
  \hre{hep-th}{0411032}.


\bibitem{mald1}
J.~M.~Maldacena,
{\em ``The large N limit of superconformal field theories and supergravity,''}
Adv.\ Theor.\ Math.\ Phys.\  {\bf 2} (1998) 231
[Int.\ J.\ Theor.\ Phys.\  {\bf 38} (1999) 1113]
\hre{hep-th}{9711200}.






\bibitem{Ibanez:2006da}
L.~Ibanez and A.~Uranga,
{\em ``Neutrino Majorana Masses from String Theory Instanton Effects,''}
JHEP \textbf{03} (2007), 052
\hre{hep-th}{0609213}.

\bibitem{Ibanez:2007rs}
L.~Ibanez, A.~Schellekens and A.~Uranga,
{\em ``Instanton Induced Neutrino Majorana Masses in CFT Orientifolds with MSSM-like spectra,''}
JHEP \textbf{06} (2007), 011
\hri{0704.1079}{[hep-th]}.

\bibitem{Addazi:2014ila}
A.~Addazi and M.~Bianchi,
{\em ``Neutron Majorana mass from exotic instantons,''}
JHEP \textbf{12} (2014), 089
\hri{1407.2897}{[hep-ph]}.

\bibitem{Addazi:2015hka}
A.~Addazi and M.~Bianchi,
{\em ``Neutron Majorana mass from Exotic Instantons in a Pati-Salam model,''}
JHEP \textbf{06} (2015), 012
\hri{1502.08041}{[hep-ph]}.

\bibitem{Addazi:2015yna}
A.~Addazi, M.~Bianchi and G.~Ricciardi,
{\em ``Exotic see-saw mechanism for neutrinos and leptogenesis in a Pati-Salam model,''}
JHEP \textbf{02} (2016), 035
\hri{1510.00243}{[hep-ph]}.



\bibitem{Coriano:2007fw}
C.~Coriano, N.~Irges and S.~Morelli,
{\em ``Stuckelberg axions and the effective action of anomalous Abelian models. 1. A Unitarity analysis of the Higgs-axion mixing,''}
JHEP \textbf{07} (2007), 008
\hre{hep-ph}{0701010}.

\bibitem{Coriano:2007xg}
N.~Irges, C.~Coriano and S.~Morelli,
{\em ``Stuckelberg Axions and the Effective Action of Anomalous Abelian Models 2. A SU(3)C x SU(2)W x U(1)Y x U(1)B model and its signature at the LHC,''}
Nucl. Phys. B \textbf{789} (2008), 133-174
\hre{hep-ph}{0703127}.

\bibitem{Fucito:2008ai}
F.~Fucito, A.~Lionetto, A.~Mammarella and A.~Racioppi,
{\em ``Stueckelino dark matter in anomalous U(1)-prime models,''}
Eur. Phys. J. C \textbf{69} (2010), 455-465
\hri{0811.1953}{[hep-ph]}.



\bibitem{Angelantonj:2000hi}
C.~Angelantonj, I.~Antoniadis, E.~Dudas and A.~Sagnotti,
{\em ``Type I strings on magnetized orbifolds and brane transmutation,''}
Phys. Lett. B \textbf{489} (2000), 223-232
\hre{hep-th}{0007090}.

\bibitem{Larosa:2003mz}
M.~Larosa and G.~Pradisi,
{\em ``Magnetized four-dimensional Z(2) x Z(2) orientifolds,''}
Nucl. Phys. B \textbf{667} (2003), 261-309
\hre{hep-th}{0305224}.

\bibitem{Dudas:2005jx}
E.~Dudas and C.~Timirgaziu,
{\em ``Internal magnetic fields and supersymmetry in orientifolds,''}
Nucl. Phys. B \textbf{716} (2005), 65-87
\hre{hep-th}{0502085}.

\bibitem{Bianchi:2005yz}
M.~Bianchi and E.~Trevigne,
{\em``The Open story of the magnetic fluxes,''}
JHEP \textbf{08} (2005), 034
\hre{hep-th}{0502147}.

\bibitem{Bianchi:2005sa}
M.~Bianchi and E.~Trevigne,
{\em``Gauge thresholds in the presence of oblique magnetic fluxes,''}
JHEP \textbf{01} (2006), 092
\hre{hep-th}{0506080}.

\bibitem{Collinucci:2014qfa}
A.~Collinucci and R.~Savelli,
{\em ``T-branes as branes within branes,''}
JHEP \textbf{09} (2015), 161
\hri{1410.4178}{[hep-th]}.



\bibitem{book}
E.~Kiritsis,
{\em ``String theory in a nutshell,''}, Second Edition.
   \href{https://press.princeton.edu/books/hardcover/9780691155791/string-theory-in-a-nutshell}{ Published by Princeton University Press,  (2019)  888 p.}



\bibitem{Anastasopoulos:2011hj}
P.~Anastasopoulos, M.~Bianchi and R.~Richter,
{\em ``Light stringy states,''}
JHEP \textbf{03}, 068 (2012)
\hri{1110.5424}{[hep-th]}.

\bibitem{Anastasopoulos:2013sta}
P.~Anastasopoulos, M.~D.~Goodsell and R.~Richter,
{\em``Three- and Four-point correlators of excited bosonic twist fields,''}
JHEP \textbf{10}, 182 (2013)
\hri{1305.7166}{[hep-th]}.

\bibitem{Anastasopoulos:2014lpa}
P.~Anastasopoulos and R.~Richter,
{\em``Production of light stringy states,''}
JHEP \textbf{12}, 059 (2014)
\hri{1408.4810}{[hep-th]}.

\bibitem{Anastasopoulos:2016yjs}
  P.~Anastasopoulos, M.~Bianchi and D.~Consoli,
  {\em ``Yukawa's of light stringy states,''}
  Fortsch.\ Phys.\  {\bf 65}, no. 1, 1600110 (2017)
  \hri{1609.09299}{[hep-th]}.



\bibitem{Minahan}
J.~A.~Minahan,
{\em``One Loop Amplitudes on Orbifolds and the Renormalization of Coupling Constants,''}
\href{https://doi.org/10.1016/0550-3213(88)90303-3}{Nucl. Phys. B \textbf{298}, 36-74 (1988)}.

\bibitem{Kaplunovsky:1985yy}
  V.~S.~Kaplunovsky,
  {\em ``Mass Scales of the String Unification,''}
  \href{https://doi.org/10.1103/PhysRevLett.55.1036}{Phys.\ Rev.\ Lett.\  {\bf 55} (1985) 1036}.

\bibitem{Kaplunovsky:1987rp}
  V.~S.~Kaplunovsky,
  {\em ``One Loop Threshold Effects in String Unification,''}
  \href{https://doi.org/10.1016/0550-3213(88)90526-3}{Nucl.\ Phys.\ B {\bf 307} (1988) 145}   Erratum: [\href{https://doi.org/10.1016/0550-3213(92)90193-F}{Nucl.\ Phys.\ B {\bf 382} (1992) 436}].

\bibitem{Klapuetc}
  L.~J.~Dixon, V.~Kaplunovsky and J.~Louis,
  {\em `Moduli dependence of string loop corrections to gauge coupling constants,''}
  \href{https://doi.org/10.1016/0550-3213(91)90490-O}{Nucl.\ Phys.\ B {\bf 355} (1991) 649}.



\bibitem{MBConsoli}
  M.~Bianchi and D.~Consoli,
  {\em ``Simplifying one-loop amplitudes in superstring theory,''}
  JHEP {\bf 1601}, 043 (2016)
  \hri{1508.00421}{[hep-th]}.

\bibitem{Bianchi:2006nf}
M.~Bianchi and A.~V.~Santini,
{\em ``String predictions for near future colliders from one-loop scattering amplitudes around D-brane worlds,''}
JHEP \textbf{12}, 010 (2006)
\hri{0607224}{[hep-th]}


\bibitem{witten}
  E. Witten,
  {\em ``Current Algebra Theorems For The U(1) Goldstone Boson,"} Nucl.
  \href{https://doi.org/10.1016/0550-3213(79)90031-2}{Phys. {\bf B 156},(1979) 269}.

\bibitem{diss}
  E.~Kiritsis,
  {\em ``Dissecting the string theory dual of QCD,''}
  Fortsch.\ Phys.\  {\bf 57} (2009) 396
  \hri{0901.1772}{[hep-th]}.



\bibitem{Napoli}
M.~De Napoli [HPS],
{\em ``The HPS experiment at JLab,''}
\href{https://s3.cern.ch/inspire-prod-files-2/233780ea478f3818abd26e90191a255a}{EPJ Web Conf. \textbf{142} (2017), 01011}.

\bibitem{Gherghetta}
T.~Gherghetta, J.~Kersten, K.~Olive and M.~Pospelov,
{\em ``Evaluating the price of tiny kinetic mixing,''}
Phys. Rev. D \textbf{100} (2019) no.9, 095001
\hri{1909.00696}{[hep-ph]}.

\bibitem{Strassler}
M.~J.~Strassler and K.~M.~Zurek,
{\em ``Echoes of a hidden valley at hadron colliders,''}
Phys. Lett. B \textbf{651} (2007), 374-379
\hre{hep-ph}{0604261}.

\bibitem{Acharya}
B.~S.~Acharya and M.~Torabian,
{\em ``Supersymmetry Breaking, Moduli Stabilization and Hidden U(1) Breaking in M-Theory,''}
Phys. Rev. D \textbf{83} (2011), 126001
\hri{1101.0108}{[hep-th]}.

\bibitem{dgp}
G.~R.~Dvali, G.~Gabadadze and M.~Porrati,
{\em ``4-D gravity on a brane in 5-D Minkowski space,''}
\hrj{10.1016/S0370-2693(00)00669-9}{Phys. Lett. B \textbf{485} (2000), 208-214};
\hre{hep-th}{0005016}.

\bibitem{Luk}
J.~K.~Ghosh, E.~Kiritsis, F.~Nitti and L.~T.~Witkowski,
{\em ``Back-reaction in massless de Sitter QFTs: holography, gravitational DBI action and f(R) gravity,''}
\hrj{10.1088/1475-7516/2020/07/040}{JCAP \textbf{07} (2020), 040};
\hri{2003.09435}{ [hep-th]}.



\bibitem{RS}
L.~Randall and R.~Sundrum,
{\em ``A Large mass hierarchy from a small extra dimension,''}
\hrj{10.1103/PhysRevLett.83.3370}{Phys. Rev. Lett. \textbf{83} (1999), 3370-3373};
\hre{hep-ph}{9905221};\\
{\em ``An Alternative to compactification,''}
\hrj{10.1103/PhysRevLett.83.4690}{Phys. Rev. Lett. \textbf{83} (1999), 4690-4693};
\hre{hep-th}{9906064}.


\end{thebibliography}
\end{document}